\documentclass[reprint,aps,prx]{revtex4-2}

\newcommand{\be}{\begin{equation}}
\newcommand{\ee}{\end{equation}}
\newcommand{\w}{\omega}
\newcommand{\wLR}{\omega_{LR}}

\newcommand{\e}{\epsilon}

\newcommand{\eref}[1]{Eq.~(\ref{#1})}

\newcommand{\qp}{\mathrm{qp}}

\newcommand{\dL}{\Delta_L}
\newcommand{\dR}{\Delta_R}
\newcommand{\xL}{x_L}

\newcommand{\xRp}{x_{R>}}
\newcommand{\xRm}{x_{R<}}

\usepackage{hyperref} 
\usepackage{graphicx}
\usepackage{amsmath, amsthm, amssymb}
\usepackage{epstopdf}
\usepackage{epsfig}
\usepackage{epsf}
\usepackage{array}
\usepackage{mathtools}
\usepackage{circuitikz}
\usepackage{dcolumn}
\usepackage[tight]{subfigure}
\usepackage[normalem]{ulem}
\usepackage{verbatim}
\usepackage{inputenc}
\usepackage{color}
\usepackage{bm} 
\usepackage{natbib}
\usepackage{xspace}
\usepackage{mathbbol}
\usepackage{cleveref}
\usepackage{mathcomp}

\begin{document}

\title{Quasiparticles in superconducting qubits with asymmetric junctions}

\author{Giampiero Marchegiani}
\email{giampiero.marchegiani@tii.ae}
\affiliation{Quantum Research Centre, Technology Innovation Institute, Abu Dhabi, UAE}

\author{Luigi~Amico}
\affiliation{Quantum Research Center, Technology Innovation Institute, Abu Dhabi, UAE}
\affiliation{Centre for Quantum Technologies, National University of Singapore, 3 Science Drive 2, Singapore 117543}
\affiliation{INFN-Sezione di Catania, Via S. Sofia 64, 95127 Catania, Italy}
\affiliation{LANEF `Chaire d’excellence’, Universit\'{e} Grenoble-Alpes \& CNRS, F-38000 Grenoble, France}
\affiliation{MajuLab, CNRS-UNS-NUS-NTU International Joint Research Unit, UMI 3654, Singapore}

\author{Gianluigi Catelani}
\affiliation{JARA Institute for Quantum Information (PGI-11),Forschungszentrum J\"ulich, 52425 J\"ulich, Germany}
\affiliation{Quantum Research Centre, Technology Innovation Institute, Abu Dhabi, UAE}

\date{\today}

\setlength{\arraycolsep}{2pt}

\begin{abstract}
  Designing the spatial profile of the superconducting gap -- gap engineering -- has long been recognized as an effective way of controlling quasiparticles in superconducting devices. In aluminum films, their thickness modulates the gap; therefore, standard fabrication of Al/AlOx/Al Josephson junctions, which relies on overlapping a thicker film on top of a thinner one, always results in gap-engineered devices. Here we reconsider quasiparticle effects in superconducting qubits to explicitly account for the unavoidable asymmetry in the gap on the two sides of a Josephson junction. We find that different regimes can be encountered in which the quasiparticles have either similar densities in the two junction leads, or are largely confined to the lower-gap lead. Qualitatively, for similar densities the qubit's excited state population is lower but its relaxation rate higher than when the quasiparticles are confined; therefore, there is a potential trade-off between two desirable properties in a qubit.
\end{abstract}

\date{\today}


\maketitle

\section{Introduction}
\label{sec:intro}
\begin{figure}
    \centering
    \includegraphics{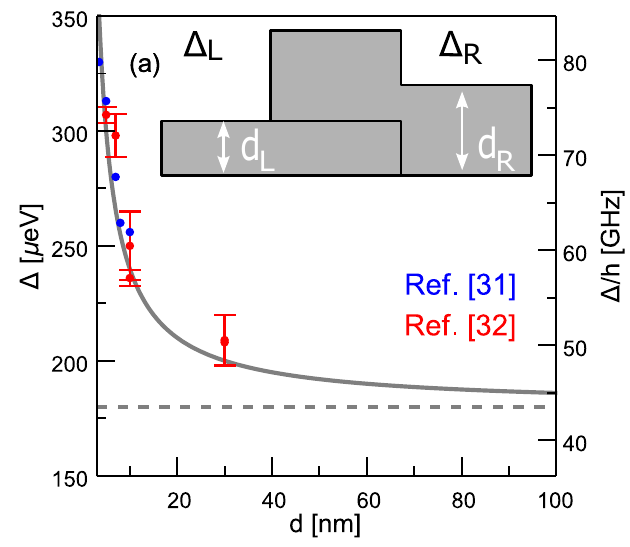}
    \caption{Superconducting gap ($\Delta$) vs thickness ($\mathrm{d}$) in a typical Al thin film. For $\mathrm{d}\gtrsim 3\,$nm, the phenomenological dependence of the Al gap on thickness $\mathrm{d}$ has been established to be $\Delta (\mathrm{d})=\Delta_{\rm bulk}+a \mathrm{d}^{-1}$, where $\Delta_{\rm bulk}\simeq 180~\mu$eV is the gap in the bulk limit, and $a$ is a parameter whose value depends on the details of the deposition process~\cite{CherneyThinAl,ChubovThinAl,MeserveyThinAl,Court_2007,IvryPRB90} ($a=600~\mu$eV$\cdot$nm for the solid curve). Inset: schematic representation of an Al junction deposited through electron beam evaporation. To ensure a good overlap in the junction region, the thickness of the second layer ($\mathrm{d}_R$) is larger than that of the first evaporated layer ($\mathrm{d}_L$), hence resulting in a gap difference between the two leads of the junction ($\dL>\dR$).}
    \label{fig:Figure1}
\end{figure}

Superconducting technology offers an ideal platform for the implementation of quantum devices~\cite{BlaisRMP93,DevoretScience339,GaoPRXQuantum2,EsmaeilAPL118}, and is currently in the spotlight for the future development of quantum processors~\cite{Arute2019,ZuchongzhiPRL127}. Superconducting qubits~\cite{OliverReview} have demonstrated an impressive technological advancement in the last few decades, with state-of-the art lifetimes five orders of magnitude larger~\cite{Wang2022} than the pioneering realizations~\cite{Nakamura}. This tremendous progress was the result of clever engineering of the qubit design, combined with improved filtering techniques, and quasiparticle poisoning suppression. Over the last few years, better qubit's performance was achieved by exploiting different materials~\cite{SiddiqiReview}. Various superconducting materials have been investigated over the last several years to manufacture components of superconducting circuits: resonators can be made out of TiN~\cite{TiN2010}, NbTiN~\cite{NbTiN2010,NbTiN2016}, and granular Al~\cite{Rotzinger_2016,PRL121Res,Maleeva}, capacitor pads out of TiN~\cite{TiN2013}, Nb~\cite{McDermott2019,GordonAPL120}, and Ta~\cite{Place2021,Wang2022,tennant2021low}, and inductances out of NbN~\cite{Annunziata_2010,Bylander2019}, NbTiN~\cite{Hazard2019,Kou2020}, and granular Al~\cite{grAl2019}. However, the central elements of qubits, the Josephson junctions, are generally based on aluminum/aluminum oxide/aluminum structures, characterized by high-quality and reproducible tunnel barriers~\cite{Kreikebaum_2020}. Aluminum displays an almost ideal behaviour in comparison to the original Bardeen-Cooper-Schrieffer (BCS) model~\cite{tinkham}, and effects due to, for instance, Fermi-liquid renormalization of the spin susceptibility and spin-orbit scattering can be incorporated in the modeling of the superconducting properties (see \textit{e.g.}~\cite{Adams2008} and references therein). In the typical fabrication process, a junction is realized by first depositing a layer of thickness $\mathrm{d}_L$ on the substrate; after oxidation, a second layer is evaporated on top of the structure (see inset of Fig.~\ref{fig:Figure1}), with thickness $\mathrm{d}_R>\mathrm{d}_L$ to ensure that the top layer is electrically continuous. For thin Al films, it is well known that the superconducting critical temperature $T_c$, and hence the energy gap $\Delta=1.764 k_B T_c$, depend on the film thickness in a substantial way~\cite{IvryPRB90}. In Fig.~\ref{fig:Figure1}, we display measurements of thin-film Al gap reported in the literature~\cite{CherneyThinAl,Court_2007} (points). In actual devices, the frequency associated with the gap asymmetry $\w_{LR}/(2\pi)=|\dL-\dR|/h$ can be of the order of a few GHz (here $h$ is the Planck's constant). For instance, in a Josephson junction with $\mathrm{d}_L=20\,$nm and $\mathrm{d}_R=60\,$nm, we estimate $\w_{LR}/(2\pi)\approx 5\,$GHz (see Fig.~\ref{fig:Figure1}). Such value is comparable to typical qubit frequencies, and hence potentially harmful to the qubit operation. In this work, we extensively discuss the effect of gap asymmetry onto the qubit properties.  

The gap asymmetry introduces the new energy scale $\omega_{LR}$ in the description of the system, in addition to the qubit frequency $\omega_{10}$ [associated with the energy difference between excited (1) and ground (0) states of the qubit] and the (effective) temperature $T$; hereinafter we set $\hbar=k_B=1$, where $\hbar$, $k_B$ are the reduced Planck's constant and the Boltzmann constant, respectively. If $\omega_{LR}$ is the smallest scale, the gap asymmetry can be neglected. Since the condition $\omega_{10} > T$ is typically fulfilled for the proper qubit operation, the only two possibilities to address are: I. $\omega_{10}>\omega_{LR}>T$ or II. $\omega_{LR}>\omega_{10}>T$. The case I will turn out to be the most complex. In this regime, the quasiparticles can have similar densities in both leads, or be trapped in the lead with the lowest gap, depending on the interplay between their generation at the junction (via the absorption of pair-breaking photons), recombination, and relaxation. Qualitatively, for slow generation the quasiparticles can be trapped, while for sufficiently fast generation and/or sufficiently small gap asymmetry the quasiparticles populate both leads with similar densities. Clearly case II is the farthest one from the equal gap (or symmetric) situation. As we will show below, the quasiparticles are largely trapped in the lead with the lowest gap. Indeed, the qubit transition cannot give sufficient energy to the quasiparticles to escape to the other lead, due to the large gap asymmetry. On the one hand, the trapped quasiparticles cannot contribute to the qubit relaxation rate, so the qubit lifetime can be longer. On the other hand, the exclusion of a possible relaxation mechanism is associated with an increased, and possibly high, residual excited state population of the qubit.
In other words, in case I
the condition $\omega_{10}>\omega_{LR}$ ensures the qubit relaxation can be mediated by quasiparticle tunneling, leading to higher qubit decay rate and lower excited state population than for $\omega_{10}<\omega_{LR}$ (if other parameters are not changed). Our main goal is the quantitative analysis of the two cases, using realistic parameters; in particular, we re-examine a previously published experiment~\cite{Wang2014} in light of our results for case I, before extending our study to encompass case II.

The article is organized as follows. In Sec.~\ref{sec:micromodel} we review the microscopic model for the qubit-quasiparticle interaction, generalizing the main results to the case of leads with asymmetric gaps. The model for the dynamics of the quasiparticle density is developed in Sec.~\ref{sec:model}, accounting for the gap difference in the leads. The steady state for the qubit-quasiparticle system is then analyzed in Sec.~\ref{sec:steady}. The case of a split transmon,  whose frequency can be controlled on chip, is discussed in Sec.~\ref{sec:split}. We finally summarize our results in Sec.~\ref{sec:conclusions}.

\section{Microscopic model}
\label{sec:micromodel}

In this section, we review the microscopic model describing a superconducting qubit in the presence of quasiparticle excitations. Even though our approach can be applied to arbitrary superconducting qubit designs (see Refs.~\cite{CatelaniPRL106,CatelaniPRB84}), for concreteness here
we consider transmon qubits~\cite{transmonPhysRevA76}. In the single-junction case, the low-energy Hamiltonian is
\be
\label{Htot}
\hat{H} = \hat{H}_{\varphi} + \hat{H}_{\qp} + \hat{H}_T \, .
\ee
In the absence of quasiparticles, $\hat{H}_{\varphi}= 4E_C (\hat{N}-n_g)^2 -E_J \cos \hat\varphi$ describes the qubit dynamics, where $E_C,E_J$ are the charging and the Josephson energy, respectively. Here, $\hat{N}=-id/d\varphi$ is the number operator of Cooper pairs tunneling across the junction, and $n_g$ is the dimensionless gate voltage. 

\begin{figure}
    \centering
    \includegraphics{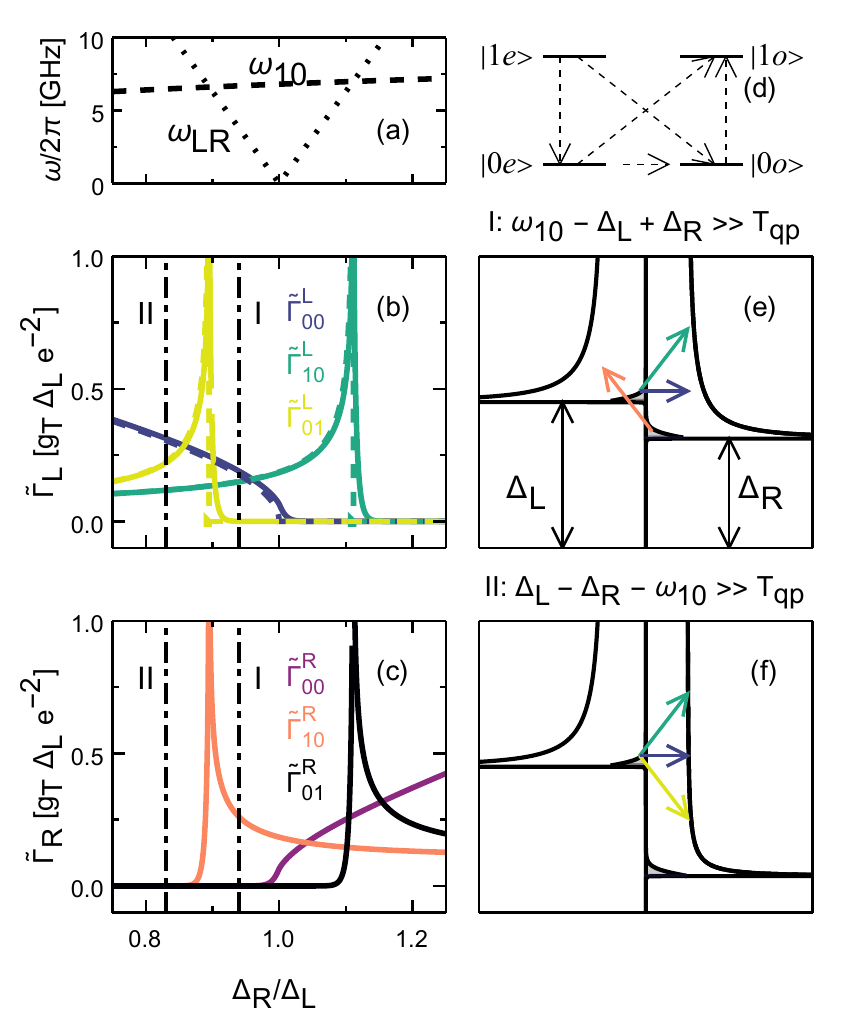}
    \caption{Gap asymmetry and qubit transition rates in thermal equilibrium. (a) Qubit frequency and gap-difference frequency as a function of the right electrode gap. Qubit transition rates [obtained from Eq.~\eqref{wif_gen} using the matrix elements in Appendix~\ref{app:matTra}] associated with quasiparticle tunneling (b) from the left and (c) from the right electrodes as a function of the right electrode gap. The rates can diverge at $\dL - \dR = \pm \w_{10}$, see text. The analytical expressions for the rates [dashed curves in panel (b)] are given in Appendix~\ref{app:qubitrates}. Parameters: $\dL=260~\mu$eV, $E_C/h=300$~MHz, $E_J^0=70 E_C$, $T=20$~mK. (d) Schematics of the lowest energy states of the qubit with logical state $i=\{0,1\}$, parity $j=\{e,o\}$, and the corresponding qubit
    transitions (arrows).  Panels (e) and (f) depict the relevant tunneling events for two choices of gap ratio exemplifying cases I and II, respectively [cf. dotted-dashed vertical lines in panels (b) and (c)].}
    \label{fig:Figure2}
\end{figure}

The quasiparticle term $\hat H_\qp$ is the sum of the mean-field BCS
quasiparticle Hamiltonians for the left and right electrodes,
\be
\hat{H}_{\qp} = \sum_{\alpha=L,R} \hat{H}_{\qp}^\alpha \, , \quad \hat{H}_{\qp}^\alpha = \sum_{n,\sigma} \epsilon_{n}^\alpha
\hat\gamma^{\alpha\dagger}_{n\sigma} \hat\gamma^\alpha_{n\sigma}.
\ee
Here $\hat\gamma^\alpha_{n\sigma}$($\hat\gamma^{\alpha\dagger}_{n\sigma}$) is the quasiparticle annihilation (creation) operator, $\sigma=\uparrow, \downarrow$ accounts for spin, and the
quasiparticle energies are $\epsilon^\alpha_{n} =
\sqrt{(\xi_{n}^\alpha)^2+\Delta_\alpha^2}$, with $\xi_{n}^\alpha$ and $\Delta_\alpha$ being the
single-particle energy level $n$ in the normal state of electrode $\alpha$, and
the gap parameter in that electrode, respectively.  Finally, the
tunneling term $\hat{H}_T$
describes single quasiparticle tunneling across the junction,
\begin{align}
\label{HT}
\hat{H}_{T} = t\!\!\sum_{n,m,\sigma}\!\!
\left(e^{i\frac{\hat\varphi}{2}} u_{n}^L u_{m}^R
- e^{-i\frac{\hat\varphi}{2}} v_{m}^R v_{n}^L
\right)\hat\gamma_{n\sigma}^{L\dagger} \hat\gamma^R_{m\sigma}
+ \text{H.c.}\, ,
\end{align}
while pair tunneling has already been taken into account by including the Josephson energy term in $\hat H_\varphi$ (see Appendix~A in Ref.~\cite{CatelaniPRB84} for more details). The tunneling amplitude $t$ is related to the normal-state junction conductance by
$g_T = 4\pi e^2 \nu^L \nu^R \mathcal V_L\mathcal V_R t^2$~\cite{Barone}, with $e$ the electron's charge and $\mathcal V_L (\mathcal V_R)$ volume of the left (right) lead. We assume identical
densities of states per spin direction and unit volume in the leads,
$\nu^L=\nu^R=\nu_0$. The tunneling term, \eref{HT}, couples the qubit
to the quasiparticles; therefore, a transition between initial,
$|i\rangle$, and final, $|f\rangle$, qubit states of energies $\omega_i$ and $\omega_f$ is possible when a quasiparticle tunnels through the junction. We will restrict our considerations to the lowest energy states of the transmon system. These states can be classified in terms of their logical value being 0 or 1 (with the 1 state having higher energy), and their parity (number of individual quasiparticles crossing the junction) being even $(e)$ or odd $(o)$~\cite{CatelaniPRB89}, as schematically depicted in Fig.~\ref{fig:Figure2}d. We recall that in transmons the energy difference between states of different parity but same logical value turns out to be small when $E_J/E_C\gg 1$ and will be neglected~\cite{CatelaniPRB89}.

The qubit transition rate $\Gamma_{if}$ can be calculated
using Fermi's Golden Rule, yielding~\cite{CatelaniPRB89,CatelaniSciPostReview}
\be
\label{wif_gen}
\Gamma_{if} = \left|\langle f|\sin \frac{\hat\varphi}{2}|i\rangle\right|^2
S^{+}_\qp\left(\w_{if}\right)+\left|\langle f|\cos \frac{\hat\varphi}{2}|i\rangle\right|^2
 S^{-}_\qp\left(\w_{if}\right),
\ee
where $\omega_{if}=\omega_i-\omega_f$. The normalized spectral current density $S_\qp^\pm$ is the sum of the contributions from the two electrodes $S_\qp^\pm(\w)=\sum_{\alpha=L,R} S_{\alpha}^\pm(\w)$:  
\begin{align}
    &  S_{\alpha}^\pm(\w) =\frac{2g_T}{e^2}\int_{\Delta_\alpha(\w)}^{+\infty}\! d\e \, 
  \kappa_\alpha^\pm(\e,\e+\w)
    f_\alpha(\e)[1-f_{\bar\alpha}(\e+\w)].
\label{eq:Sqp_general}
\end{align}
where $\alpha$ identifies the initial location of the quasiparticle involved in the transition. In Eq.~\eqref{eq:Sqp_general}, $f_\alpha$ denotes the occupation of the quasiparticle states $f_\alpha(\xi^\alpha_{n})= \langle\!\langle \hat\gamma_{n\uparrow}^{\alpha\dagger} \hat\gamma^\alpha_{n\uparrow}\rangle\!\rangle_\qp
= \langle\!\langle \hat\gamma_{n\downarrow}^{\alpha\dagger} \hat\gamma^\alpha_{n\downarrow}\rangle\!\rangle_\qp$
($\alpha=L,R$), double angular brackets indicating the averaging over quasiparticle states, assumed to be independent of spin. In Eq.~\eqref{eq:Sqp_general}, $\bar\alpha=R$ for $\alpha=L$ (and \textit{vice versa}),  $\Delta_\alpha(\w)=\max\{\Delta_\alpha,\Delta_{\bar\alpha}-\w\}$, and we introduced the function
\be
\kappa_\alpha^\pm(\e,\e')=  \frac{\e \e' \pm\Delta_\alpha\Delta_{\bar\alpha}}
    {\sqrt{\e^2-\Delta_\alpha^2}\sqrt{\e'^2-\Delta_{\bar\alpha}^2}}.
\label{eq:nu+-}
\ee
which accounts for the coherence factors multiplied by the normalized superconducting density of states.
The expression in Eq.~\eqref{wif_gen} can be used for qubit relaxation processes ($\w=\w_{10}>0$, with $\w_{10}$ the qubit frequency, downward diagonal arrow in Fig.~\ref{fig:Figure2}d), excitation ($\w=-\w_{10}<0$, upward diagonal arrow in Fig.~\ref{fig:Figure2}d) or charge-parity jumps ($|\w| \ll \w_{10}$, also known as parity switching events, horizontal arrow in Fig.~\ref{fig:Figure2}d). The matrix elements were derived in previous works~\cite{CatelaniPRB84,CatelaniPRB89} and are reported in Appendix~\ref{app:matTra}.

We consider a typical transmon qubit where the superconducting gaps in the leads of the Josephson junction are different, $\dL\neq\dR$, and discuss the impact of the gap asymmetry on the qubit's transition rates due to quasiparticles. 
Figure~\ref{fig:Figure2}a characterizes the relevant frequency scales in our system as a function of the gap ratio $\delta=\dR/\dL$. The transmon frequency $\w_{10}=\sqrt{8E_JE_C}-E_C$  depends relatively weakly on $\delta$ near $\delta=1$ through the Josephson energy~\footnote{This expression is derived from the critical current ($I_c$) of an asymmetric Josephson junction~\cite{Barone}, through the expression $E_J=\hbar I_C/(2e)$, and it is valid in the low quasiparticle density limit.} 
\be 
E_J=\frac{4E_{J}^0}{\pi}\frac{\delta}{1+\delta}K\left (\frac{|1-\delta|}{1+\delta}\right).
\label{Ej}
\ee
Here, $E_{J}^0=E_J(\delta=1)=g_T\dL/(8g_K)$ where $g_K = e^2/(2\pi)$ is the conductance quantum, and $K(x)$ is the complete elliptic integral of the first kind~\cite{PrudnikovBook}. In contrast, the gap mismatch frequency $\omega_{\rm LR}$ increases linearly from zero with the deviation of $\delta$ from unity, and can eventually reach the qubit frequency $\omega_{\rm 10}$. 

We now turn to the transition rates considering separately tunneling from the left (right) to the right (left) electrode. We assume the quasiparticle system in each electrode to be at thermal equilibrium, $f_L(\epsilon)=f_R(\epsilon)=(e^{\epsilon/T}+1)^{-1}$, with temperature $T=20\,$mK. For visualization purposes, the rates are normalized to the quasiparticle density $x_\alpha
\simeq \sqrt{2 \pi T/\Delta_\alpha} e^{-\Delta_\alpha /T}$, \textit{i.e.}, $\tilde\Gamma_\alpha=\Gamma_\alpha/x_\alpha$ with $\alpha=L,R$; the quasiparticle densities are themselves normalized to the Cooper pair densities in each electrode $n_{\mathrm{Cp}
\alpha} = 2\nu_0\Delta_\alpha$. Consider first the left rates, Fig.~\ref{fig:Figure2}b: at fixed $\dL$ the parity switching rate $\tilde\Gamma_{00}^L$ (blue) decreases monotonically with increasing $\dR$, and is exponentially 
suppressed for $\dR>\dL$, due to the decrease in available states in the right electrode. In contrast, both the qubit relaxation rate $\tilde\Gamma_{10}^L$ (green) and the excitation rate $\tilde\Gamma_{01}^L$ (yellow) display a non-monotonic evolution with $\delta$. In particular, $\tilde\Gamma_{01}^L$ ($\tilde\Gamma_{10}^L$) increases monotonically for $\dR\ < \dL-\w_{10}$ ($\dR < \dL+\w_{10}$), and diverges logarithmically due to the gap difference matching the qubit frequency [the matching condition makes the square root singularities in Eq.~\eqref{eq:nu+-} overlap]; both rates are 
exponentially 
suppressed at higher $\delta$. Thus, for gap difference $\wLR\sim \pm \w_{10}$, the transition rates can be significantly larger than for equal gaps, resulting in a reduced lifetime of the qubit. By symmetry, similar considerations apply to the transition rates from the right electrode upon replacing $L\leftrightarrow R$, see Fig.~\ref{fig:Figure2}c.

From now on, we set $\dL>\dR$ with no loss of generality. We introduce the quantity $T_\qp$ to denote the characteristic width of the quasiparticle distribution functions. This effective temperature does not necessarily coincide with the temperature at which samples are cooled in a refrigerator~\cite{CatelaniSciPostReview,CatelaniSciPost6}, and we assume it to be the smallest energy scale in the system, $T_\qp \ll \wLR,\,\omega_{10}$; we discuss possible values of $T_\qp$ and their impact in Sec.~\ref{sec:model}. Motivated by the previous analysis, we focus on two different regimes, schematically pictured in Fig.~\ref{fig:Figure2}e-f using quasiparticle band diagrams. In case I (see Fig.~\ref{fig:Figure2}e), $\w_{10}-\wLR \gg T_\qp$, the rate $\tilde\Gamma_{01}^L$ is suppressed, while qubit relaxation due to quasiparticles tunneling from the right electrode becomes relevant. In case II (see Fig.~\ref{fig:Figure2}f) we have $\wLR-\w_{10} \gg T_\qp$, and all the tunneling processes from the right electrode can be safely neglected. In the next section we expand this picture, based on (effective) thermal equilibrium, to allow for a more general non-equilibrium situation pertinent to realistic setups. 

\section{Quasiparticle-qubit dynamics}
\label{sec:model}
\begin{figure*}
    \centering
    \includegraphics{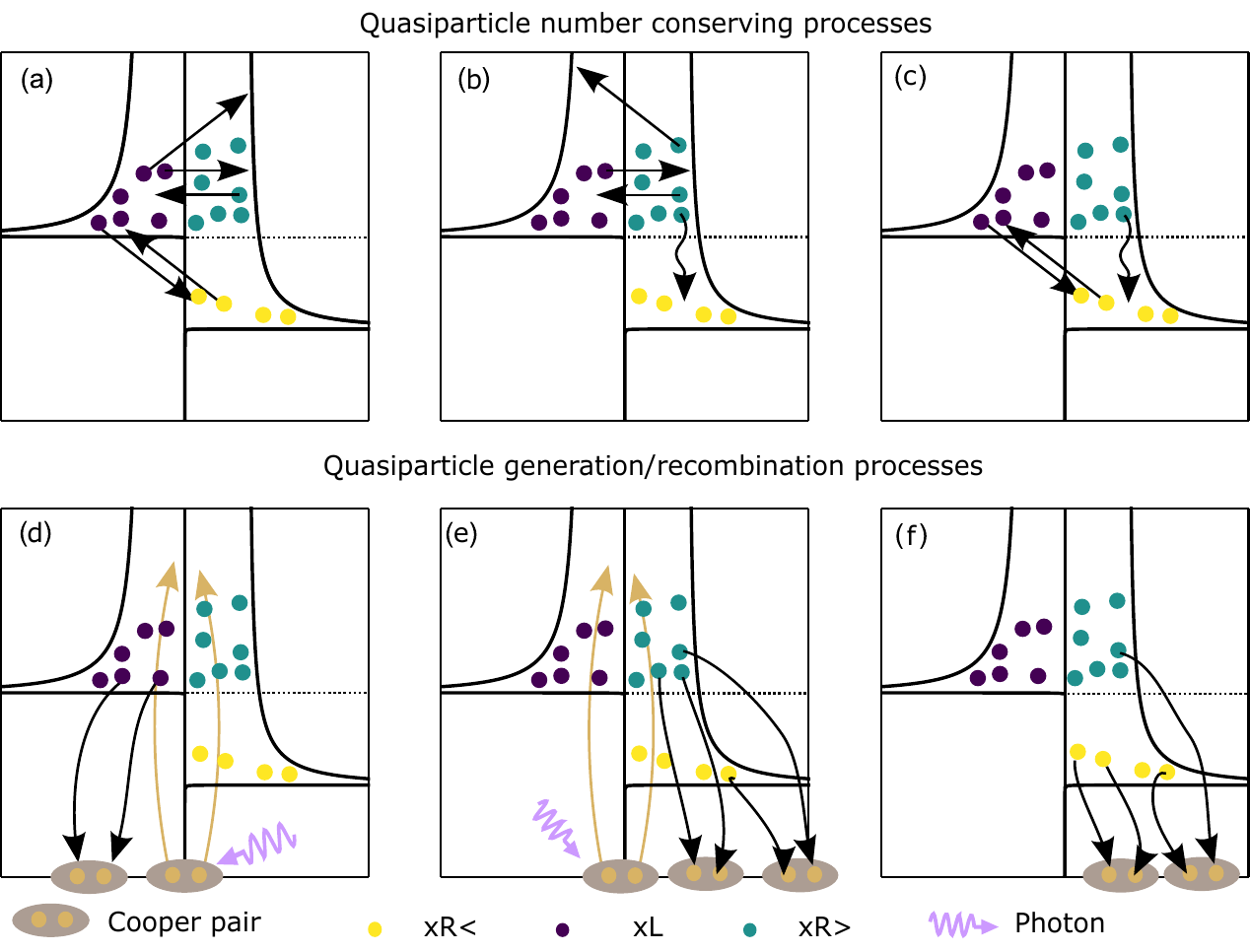}
    \caption{Representations of the processes involving quasiparticles. Top row: processes conserving the total quasiparticle number entering the equations for (a) $\xL$ [\eref{eq:dotxL}], (b) $\xRp$ [\eref{eq:dotxR>}], and (c) $\xRm$ [\eref{eq:dotxR<}]. Horizontal arrows depict parity switching transitions, upward diagonal arrows qubit relaxation, and downward diagonal arrows qubit excitation; the downward vertical wavy arrows account for quasiparticle relaxation by phonon emission. Bottom row: processes associated with generation (upward arrows) due to pair-breaking photons (wavy arrows) or recombination (downward arrows) of quasiparticles entering the equations for (d) $\xL$, (e) $\xRp$, and (f) $\xRm$.}
    \label{fig:Figure3}
\end{figure*}

Here, we present the  rate equations governing  the coupled dynamics of the qubit and quasiparticles. Their microscopic origin is discussed in Appendix~\ref{app:Micro}. We note that the quasi-elastic transitions from left to right (blue horizontal arrows in Figs.~\ref{fig:Figure2}e and f can lead to accumulation of quasiparticles in the right electrode at energies $\gtrsim \Delta_L$, if their relaxation to energies $\sim \Delta_R$ is slow. Moreover, depending on the state the qubit, different transitions can take place - compare \textit{e.g}. the yellow and green arrows in Fig.~\ref{fig:Figure2}e. 

We begin by considering the evolution equation for the qubit state occupation probability $p_{i}^j(t)$, with logical state $i=\{0,1\}$ and parity $j=\{e,o\}$. The rate equation for the probability reads~\cite{CatelaniPRB89,PRL121Exp}
\be
\dot{p}_i^j=-(\Gamma_{i\bar i}^{j\bar j}+\Gamma_{ii}^{j\bar j}+\Gamma_{i\bar i}^{jj})p_i^j
+\Gamma_{\bar i i}^{\bar j j}p_{\bar i}^{\bar j}
+\Gamma_{i i}^{\bar j j}p_{i}^{\bar j}
+\Gamma_{\bar i i}^{j j}p_{\bar i}^{j}
\label{eq:Qubit_rate_parity}
\ee
where the dot denotes derivative with respect to time, $\bar i=i+1$ (mod 2), and $\bar e=o$, $\bar o=e$. In the transmon regime $E_J/E_C\gg 1$, the energy difference between even and odd states is small compared to temperature~\cite{CatelaniPRB89}, and the transition rates are approximately symmetric under even-odd exchange: $\Gamma_{ii'}^{j\bar j}=\Gamma_{ii'}^{\bar j j}$ and $\Gamma_{i i'}^{j j}=\Gamma_{i i'}^{\bar j  \bar j}$. Summing over the two possible parities, we get the rate equations for the logical state occupation probability $p_i=\sum_{j=\{e,o\}}p_{i}^j$,
\be
\dot{p}_i=-(\Gamma_{i\bar i}^{eo}+\Gamma_{i\bar i}^{ee})p_i
+(\Gamma_{\bar i i}^{eo}+\Gamma_{\bar i i}^{ee})p_{\bar i} \, .
\label{eq:Qubit_rate}
\ee
The rates in Eq.~\eqref{eq:Qubit_rate} can be grouped in two categories. The parity conserving terms $\Gamma_{10}^{ee}$ and $\Gamma_{01}^{ee}$ (cf. vertical arrows in Fig.~\ref{fig:Figure2}d) corresponds to a ``cold'' dissipative environment,  $\Gamma_{01}^{ee}\ll\Gamma_{10}^{ee}$~\cite{PRL121Exp}, and are independent of quasiparticle density. Conversely, parity-switching events always involve quasiparticles, and the rates have the general expression
\be
\label{eq:Geo}
\Gamma_{ii'}^{eo}=\Gamma_{ii'}^{ph}
+\tilde\Gamma_{ii'}^{L}\xL
+\tilde\Gamma_{ii'}^{R<}\xRm
+\tilde\Gamma_{ii'}^{R>}\xRp \,.
\ee
Here $\Gamma_{ii'}^{ph}$ accounts for pair-breaking at the junction due to photon-assisted tunneling~\cite{PRL123Photon}, while the second term in the right-hand side, proportional to the quasiparticle density $\xL$ in the left electrode, accounts for transitions in which a quasiparticle tunnels from the left to the right electrode. The last two terms in Eq.~\eqref{eq:Geo} originate from quasiparticles tunneling in the opposite direction (right to left), and we have separated the density in two parts: $\xRm$ includes excitations with energy between $\Delta_R$ and $\Delta_L$, while $\xRp$ excitations with energy above $\Delta_L$, see Appendix~\ref{app:xRdef}. Explicit formulas for the rates $\tilde\Gamma_{ii'}^\alpha$ are presented in Appendix~\ref{app:qubitrates} (here and below, $\alpha$ denotes $L$, $R>$ and $R<$ when labeling the rates $\tilde\Gamma_{ii'}^\alpha$).

The time evolution equations for the quasiparticle densities are a generalization of those introduced by Rothwarf and Taylor~\cite{RothwarfTaylor} to phenomenologically account for the interplay between generation and recombination in determining the density of quasiparticles out of equilibrium,
\begin{align}
\dot x_L&=g^{L}-r^L\xL{}^2
-\delta[(\bar\Gamma_{00}^{L}+\bar\Gamma_{01}^L)p_0
+(\bar\Gamma_{11}^{L}+\bar{\Gamma}^L_{10})p_1]\xL 
\nonumber\\
&+\delta\bar\Gamma_{10}^{R<}p_1 \xRm +\delta[\bar\Gamma_{00}^{R>}p_0+(\bar\Gamma_{11}^{R>}+\bar\Gamma_{10}^{R>})p_1]\xRp 
\label{eq:dotxL}\\
\dot x_{R>}&=g^{R>}-r^{R>} \xRp^2
+[\bar\Gamma_{00}^{L}p_0+(\bar\Gamma_{11}^{L} + \bar\Gamma_{10}^L)p_1]\xL
\nonumber\\
& - [\bar\Gamma_{00}^{R>}p_0+(\bar\Gamma_{11}^{R>}
+\bar\Gamma_{10}^{R>})p_1+\tau_R^{-1}]\xRp 
\nonumber \\ & -r^{<>}\xRm\xRp
\label{eq:dotxR>}\\
\dot x_{R<}&=g^{R<}-r^{R<} \xRm^2+\tau_R^{-1}\xRp
+\bar\Gamma_{01}^Lp_0 \xL \nonumber \\ & -\bar\Gamma_{10}^{R<}p_1 \xRm
-r^{<>}\xRm\xRp
\label{eq:dotxR<}
\end{align}
where $g^\alpha$ and $r^\alpha$ denotes quasiparticle generation and recombination rates, respectively. Equations~\eqref{eq:dotxL}-\eqref{eq:dotxR<} can be obtained through energy integration of the kinetic equations for the quasiparticle distribution functions $f_\alpha(\xi^\alpha)$. The terms linear in the densities account for removal/addition of single quasiparticles from/to a given electrode due to tunneling to/from the other electrode; the terms proportional to $\tau_R^{-1}$ describe relaxation in the right electrode from energies above $\Delta_L$ to below it by phonon emission (see Appendix~\ref{app:RelaxationRecombination}). To convert the qubit transition rates $\tilde\Gamma_{ii'}^{\alpha}$ into quasiparticle rates for the (normalized) densities $x_\alpha$, we divide those rates by a factor proportional to the number of Cooper pairs in the electrodes $n_{\mathrm{Cp}\alpha}\mathcal V_\alpha$ (with $\alpha=L,R$),  $\mathcal V_\alpha$ and $n_{\mathrm{Cp}\alpha}$ being the volume and the Cooper pair density in the $\alpha$-electrode respectively. 
Concretely, the barred rates are defined as $\bar\Gamma_{ii'}^{\alpha}=\tilde\Gamma_{ii'}^{\alpha}/(2\nu_0 \Delta_R \mathcal V_\alpha)$; for simplicity, we assume hereafter $\mathcal V_L=\mathcal V_R$.
Note that $\bar\Gamma_{01}^{R<}$, $\bar\Gamma_{01}^{R>}$, and $\bar\Gamma_{00}^{R<}$, do not appear in the equations. Indeed, the rates $\bar\Gamma_{01}^{R<}$ and $\bar\Gamma_{00}^{R<}$ vanish, since there are no states in the left electrode at energies smaller than $\Delta_L$. The term $\bar\Gamma_{01}^{R>}$ is neglected based on the assumption $T_{\rm qp}\ll \omega_{10}$, which implies that there are no quasiparticles with sufficient energy (that is, energy $> \Delta_L+\omega_{10}$) to make the transition possible. A schematic representation of the various terms on the right-hand sides of Eqs.~(\ref{eq:dotxL})-(\ref{eq:dotxR<}) is given in Fig.~\ref{fig:Figure3}.

We remind that in writing Eqs.~(\ref{eq:dotxL})-(\ref{eq:dotxR<}) we assumed $T_\qp \ll \wLR, \, \w_{10}, \, |\wLR-\omega_{10}|$~--~see the end of Sec.~\ref{sec:micromodel}. Under these assumptions, the rates $\bar\Gamma^\alpha_{ii'}$ are proportional to the junction conductance $g_T$ and, with one exception, are functions of the gaps ($\dL$, $\dR$) and the qubit frequency ($\w_{10}$), but not of $T_\qp$. The exception is the rate $\bar\Gamma^{R>}_{00}$, which is proportional to $\sqrt{\wLR/T_\qp}$. In experiments, typical fridge base temperatures are in the range 10-20~mK, while chip and/or effective qubit temperatures are estimated to be of order 35-60~mK~-~see for instance~\cite{PRL114MIT,Vool2014,Fedorov2020}. The effective temperature $T_\qp$ is in general determined by the competition between the interaction of the quasiparticles with phonons (presumably in equilibrium at the fridge temperature, or at least with the qubit chip, if the latter effective temperature is higher than the fridge one) and with photons (for instance, the qubit excitation); in qubit devices the number of photons is low, which implies that $T_\qp \ll \omega_{10}$; moreover, for devices with electrodes of sufficiently large volume, such as those we will consider in the next Sections, the effective temperature is expected to be close to that of the phonons (we refer the reader to Ref.~\cite{Basko} for a more detailed discussion). Since in Eqs.~(\ref{eq:dotxL})-(\ref{eq:dotxR<}) only a single rate depends (weakly) on $T_\qp$, the quasiparticle-qubit dynamics and our results are not significantly affected by the specific temperature considered (either fridge or effective qubit temperature). 

\begin{figure}
    \centering
    \includegraphics{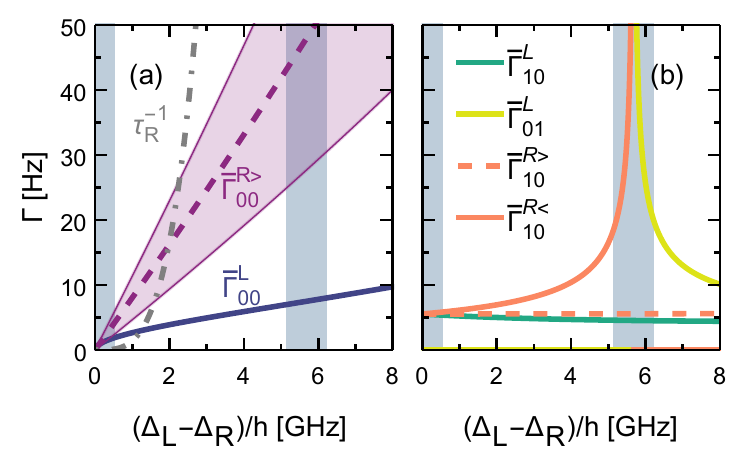}
\caption{Quasiparticle number conserving rates for a single junction transmon vs gaps asymmetry. (a) Quasiparticle rates for transitions that do not change the qubit logical state. The dot-dashed curve for the relaxation rate grows rapidly as $(\dL - \dR)^{7/2}$ (see Appendix~\ref{app:RelaxationRecombination}).  (b) Quasiparticle rates for transition changing the logical state of the qubit. Analytical expressions for the plotted rates can be found in Appendix~\ref{app:qubitrates}. In both panels, the vertical grey shadings identify the regions outside the model applicability. Parameters: $\dL/h=52$~GHz, $\omega_{10}/(2\pi)=5.7$~GHz, $E_J/h=15.5$~GHz, $E_C/h=290$~MHz, $\mathcal V_L=\mathcal V_R=940~\mu$m$^3$, $r^L=r^{R<}=(160~$ns$)^{-1}$, and $\nu_0=0.73\times10^{47}$J$^{-1}$m$^{-3}$. For the dashed line in the left panel, we also set $T_\qp=20\,$mK, while the shaded region corresponds to temperatures between 10 and 60~mK.}
    \label{fig:Figure4}
\end{figure}

In Fig.~\ref{fig:Figure4}, we display the estimates for the rates that do not change the total number of quasiparticles (cf. top row in Fig.~\ref{fig:Figure3}) as a function of the gap asymmetry $\wLR$. Even though the displayed values are computed considering parameters from a particular experiment that will be discussed in more detail in Sec.~\ref{sec:small_asymm}, the estimates are reasonably representative for aluminum devices. The rates preserving the qubit logic state increase monotonically with $\wLR$ (see Fig.~\ref{fig:Figure4}a). For the temperature-dependent rate $\bar\Gamma_{00}^{R>}$, we consider the range from 10 to 60~mK (red shaded area in Fig.~\ref{fig:Figure4}a). The relaxation rate grows rapidly $\tau_R^{-1}\propto\wLR^{7/2}$ (see Appendix~\ref{app:RelaxationRecombination}) and is dominant over the parity switching rates $\tau_R^{-1}\gg\bar\Gamma_{00}^{L}$, $\bar\Gamma_{00}^{R>}$, for sizeable asymmetries $\wLR/(2\pi)\geq2$~GHz. In Fig.~\ref{fig:Figure4}b, we show the tunneling rates changing the qubit's logical state.  While $\bar\Gamma_{10}^{R>}$, $\bar\Gamma_{10}^{L}$ are only weakly affected by $\wLR$, the rates $\bar\Gamma_{01}^{L}$ and $\bar\Gamma_{10}^{R<}$ are strongly enhanced around the resonant condition $\wLR=\w_{10}$ and zero for $\wLR<\w_{10}$ and $\wLR>\w_{10}$, respectively.

Quasiparticle generation may be caused by pair-breaking due to both photon-assisted tunneling~\cite{PRL123Photon} and bulk mechanisms, $g^\alpha=g^{\alpha,ph}+g^{\alpha,b}$. The bulk generation rate comes from both photons and phonons of energy larger than $2\Delta_\alpha$, while the photon-assisted tunneling corresponds to photons of energy larger than $\dL+\dR$; since the junction has much higher impedance than the bulk, we expect that pair-breaking photons are primarily absorbed at the junction. As for pair-breaking phonons, recent works show that they can be generated in the substrate by absorption of environmental radiation~\cite{PRL121Res,Vepslinen2020,Cardani2021,McEwen2021,Wilen2021,GordonAPL120}. While these non-equilibrium phonons can lead to temporary increases in the densities, the rate at which large density increases happen is low ($\lesssim 0.1\,$Hz) compared to the relevant time-scales, and can be further suppressed using phonon traps~\cite{APLphonon,PlourdePhonons}. So for simplicity we neglect the bulk mechanisms, $g^{\alpha,b}=0$.

Turning to the photon-assisted tunneling generation, the total rate at which quasiparticles are added to left electrode is $\left(\Gamma^{ph}_{10}+\Gamma^{ph}_{11}\right)p_1+\left(\Gamma^{ph}_{01}+\Gamma^{ph}_{00}\right)p_0$; in the transmon regime, assuming that the photon energy is much larger than $\Delta_L+\Delta_R+\omega_{10}$ we have $\Gamma^{ph}_{01} \simeq \Gamma^{ph}_{10} \equiv \Gamma^{ph}$ and $\Gamma^{ph}_{11}\simeq \Gamma^{ph}_{00} \approx \Gamma^{ph}\sqrt{8E_J/E_C}$~\cite{PRL123Photon}, so that the total rate in each lead is approximately $\Gamma^{ph}\left(1+\sqrt{8E_J/E_C}\right)$. To convert this rate into a generation rate for the (normalized) density we divide it by $2\nu_0\Delta_L \mathcal V_L$,
\begin{equation}\label{eq:gL}
g^{L} = \frac{\Gamma^{ph}}{2\nu_0\Delta_L \mathcal V_L} \left(1+\sqrt{\frac{8E_J}{E_C}}\right) .
\end{equation}
For the generation in the right electrode, we distinguish two possibilities: first, if the photon energy is large compared to $\Delta_L+\Delta_R+\omega_{10}$, we can neglect the generation rate at energies between $\Delta_R$ and $\Delta_L$ compared to that at energies above $\Delta_L$ and set $g^{R<}=0$, see Appendix~\ref{app:gen}; then, assuming equal volumes ($\mathcal V_R=\mathcal V_L$), we have $g^{R>} = g^L/\delta$. This first possibility is supported by the analysis~\cite{PRL123Photon} of experimental measurements~\cite{PRL121Exp}. Similarly, we can neglect $g^{R<}$ if thermal photons of temperature $T_{ph}$ reach the device from higher temperature stages of the fridge, so that $T_{ph}$ is large compared to $\omega_{10}$ and $\wLR$ (but still smaller than $\Delta_R$); then we find $\Gamma^{ph}_{11}\simeq \Gamma^{ph}_{00} \approx \Gamma^{ph}T_{ph}(\Delta_L+\Delta_R)/(4\Delta_L\Delta_R)\sqrt{8E_J/E_C}$ and, consequently, one should replace the terms in brackets in Eq.~\eqref{eq:gL} with $1+T_{ph}(\Delta_L+\Delta_R)/(4\Delta_L\Delta_R)\sqrt{8E_J/E_C}$. Note that although the relationships between the rates $\Gamma_{11}^{ph}$ and $\Gamma_{10}^{ph}$ are different, the structure of Eqs.~\eqref{eq:dotxL}-\eqref{eq:dotxR<} remains the same as for photons of large energy. 
A second possibility is that the photons have
energy smaller than $2\Delta_L-\omega_{10}$;
then 
$g^{R>}=0$ by energy conservation. The detailed study of this second possibility is beyond the scope of the present work. Henceforth we focus on the situation captured by Eq.~\eqref{eq:gL}.

\section{Steady state}
\label{sec:steady}

Here we focus on the steady state of the quasiparticle-qubit system, obtained by setting to zero the left-hand sides of Eqs.~\eqref{eq:Qubit_rate}, \eqref{eq:dotxL}-\eqref{eq:dotxR<}. 
The analysis can be simplified in the limit of small qubit's excited steady-state population $p_1\ll1$. In the following, we discuss how such limit provides a good representation of the experimental implementations. 
Using $p_0=1-p_1$, the population $p_1$ can be formally written as 
\begin{equation}
    p_1 = \frac{\Gamma_{01}^{ee}+\Gamma_{01}^{eo}}{\Gamma_{01}^{ee}+\Gamma_{01}^{eo}+\Gamma_{10}^{ee}+\Gamma_{10}^{eo}} \, .
    \label{eq:p1_steady}
\end{equation}
As remarked in the previous section, parity-conserving rates are of non-quasiparticle origin and usually associated to a ``cold'' environment~\cite{PRL121Exp}, resulting in $\Gamma^{ee}_{01}\ll\Gamma^{ee}_{10}$. For small quasiparticle densities, we expect the rates proportional to $x_\alpha$ to be small compared to the corresponding rates due to photon pair-breaking (that is, $\Gamma_{i\bar i}^{eo}\simeq \Gamma^{ph}$), so that
\begin{equation}
    p_1 \simeq \frac{\Gamma^{ph}+\Gamma^{ee}_{01}}{\Gamma^{ee}_{10}+2\Gamma^{ph}} \, .
    \label{eq:p1_steadyCaseI}
\end{equation} 
According to recent experiments on state-of-the-art transmons~\cite{PRL121Exp,spring2021high,GordonAPL120,kurter2022quasiparticle}, the main qubit relaxation mechanism is of non-quasiparticle origin, $\Gamma^{ee}_{10} > \Gamma^{eo}_{10}\sim\Gamma^{ph}$. Indeed, the relaxation time $T_1=(\Gamma_{10}^{ee}+\Gamma_{10}^{eo}+\Gamma_{01}^{ee}+\Gamma_{01}^{eo})^{-1}$ results to be of the order of few hundred $\mu$s, hence $\Gamma^{ee}_{10}\sim T_1^{-1}$ is a few kHz~\cite{Place2021,spring2021high,GordonAPL120}. On the other hand, recent measurements reports $\Gamma^{ph}$ ranging from $100\,$Hz down to less than 1~Hz, so that we can assume $\Gamma^{ph}\ll \Gamma^{ee}_{10}$ and $p_1 \ll 1$. Thus, well-shielded transmons are ``cold''~\cite{PRL114MIT}. One can then take $p_1$ as a small parameter in solving the rate equations in the steady-state, and look for the zeroth-order solution by setting $p_1 = 0$ and $p_0 = 1-p_1 =1$. In this limit, the steady-state equations for $x_\alpha$ are
\begin{align}
0 &= g-\delta^{-1}r^L\xL{}^2
-(\bar\Gamma_{00}^{L}+\bar\Gamma_{01}^L) \xL +\bar\Gamma_{00}^{R>}\xRp
\label{eq:xLsteady}
\\ 
0 &=g-r^{R>} \xRp^2-(\bar\Gamma_{00}^{R>}+\tau_R^{-1})\xRp+\bar\Gamma_{00}^{L}\xL
\nonumber
\\
& - r^{<>}\xRm\xRp
\label{eq:xR>steady}
\\
0 &=-r^{R<} \xRm^2- r^{<>}\xRm\xRp +\tau_R^{-1}\xRp
+\bar\Gamma_{01}^L \xL
\label{eq:xR<steady}
\end{align}
where, as discussed above, we neglect generation in the right electrode at energy below $\Delta_L$, assume equal volume electrodes, and set $g=g^{R>}$. Note that in transmons, the excited state qubit's population weakly affects the terms $\bar\Gamma^\alpha_{00}p_0 + \bar\Gamma^\alpha_{11}p_1=\bar\Gamma^\alpha_{00} + (\bar\Gamma^\alpha_{11}-\bar\Gamma^\alpha_{00})p_1$ present in Eqs.~\eqref{eq:dotxL} and \eqref{eq:dotxR>} for $p_1>0$. Indeed, the prefactor of the $p_1$ term is small compared to $\bar\Gamma^\alpha_{00}$ by the factor $(E_C/8E_J)^{1/2} \ll 1$, since this parameter determines the difference between the matrix elements of the parity switching transitions in the two logical states, see Appendix~\ref{app:matTra} (in other words, $\bar\Gamma^\alpha_{11} \approx \bar\Gamma^\alpha_{00}$). The smallness of the prefactor makes for less stringent conditions for the approximation $p_1 = 0$ to be applicable. Finally, summation of Eqs.~\eqref{eq:xLsteady}-\eqref{eq:xR<steady} yields the relation
\be
 \delta^{-1}r^L x_L^2+r^{R>}\xRp^2 +r^{R<} \xRm^2+2r^{<>}x_{R<}x_{R>}=2g,
\label{eq:qpBalance}
\ee
which in fact holds for generic values of $p_0,\, p_1$ in the steady state, showing that quasiparticle generation is balanced by the recombination processes. 

\begin{figure}
    \centering
    \includegraphics{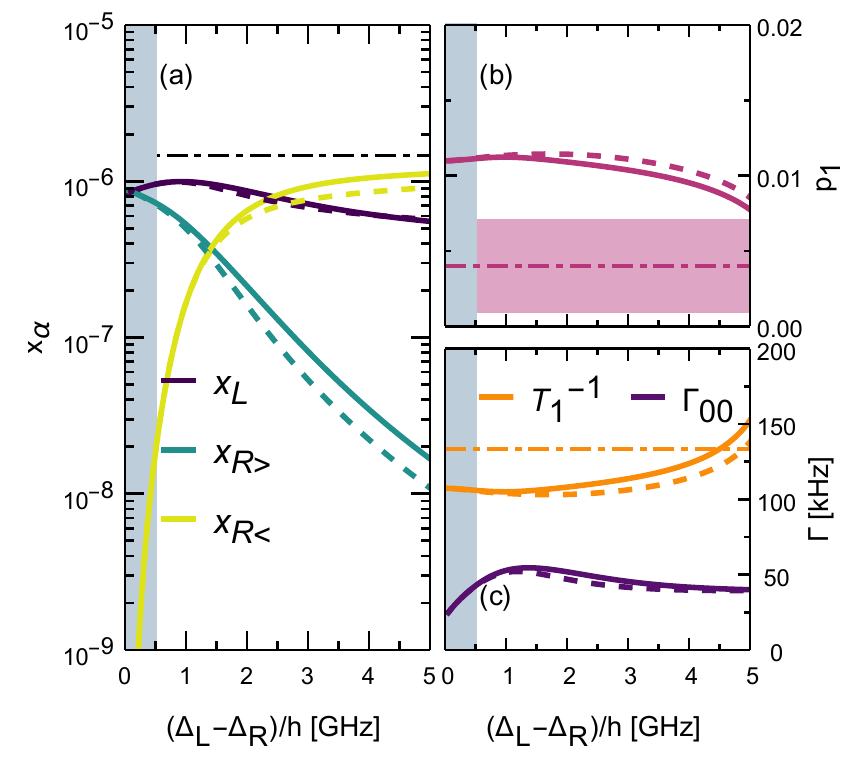}
    \caption{Steady state for the quasiparticle-qubit system as a function of the gap difference $(\Delta_L-\Delta_R)/h$ in the modeling of the experiment of Ref.~\onlinecite{Wang2014}, whose results are represented by the horizontal dot-dashed lines. (a) Quasiparticle densities in the steady state. (b) Population of the excited state of the qubit in the steady state. (c) Qubit inverse relaxation time and parity switching rates in the steady state. The dashed curves are given by Eqs.~\eqref{eq:xRmss-I}-\eqref{eq:xLss-I} for panel (a), Eq.~\eqref{p1ssI} for panel (b), and Eqs.~\eqref{eq:Geo} and \eqref{eq:T1inv} for panel (c). Parameters: $\dL/h=52$~GHz, $\omega_{10}/(2\pi)=5.7$~GHz, $E_J/h=15.5$~GHz, $E_C/h=290$~MHz, $T_\qp=20$~mK, $\nu_0=0.73\cdot10^{47}~$J$^{-1}$m$^{-3}$, $\mathcal V_L=\mathcal V_R=940~\mu$m$^3$, $r^L=r^{R<}=(160~{\rm ns})^{-1}$, $\Gamma^{ph}=1$ kHz, $\Gamma_{\rm ex}=(16~\mu$s$)^{-1}$.}
    \label{fig:Figure5}
\end{figure}
\subsection{Small gap asymmetry and comparison with experiment}
\label{sec:small_asymm}

We start the discussion of the qubit-quasiparticle steady state for a qubit of frequency larger than gap difference $\w_{10}>\wLR$ (case I). This situation is likely the most representative for transmon qubits, with typical frequency $\w_{10}/(2\pi)\simeq 6$~GHz, and an estimated gap asymmetry smaller than $5$~GHz. The goal is to solve Eqs.~\eqref{eq:xLsteady}-~\eqref{eq:xR<steady} for $\xL$, $\xRp$, and $\xRm$. In case I, the qubit cannot be excited by a quasiparticle tunneling transition originating from the left lead, $\tilde\Gamma_{01}^L=0$ (see also Fig.~\ref{fig:Figure4}b) since there are no quasiparticle states available at energies $\sim \dL-\w_{10}<\dR$ in the right lead. Then we can formally express the density $\xRm$ in terms of $\xRp$ by solving the quadratic equation Eq.~\eqref{eq:xR<steady},
\be
\xRm^{I}=\frac{\sqrt{(r^{<>}\xRp^{I})^2+4\tau_R^{-1}r^{R<}\xRp^{I}}-r^{<>}\xRp^{I}}{2r^{R<}},
\label{eq:xRmss-I}
\ee
where the superscript in $x^{I}_\alpha$ denotes that the given formulas apply to case I (and, in what follows, sufficiently close to the gaps having the same value, $\dL\approx\dR$).

For small values of the gap asymmetry $\eta\equiv 1-\delta\ll 1$, the quasiparticle relaxation rate in the right lead $\tau_R^{-1}$ is small ($\tau_R^{-1} \propto \eta^{7/2}$, see Appendix~\ref{app:RelaxationRecombination} and dot-dashed curve in Fig.~\ref{fig:Figure4}a), since quasiparticle states available for relaxation are constrained in the narrow energy window between $\dR$ and $\dL$. Thus, quasiparticles generated with rate $g$ at high energies, as in our modeling, accumulate at energies $>\dL$ in the right lead and $\xRm$ is expected to be small compared to $\xRp$~\footnote{Note that the approximation $g^{R<}\approx 0$ is particularly accurate in this case, due to the limited width of the quasiparticle energy window between $\dR$ and $\dL$}. In fact, in the limit $\tau_R^{-1} \to 0$ the leading order approximation to Eq.~(\ref{eq:xRmss-I}) gives $\xRm^{I} \simeq \tau_R^{-1}/r^{<>}$, a formula valid for $\xRm \ll \xRp$ (in writing this condition we used that $r^{R<} \sim r^{<>}$, see Appendix~\ref{app:RelaxationRecombination}). Using this approximation in Eq.~\eqref{eq:xR>steady}, we can solve for $\xRp$ in terms of $\xL$:
\begin{equation}
\xRp^{I}=\sqrt{\left(\frac{\bar\Gamma_{00}^{R>}+2\tau_R^{-1}}{2r^{R>}}\right)^2+\frac{g+\bar\Gamma_{00}^L\xL^{I}}{r^{R>}}
}-\frac{\bar\Gamma_{00}^{R>}+2\tau_R^{-1}}{2r^{R>}}.
\label{eq:xRpss-I}
\end{equation}
Finally, we use Eq.~\eqref{eq:xLsteady} to compute $\xL$ in a perturbative way in $\eta$; that is, we write $\xL^{I}=\xL^{(0)}+\xL^{(1)}$, with $\xL^{(1)} \ll \xL^{(0)}$ for $\eta \ll 1$. At lowest order, we can neglect the last two terms in Eq.~(\ref{eq:xLsteady}), since $\bar\Gamma_{00}^\alpha \to 0$ when $\eta \to 0$, and we find $\xL^{(0)} = \sqrt{\delta g/r^L}$ (we remind that $\bar\Gamma^L_{01} = 0$ in case I). Then, neglecting terms quadratic in $\xL^{(1)}$, we finally arrive at
\be
x_L^{I}=\frac{2\xL^{(0)2}r^L/\delta+\bar\Gamma_{00}^{R>}\xRp^{I}}{\bar\Gamma_{00}^L+2\xL^{(0)} r^L/\delta}
\label{eq:xLss-I}
\ee
where, for consistency, Eq.~\eqref{eq:xRpss-I} for $\xRp^{I}$ should be used with the substitution $\xL^{I} \to \xL^{(0)}$. An approximate expression for the population of the excited state of the qubit can be subsequently obtained by substituting the analytical expressions Eqs.~\eqref{eq:xRmss-I}-\eqref{eq:xLss-I} into Eq.~\eqref{eq:p1_steady}, reading
\be
p_1^{I}=\frac{\Gamma^{ph}+\Gamma_{01}^{ee}}{2\Gamma^{ph}+\Gamma_{10}^{ee}+\Gamma_{01}^{ee}+\sum_\alpha \tilde\Gamma_{10}^\alpha x_\alpha^{I}}.
\label{p1ssI}
\ee
Similarly, the parity switching, relaxation, and excitation rates (due to pair breaking plus quasiparticle tunneling) are obtained by substituting the same expressions into Eq.~\eqref{eq:Geo}, while the qubit inverse lifetime is given by
\be\label{eq:T1inv}
T_1^{-1} = \Gamma_{10}^{ee}+\Gamma_{10}^{eo}+\Gamma_{01}^{ee}+\Gamma_{01}^{eo}\, .
\ee

As a concrete application of our reasoning, we refer to the experimental results reported in Ref.~\onlinecite{Wang2014}. In that work, quasiparticles are injected in 3D-transmons devices and their dynamics is monitored; in particular, evidence for quasiparticle recombination is reported (for the specific design referred to, in Ref.~\onlinecite{Wang2014} and hereinafter, as type B). Figure~\ref{fig:Figure5}a displays the steady-state values of the quasiparticle densities $x_\alpha$ as a function of the gap difference. In the calculation, we consider a realistic value for the higher gap lead, $\dL/2\pi=52\,$GHz, a photon pair-breaking rate $\Gamma^{ph}=1$~kHz (consistent with the measurements in Ref.~\cite{PRL121Exp}, see also \cite{CatelaniSciPost6}), and the reported values of $E_J$, $E_C$, $\w_{10}$, $r=r^L\approx r^{R<}$ (see Appendix~\ref{app:RelaxationRecombination}), and $\Gamma_{\rm ex}$~\footnote{In our modeling $\Gamma_{\rm ex}=2\Gamma^{ph}+\Gamma_{10}^{ee}$ is the inverse lifetime of the qubit in the absence of nonequilibrium quasiparticles.} for the device B2 in Ref.~\onlinecite{Wang2014}. The electrode volume $\mathcal V$ of the Type B qubit design is estimated from the dimensions provided in Supplementary Fig.~1 of Ref.~\cite{Wang2014}~\footnote{In the calculations, we consider a fixed electrode volume $\mathcal V$ while decreasing the smaller gap $\Delta_R$. This variation can be obtained by increasing the thickness of the right junction lead. By limiting the change of thickness to the junction region, as done for example in Ref.~\cite{Fei2022}, the total electrode volume is not significantly affected.}.
The quasiparticle densities obtained through numerical solution of Eqs.~\eqref{eq:Qubit_rate} and \eqref{eq:dotxL}-\eqref{eq:dotxR<} (solid curves)~\footnote{The solutions are obtained through a find-root procedure, setting to zero time derivative terms} are compared to the approximate expressions Eqs.~\eqref{eq:xRmss-I}-\eqref{eq:xLss-I} (dashed curves). We observe that the analytical approximation for $\xL$ is in good agreement with the numerical results in all the relevant range for our parameter values. The approximations for $\xRm$ and $\xRp$ correctly reproduce the trend of the numerical curves but, as expected, the quantitative agreement worsens with increasing gap difference when the two densities become comparable.

Notably, our results are in general agreement with the experimental measurements. First, the single-lead quasiparticle density $x_\qp\simeq\max\{\xL,\xRp,\xRm\}\approx 10^{-6}$ is within a factor of less than 2 from the value extracted from the experimental data
(dot-dashed black line)~\footnote{This value is computed using Eq.~(3) in Ref.~\cite{Wang2014}, using the parameters $\w_{10}$, $T_1=T_1(B\sim 0$~mG) given in the supplementary material for device B2, and $\Gamma_{\rm ex}= T_1^{-1}(B\sim 100$~mG)}. Moreover, the theoretical calculations appropriately capture the impact of nonequilibrium quasiparticles on the qubit. Indeed, the computed population of the qubit's excited state, displayed in Fig.~\ref{fig:Figure5}b, is 
within a factor of three of the reported value $p_1 \simeq (0.4\pm0.3)\%$ and weakly dependent on the gap asymmetry. Finally, the computed inverse relaxation time of the qubit $T_1^{-1}$ (solid orange) matches
the experimental measurement (dot-dashed orange) for a gap asymmetry $\wLR/(2\pi)\simeq 4.5$~GHz (see Fig.~\ref{fig:Figure5}c). Over the whole model applicability range in gap asymmetry, nonequilibrium quasiparticles give a significant contribution (of order 40\% to 60\%) to the total relaxation rate. For completeness, we computed the total parity-switching rate $\Gamma_{11}\approx\Gamma_{00}=\Gamma^{ph}\sqrt{8E_J/E_C}+\tilde\Gamma_{00}^{L}\xL+\tilde\Gamma_{00}^{R>}\xRp$; again, a considerable part (45\% to 60\% in the region of validity of our model) of this rate originates from nonequilibrium quasiparticles. 
Parity switching can contribute to qubit dephasing~\cite{CatelaniPRB89}. Yet, the charge dispersion is small for the experimental parameters ($E_J/E_C\approx 50$) and the contribution to dephasing is also small: we estimate it to be of order 60~Hz. We note that the generation rate estimated in Ref.~\onlinecite{Wang2014} is an order of magnitude larger than the one used in the plots of Fig.~\ref{fig:Figure5} (a small discussion on this point is given at the end of Appendix.~\ref{app:Vortex}, where we discuss possible trapping effects). 

We close the analysis by commenting that sufficient conditions for our assumptions $p_1\ll 1$ to hold are $p_1 \ll \bar\Gamma^L_{00}/\bar\Gamma^L_{10},\,\bar\Gamma^{R>}_{00}/\bar\Gamma^{R>}_{10}$ in the equations for $\xL$ and $\xRp$, and $p_1 \ll (r^{R<}\xRm + r^{<>}\xRp)/\bar\Gamma_{10}^{R<}$ in the equation for $\xRm$. In the region of validity for case I, using the values plotted in Figs.~\ref{fig:Figure4} and \ref{fig:Figure5} together with $r^{R<}\approx r^{<>} \approx (160~{\rm ns})^{-1}\,$, one can check that the conditions are in fact met.

\subsection{Large gap asymmetry and state-of-the-art qubits}
\label{sec:strong_asymm}

We now consider the situation in which the qubit frequency is smaller than the gap difference, $\w_{10}<\wLR$ (case II). This regime may be relevant to low-frequency qubits and split transmons (see Sec.~\ref{sec:split} for a more detailed discussion of the latter). In this case, the strong gap asymmetry effectively traps quasiparticles in lower gap electrode at energy $\sim\dR$. Indeed, quasiparticles initially located in the right lead at energy $\sim\Delta_R$ cannot tunnel into the left lead deexciting the qubit, $\tilde\Gamma_{10}^{R<}=0$, since $\w_{10}<\wLR$. Conversely, quasiparticles initially located in the left electrode (with density $\xL$) can lose energy and excite the qubit ($\tilde\Gamma_{01}^L\neq 0$), populating low-energy states (thus increasing $\xRm$). Moreover, quasiparticles in the right lead with energy above $\dL$ relax more rapidly with respect to the small-asymmetry case, due to an increased number of available quasiparticle states at lower energies (see Fig.~\ref{fig:Figure4}). Thus, we expect the highest quasiparticle density in the low-energy states of the right lead, $\xRm\gg\xL,\,\xRp$.

In order to obtain approximate analytical formulas for the densities, we proceed as follows. Assuming other outgoing (negative) terms are dominant, the conditions being $\delta^{-1}r^L x_L \ll \bar\Gamma_{00}^{L}+\bar\Gamma_{01}^L$ and $r^{R>} x_{R>}+ r^{<>} x_{R<}\ll \bar\Gamma_{00}^{R>}+\tau_R^{-1}$, we neglect recombination terms quadratic in the densities in Eqs.~\eqref{eq:xLsteady}-\eqref{eq:xR>steady}. Note that the term proportional to $r^{R<}$ in Eq.~\eqref{eq:xR<steady} cannot be neglected~\footnote{For the consistency of the set of equations, the term $r^{R<>}$ must be dropped also in the last equation. This approximation works well for
$\xRp\ll \xRm$.}, as otherwise there would not be a quasiparticle sink and hence no steady state. With these approximations, the quasiparticle densities $\xL$ and $\xRp$ are obtained by solving the linear system formed by Eqs.~\eqref{eq:xLsteady} and \eqref{eq:xR>steady}, while $\xRm$ is found after substituting the solution to the linear system into Eq.~\eqref{eq:xR<steady} (where the term proportional to $r^{<>}$ is neglected), or alternatively from Eq.~\eqref{eq:qpBalance} (where only the third term in the left-hand side is retained). Explicitly, we have
\begin{align}
    \xL^{II} & =
    g \,\frac{2\bar\Gamma_{00}^{R>}+ \tau_R^{-1}}{\bar\Gamma_{00}^L\tau_R^{-1}+\bar\Gamma_{01}^L(\bar\Gamma_{00}^{R>} +\tau_R^{-1})} \, ,\label{eq:xlss-II} \\
    \xRp^{II} & =  g \,\frac{2\bar\Gamma_{00}^L+\bar\Gamma_{01}^L}{\bar\Gamma_{00}^L\tau_R^{-1}+\bar\Gamma_{01}^L(\bar\Gamma_{00}^{R>} +\tau_R^{-1})} \, , \label{eq:xrpss-II} \\
    \xRm^{II}
    & = \sqrt{\frac{2g}{r^{R<}}} \, .
    \label{eq:xrmss-II}
\end{align}
The superscript in $x^{II}_\alpha$ (with $\alpha=L,R<,R>$) indicates that the given formulas apply not only in case II of gap asymmetry larger than qubit frequency, but more generally for all the situations where $\xRm\gg \xL,\xRp$. Correspondingly, the population of the excited state of the qubit reads
\be
p_1^{II}=\frac{\Gamma^{ph}+\Gamma_{01}^{ee}+
\tilde\Gamma_{01}^L\xL^{II}
}{2\Gamma^{ph}+\Gamma_{10}^{ee}+\tilde\Gamma_{01}^L\xL^{II}+\sum_\alpha \tilde\Gamma_{10}^\alpha x_\alpha^{II}} \, .
\label{p1ssA}
\ee

\begin{figure}
    \centering
    \includegraphics{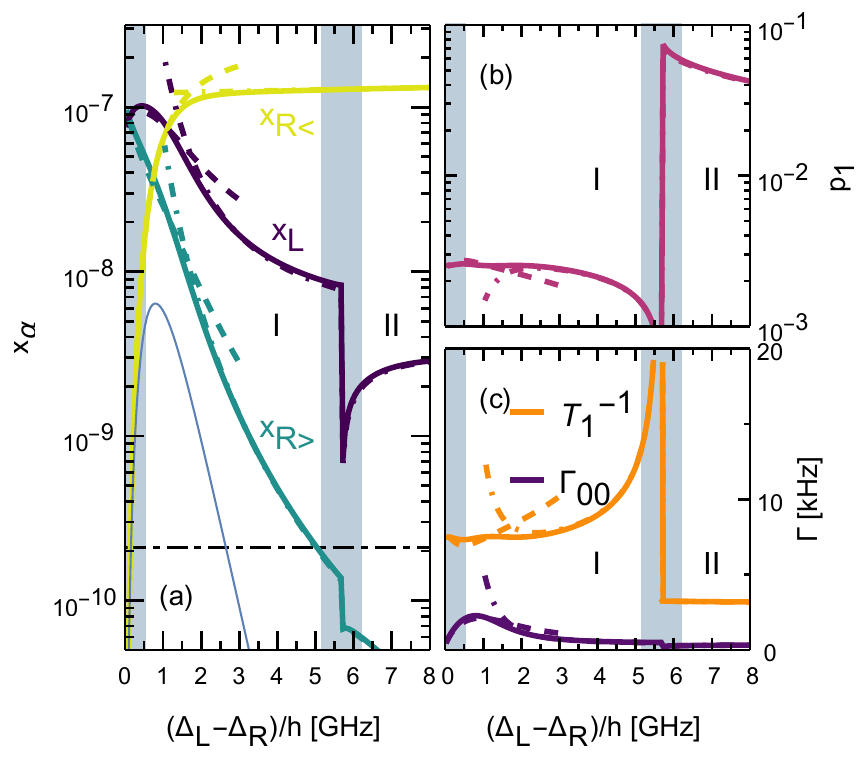}
    \caption{Steady state for the quasiparticle-qubit system as a function of the gap difference $(\Delta_L-\Delta_R)/h$ for a state-of-the-art transmon. (a) Quasiparticle densities in the steady state. The horizontal dot-dashed line (black) denotes the quasiparticle density corresponding to a single quasiparticle. The thin blue curve shows the density $\xRp$ that would be expected in thermal equilibrium at temperature $T_\qp$ given the density $\xRm$. (b) Population of the excited state of the qubit in the steady-state.
    (c) Qubit inverse relaxation time and parity switching rate in the steady-state. In all panels, the vertical grey shadings identify the regions outside the model applicability. Parameters: $\dL/h=52$~GHz, $\omega_{10}/(2\pi)=5.7$~GHz, $E_J/h = 15.5$~GHz, $E_C/h = 290$~MHz, $\Gamma^{ph}=10$~Hz, $T_\qp=20$ mK, $\mathcal V_L=\mathcal V_R=940~\mu$m$^3$, $r^L=r^{R<}=(160$~ns$)^{-1}$, $\nu_0=0.73\times 10^{47}$J$^{-1}$m$^{-3}$, $\Gamma_{10}^{ee}=3$~kHz, $\Gamma_{01}^{ee}=0.003~\Gamma_{10}^{ee}$.
    The dashed curves are given by Eqs.~\eqref{eq:xRmss-I}-\eqref{eq:xLss-I} for panel (a) and Eq.~\eqref{p1ssI} for panel (b), while the dot-dashed curves by Eqs.~\eqref{eq:xlss-II}-\eqref{eq:xrmss-II} for panel (a) and Eq.~\eqref{p1ssA} for panel (b). In panel (c), the dashed [dot-dashed] curves are found using Eqs.~\eqref{eq:xRmss-I}-\eqref{eq:xLss-I} [\eqref{eq:xlss-II}-\eqref{eq:xrmss-II}] in  Eqs.~\eqref{eq:Geo} and \eqref{eq:T1inv}.}
    \label{fig:Figure6}
\end{figure}

Using the expressions Eqs.~\eqref{eq:xlss-II}-\eqref{p1ssA} and those in the previous subsection, we now focus on the steady-state of the qubit-quasiparticle system in state-of-the-art transmons. In terms of parameters, we take $\Gamma^{ph}=10\,$Hz, $\Gamma_{10}^{ee}=3\,$kHz, and the remaining parameters as in the plots of Fig.~\ref{fig:Figure5}. The lower values of $\Gamma^{ph}$ and $\Gamma_{10}^{ee}$ in comparison to those used so far reflect improved filtering and device design and fabrication~\cite{SerniakPRApp,Fei2022,McDermott2022}. Figure~\ref{fig:Figure6}(a) displays $x_\alpha$ obtained finding numerically the steady-state of Eqs.~\eqref{eq:Qubit_rate}-\eqref{eq:dotxR<} (solid). The shaded areas in the plot denote the regions $\wLR\lesssim T_\qp$ and $|\wLR-\w_{10}|\lesssim T_\qp$ where our model is not applicable. The analytical approximations of Eqs.~\eqref{eq:xRmss-I}-\eqref{eq:xLss-I} (dashed curves) are now reliable only for small values of the gap difference, because the crossover from $\xRm < \xRp$ to $\xRm > \xRp$ takes place at a lower gap difference than in Fig.~\ref{fig:Figure5} due to the smaller generation rate. On other hand, the approximate expressions of Eqs.~\eqref{eq:xlss-II}-\eqref{eq:xrmss-II} (dotted-dashed curves) show excellent agreement with the numerical results starting from $\wLR/(2\pi)\approx 2\,$GHz. 
Noting that as the asymmetry increases the relaxation rate becomes quickly the dominant one, see Fig.~\ref{fig:Figure4}a, we can further approximate Eqs.~\eqref{eq:xlss-II} and \eqref{eq:xrpss-II} as $\xL^{II} \approx g/(\bar\Gamma_{00}^L+\bar\Gamma_{01}^L)$ and $\xRp^{II} \approx g \tau_R (2\bar\Gamma_{00}^L+\bar\Gamma_{01}^L)/(\bar\Gamma_{00}^L+\bar\Gamma_{01}^L)$; these formulas account for the jump in the densities when the gap asymmetry becomes larger than the qubit frequency, and for the fact that $\xL$ increases after the jump.

At low generation rate, the gap difference has a stronger impact on the quasiparticle populations than in the high-generation case: already at moderate values of gap difference almost all the quasiparticles are trapped in the lower gap electrode, and the densities $\xL$ and $\xRp$ are significantly smaller than $\xRm$, with $\xRp\ll\xL\ll\xRm$. Interestingly, the ratio $\xRp/\xRm$ strongly deviates from the Boltzmann factor $e^{-\wLR/T_\qp}$ expected in thermal equilibrium (see thin blue curve in Fig.~\ref{fig:Figure6}a): in the right lead there are more quasiparticles than expected at energies above $\dL$. 
However, as temperature increases the relative excess in $\xRp$ decreases because $\bar\Gamma_{00}^{R>}$ is correspondingly reduced, see Fig.~\ref{fig:Figure4}. Therefore, at these higher temperatures, it is more appropriate to assume the quasiparticle distribution to be of thermal (quasi)equilibrium form, $f_{L,R}=1/[e^{(\epsilon-\mu_{L,R})/T_\qp}+1]$ with effective chemical potentials $\mu_{L,R}$, rather than treat separately $\xRp$ and $\xRm$. 
In other words, at higher temperatures
we cannot neglect 
in Eqs.~\eqref{eq:dotxR>} and \eqref{eq:dotxR<} 
the thermal excitation of quasiparticles back to higher energies by phonon absorption. For the parameters in Fig.~\ref{fig:Figure6}, the cross-over between the two limits of strong non-equilibrium and quasiequilibrium
would take place at around 35~mK, and for the higher generation rate of Fig.~\ref{fig:Figure5} at around 60~mK (or even higher temperature for gap difference below about 3~GHz).

A second signature of the non-equilibrium state of the system comes from converting the qubit excited state population into an effective qubit temperature $T_q$ via $p_1/p_0 = e^{-\omega_{10}/T_q}$. Concerning the state-of-the-art transmon (see Fig.~\ref{fig:Figure6}b), we obtain $T_q\sim$44~mK at small gap difference $\wLR/(2\pi)\leq 3~$GHz and $T_q\sim$86~mK for the largest asymmetry $\wLR/(2\pi)\sim 8~$GHz, while for the calculations of the experiment of Ref.~\cite{Wang2014} (see Fig.~\ref{fig:Figure5}b), we find $T_q \approx 59\,$mK (for a population $p_1\sim 1\%$). We stress that, in all cases, the effective qubit temperature is significantly larger than the quasiparticle temperature $T_\qp=20\,$mK used in the calculations. Therefore, in general the qubit effective temperature, $T_q$, is not a reliable proxy for the quasiparticle one, $T_\qp$.

We now consider the impact of nonequilibrium quasiparticles on experimentally measurable quantities. First, we compute the excited state population of the qubit $p_1$ (see Fig.~\ref{fig:Figure6}b). We note that 
$p_1$ is typically much larger in case II than in case I. This behaviour is due to two combined effects:
1) qubit excitation by quasiparticle tunneling from the left electrode at rate $\tilde\Gamma_{01}^L\xL$ is possible in case II but suppressed in case I, 2) qubit relaxation rates associated with the low-energy quasiparticles $\tilde\Gamma_{10}^{R<}\xRm$ are present only in case I. This interpretation is supported by the evolution of the relevant rates, visualized in Fig.~\ref{fig:Figure6}(c). More precisely, in case I (left side of the panel) nonequilibrium quasiparticle tunneling gives a sizable contribution to both the qubit relaxation rate $\Gamma_{10}^{ne}=\tilde\Gamma_{10}^L \xL + \tilde\Gamma_{10}^{R>} \xRp + \tilde\Gamma_{10}^{R<} \xRm\approx 4.5\,$kHz [for $\wLR/(2\pi)\lesssim 2.5 $~GHz] and to the parity switching rate, $\Gamma_{00}^{ne}=\tilde\Gamma_{00}^L \xL + \tilde\Gamma_{00}^{R>} \xRp \,$ in the range $0.3-2$~kHz [where the maximum is reached for $\wLR/(2\pi)\sim 0.8 $~GHz], larger than the contribution due to pair-breaking photons $\Gamma^{ph}\sqrt{8E_J/E_C} \simeq 200\,$Hz. In case II instead (right side of the panel), the parity switching rates are almost entirely determined by the photon-assisted transitions,
and the relaxation rates is mostly of nonquasiparticle nature, $\Gamma_{10}\simeq \Gamma_{10}^{ee}$.

Comparing panels (c) in Figs.~\ref{fig:Figure5} and \ref{fig:Figure6}, it is evident that the combined effect of lower $\Gamma^{ee}_{10}$ and $\Gamma^{ph}$ is to substantially reduce both relaxation and parity switching rates. Clearly, the reduction of the relaxation rate is advantageous; conversely, the reduction of the parity switching rate leads to an increased contribution to pure dephasing. For parity switching faster than the charge dispersion, a phenomenon analogous to ``motional narrowing'' leads to a small dephasing rate (see the end of Sec.~\ref{sec:small_asymm}); on the contrary, we now find that parity switching is slower than the charge dispersion, so that the parity switching rate itself contributes to pure dephasing~\cite{CatelaniPRB89}, thus limiting the pure dephasing time to be of order 1~ms. For gap asymmetry larger than qubit frequency, the nonequilibrium quasiparticle contribution to the rates is suppressed but, as discussed above, this come at the price of higher residual population in the qubit excited state.

\subsection{Steady state for the stabilized qubit}
\label{sec:stabilize}

In this subsection, we consider to what extent the manipulation of the qubit state can be exploited to control the quasiparticle densities $x_\alpha$. To this end, we analyze the stationary values of $x_\alpha$ for a qubit which is \textit{continuously stabilized} either in the excited state (with $p_1\approx 1$ and $p_0\approx0$) or the ground state (with $p_1\approx 0$ and $p_0\approx1$). Such a stabilization protocol has been recently experimentally demonstrated for fluxonium qubits~\cite{PopFluxonium2022}, yielding information on the two-level systems coupled to the qubit. For ground-state stabilization, we can still use the results for the steady state, since they were found neglecting $p_1 \ll 1$.

In the case of the qubit stabilized in the excited state, the steady-state equations for $x_\alpha$ are
\begin{align}
0 &= g-\delta^{-1}r^L\xL{}^2
-(\bar\Gamma_{11}^{L}+\bar\Gamma_{10}^L) \xL +\bar\Gamma_{10}^{R<}\xRm \label{eq:xLsteadyExc}
\\ & +(\bar\Gamma_{11}^{R>}+\bar\Gamma_{10}^{R>})\xRp
\nonumber
\\ 
0 &=g-r^{R>} \xRp^2- r^{<>}\xRm\xRp+(\bar\Gamma_{11}^{L}+\bar\Gamma_{10}^L)\xL
\label{eq:xR>steadyExc}
\\
& -(\bar\Gamma_{10}^{R>}+\bar\Gamma_{11}^{R>}+\tau_R^{-1})\xRp
\nonumber \\
0 &=-r^{R<} \xRm^2 - r^{<>}\xRm\xRp +\tau_R^{-1}\xRp
-\bar\Gamma_{10}^{R<}\xRm .
\label{eq:xR<steadyExc}
\end{align}
The density $\xRm$ can be expressed in terms of $\xRp$ as in Eq.~\eqref{eq:xRmss-I} after the substitution $r^{<>}\xRp\to r^{<>}\xRp+\bar\Gamma_{10}^{R<}$.
To find approximate solutions for the other two densities, one could proceed similarly to Sec.~\ref{sec:small_asymm} for small gap asymmetry (case I), and to  Sec.~\ref{sec:strong_asymm} for large gap asymmetry (case II).
In case I, the treatment is partially complicated by the presence of the terms proportional to $\bar\Gamma^{R<}_{10}$, which couple Eqs.~\eqref{eq:xLsteadyExc} and~\eqref{eq:xR<steadyExc}. However, the rates $\bar\Gamma^{R<}_{10}$
and $\tau_{R}^{-1}$ can become dominant when increasing the gap asymmetry (see Fig.~\ref{fig:Figure4}); this justifies dropping the terms quadratic in the densities from Eq.~\eqref{eq:xR<steadyExc} under the condition $r^{R<} \xRm + r^{<>}\xRp \ll \bar\Gamma_{10}^{R<}$, giving immediately
\begin{equation}
x_{R<}^{I,exc}=\frac{\tau_R^{-1}}{\bar\Gamma_{10}^{R<}}x_{R>}^{I,exc}\label{eq:xrmssExc-I},
\end{equation}
which implies $x_{R>}^{I,exc}\ll x_{R<}^{I,exc}$. Then the term proportional to $x_{R>}^2$ can be neglected in Eq.~\eqref{eq:xR>steadyExc} (and so the term proportional to $x_{R<}x_{R>}$ for consistency), yielding the expressions for the densities $\xL$ and $\xRp$
\begin{align}
x_L^{I,exc}&=\sqrt{\frac{\delta g}{r^L}}
\label{eq:xlssExc-I}\\
x_{R>}^{I,exc}&=\frac{g+(\bar\Gamma_{11}^L+\bar\Gamma_{10}^L)x_L^{I,exc}}{\bar\Gamma_{10}^{R>}+\bar\Gamma_{11}^{R>}+\tau_R^{-1}}\label{eq:xrpssExc-I}
\end{align}

Turning to case II, we now have $\bar\Gamma_{10}^{R<}= 0$. Repeating the analysis done for the non-stabilized steady state, we neglect all the recombination terms in Eqs.~\eqref{eq:xLsteadyExc} and \eqref{eq:xR>steadyExc} and keep the term $r^{R<} \xRm{}^2$ in Eq.~\eqref{eq:xR<steadyExc}. Within this approximation, we retrieve Eq.~\eqref{eq:xrmss-II} for $\xRm$, and find 
\begin{align}
    x_L^{II,exc} & =
    \,\frac{2 g}{\tau_R^{-1}}\frac{\bar\Gamma_{11}^{R>}+\bar\Gamma_{10}^{R>}+\tau_R^{-1}/2}{\bar\Gamma_{11}^L+\bar\Gamma_{10}^{L}}
   \label{eq:xlssExc} \\
    x_{R>}^{II,exc} & =  \,\frac{2g}{\tau_R^{-1}} \label{eq:xrpssExc} 
\end{align}

\begin{figure}
    \centering
    \includegraphics{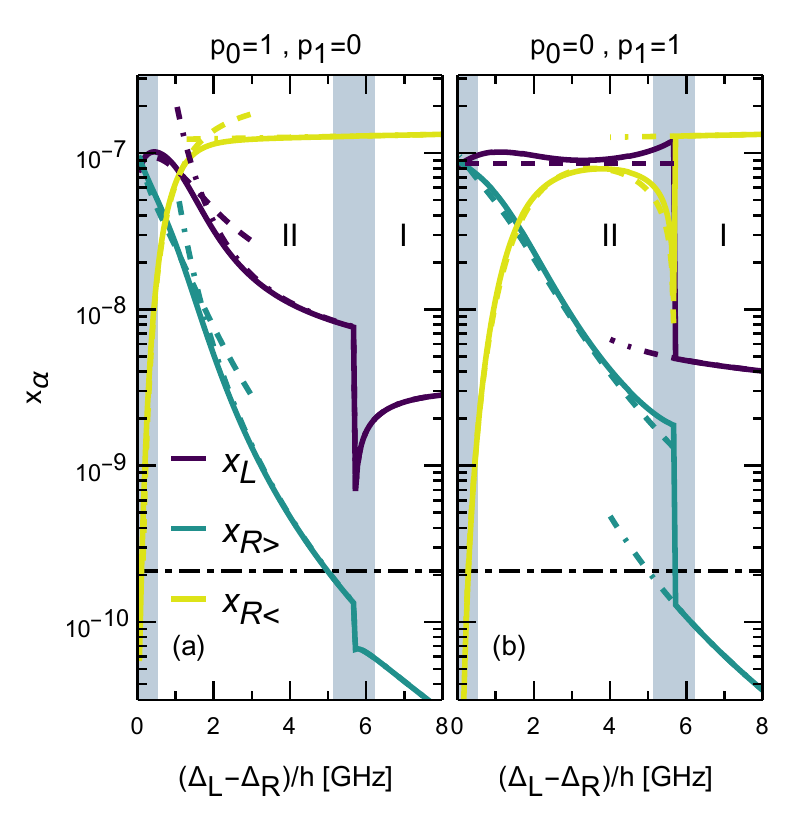}
    \caption{Quasiparticle densities in the steady state vs gap difference $(\Delta_L-\Delta_R)/h$ for a qubit stabilized in the (a) ground state ($p_1=0$) or (b) excited state ($p_1=1$). Each horizontal dotted-dashed line identifies the density corresponding to a single quasiparticle. Parameters: $\dL/h=52$~GHz, $\omega_{10}/(2\pi)=5.7$ GHz, $E_J/h = 15.5$~GHz, $E_C/h = 290$~MHz, $\Gamma^{ph}=10$~Hz, $T_\qp=20$~mK, $\mathcal V_L=\mathcal V_R=940~\mu$m$^3$, $r^L=r^{R<}=(160$~ns$)^{-1}$, $\nu_0=0.73\times 10^{47}~$J$^{-1}$m$^{-3}$. Dashed and dot-dashed curves in panel (a) are obtained as in Fig.~\ref{fig:Figure6}a. In panel (b), the dashed [dot-dashed] curves are given by Eqs.~\eqref{eq:xrmssExc-I}-\eqref{eq:xrpssExc-I} [\eqref{eq:xrmss-II},\eqref{eq:xlssExc}-\eqref{eq:xrpssExc}].}
    \label{fig:Figure7}
\end{figure}

Figure~\ref{fig:Figure7} displays the steady-state values of the quasiparticle densities for a qubit stabilized in the logical state 0 (Fig.~\ref{fig:Figure7} a) and 1 (Fig.~\ref{fig:Figure7} b) for the same parameter values as in Fig.~\ref{fig:Figure6}. For ground-state stabilization, the densities are approximately equal to the steady-state case, as the values of $p_1$ obtained through the solution of the full system are typically low  (a few percent).
The stabilization in the excited state has very little effect in case II, where the term $\bar\Gamma_{10}^{R<}$ is suppressed. In case I, stabilization is more effective. Indeed, while $\xRm$ is only slightly reduced, the quasiparticle densities $\xL$ and $\xRp$ are increased up to an a order of magnitude compared to the small $p_1$ regime for gap asymmetries $\wLR/(2\pi)\sim 4$~GHz
 close to the transition to case II. Such relatively large effect is seemingly in contrast with the expectation that qubit transitions are not effective at heating quasiparticles in devices with large electrodes~\cite{Basko}. However, the analysis in Ref.~\onlinecite{Basko} assumed equal gaps, and in the nearly symmetric case we also find negligible effect. Indeed, the enhanced influence of the qubit on the quasiparticles with increasing gap difference is a consequence of the growth of the rate $\bar\Gamma^{R<}_{10}$, see Fig.~\ref{fig:Figure4}. Future studies of the qubit-quasiparticle dynamics accounting for gap asymmetry may prove enlightening.
\subsection{Trapping effects}
\label{sec:trapping}

\begin{figure}
    \centering
    \includegraphics{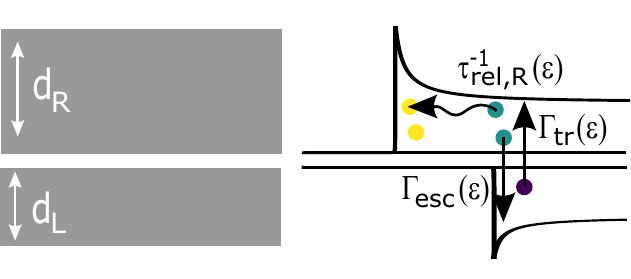}
    \caption{Left: Schematic of the typical superconducting bilayer away from the Josephson junction region. The first film is coupled to a second film (both in gray) by a large-area insulating barrier (white). Right: Quasiparticle band diagram for the two thin aluminum films. Quasiparticles in the higher gap electrode can tunnel into the lower-gap electrode at rate $\Gamma_\mathrm{tr}(\epsilon)$. In the top film, quasiparticles can relax to lower energies at rate $\tau_{rel,R}^{-1}(\epsilon)$ (see Appendix~\ref{app:RelaxationRecombination}) or tunnel back at rate $\Gamma_\mathrm{esc}(\epsilon)$. The three rates combine to give the effective trapping rate $\tau_{tr}^{-1}$ of Eq.~\eqref{eq:tautr}.}
    \label{fig:Figure8}
\end{figure}

So far, in our model for the quasiparticle dynamics we have included recombination within each electrode, but not the possible trapping of quasiparticles that can block them from tunneling through the junction and hence interacting with the qubit. As identified in Ref.~\cite{Wang2014}, we consider trapping effects arising from background and vortices. As for background trapping, it can at least in part be ascribed again to the gap difference in films of different thickness: away from the junction region, the two films again overlap, forming an electrode-wide junction in each electrode (see the schematic representation in Fig.~\ref{fig:Figure8}). Therefore the right electrode of gap $\dR$ is in contact with a film with gap $\dL > \dR$; however, this does not affect the dynamics of the densities $\xRm$ and $\xRp$. Indeed, for $\xRm$ the second film does not introduce new states that the quasiparticle could occupy; for $\xRp$, the quasiparticle can tunnel into the second film, but then they can only tunnel back, and since tunneling is fast compared to relaxation in the lower-gap film~\cite{RiwarPRB94}, the dynamics is unchanged.
For the left electrode with gap $\dL$, the situation is quite different: the quasiparticles that tunnel into the second film, which now has the lower gap $\dR$, can relax by emitting phonons; those quasiparticle do not interact with the qubit, so (at low temperature $T_\qp \ll \wLR$~\footnote{as temperature increases, quasiparticle can escape the trap due to thermal activation by phonons, see~\cite{Fei2022}.}) there is no mechanism for them to return to energy above $\dL$,  leading to their being trapped in the second film. As mentioned above, tunneling is fast compared to relaxation and therefore the latter is the bottleneck in the trapping process, as in the case of normal-metal traps~\cite{RiwarPRB94}. Adapting the derivation of the effective trapping rate for a normal trap to the present case of two superconductors with unequal gaps (see Fig.~\ref{fig:Figure8}), for the trapping rate $\tau_{tr}^{-1}$ of the quasiparticles in the left electrode we find
\begin{equation}\label{eq:tautr}
    \tau_{tr}^{-1} = \tau_R^{-1} \sqrt{\frac{T_\qp}{2\dL (1-\delta^2)}}
\end{equation}
Similarly to normal traps, the effective trapping rate is smaller than the relaxation rate in the trap (here, the second superconducting film) because of the rapid escape by tunneling from the trap. More precisely, the escape is enhanced by the large final density of states $\propto \epsilon/(\epsilon^2-\dL^2)^{1/2}$ in the first superconducting film at energies $\sim\dL$. Note that the trapping rate vanishes in the limit of equal gaps $\delta \to 1$, due to the rapid suppression of $\tau_R^{-1} \propto (1-\delta)^{7/2}$. Interestingly, with the parameters we are employing ($\dL/h =52\,$GHz, $r^L=(160~{\rm ns})^{-1}\,$, $T_\qp=20\,$mK), we estimate that $\tau_{tr}^{-1}$ can reach up to about 100~Hz, and can therefore explain most if not all of the observed background trapping that ranges between 30 and 180~Hz~\cite{Wang2014}. Moreover, in the Supplementary Note 5 to Ref.~\cite{Wang2014}, it was noted that measurements of the quasiparticle dynamics could be better modeled assuming background trapping to be present in only one electrode, as in the mechanism here considered.

This trapping process is accounted for by adding the term $-\tau_{tr}^{-1}\xL$ to the right-hand side of Eq.~(\ref{eq:dotxL}), which represents a sink for quasiparticles. Its qualitative effect is to lower the quasiparticle densities (at fixed values of other parameters). Indeed, for small gap asymmetry the only modification to the results presented in Sec.~\ref{sec:small_asymm} is to add a term $\tau_{tr}^{-1}$ in the denominator in the right-hand side of Eq.~(\ref{eq:xLss-I}). Therefore, at small asymmetry we expect little impact on $\xRp$ and $\xRm$, and a small decrease in $\xL$ which becomes more pronounced with increasing asymmetry. In fact, for large asymmetry relaxation and trapping become dominant, $\tau_R^{-1},\tau_{tr}^{-1}\gg \bar\Gamma_{00}^L,\bar\Gamma_{01}^L$. Therefore, following the approach of Sec.~\ref{sec:strong_asymm}, we find the approximate results $\xL \approx g\tau_{tr}$, $\xRp \approx g\tau_R$, and $\xRm \approx \sqrt{g/r^{R<}}\sqrt{1+\tau_{tr}\bar\Gamma_{01}^L}$. In comparison to the results in the absence of trapping, these equations predict a numerical factor reduction for $\xRp$ and $\xRm$, but a strong suppression for $\xL$.

We have verified the above results by repeating the numerical calculations of the previous two subsections. In Fig.~\ref{fig:Figure5}, the inclusion of trapping would give a sizable effect only for gap asymmetries larger than 2~GHz. In particular, $\xL$ is reduced with respect to the no-trapping case up to an order of magnitude [for $\wLR/(2\pi)\sim 5\,$GHz], while $\xRm$ and $\xRp$ are only slightly modified and still very well captured by the approximate estimates (dashed curves in Fig.~\ref{fig:Figure5}). The impact on the experimentally measurable quantities is rather limited, and characterized by reduced rates $T_1^{-1}\approx 125\,$kHz (down from 152 kHz), $\Gamma_{00}\approx 24\,$kHz (down from 40 kHz), and slightly increased qubit excited state population $p_1\approx0.009$ (up from 0.008) at the reference value $\wLR/(2\pi)\sim 5\,$~GHz. Concerning the state-of-the art transmon, similar conclusions on the rates and $p_1$ can be drawn for gap difference smaller than qubit frequency. For strongly asymmetric junctions discussed in Sec.~\ref{sec:strong_asymm} with $\wLR>\w_{10}$, the quasiparticle depletion in the higher-gap electrode is associated with a significant reduction of $p_1$ down to 1\% [for $\wLR/(2\pi)\sim 8\,$GHz], and no significant impact on the qubit rates, since even without trapping quasiparticle tunneling gives them a negligible contribution. Finally, we note that trapping by vortices has qualitatively similar, but quantitatively stronger, consequences. In particular, in the presence of a few vortices, the effects of nonequilibrium quasiparticles can be almost completely suppressed. Yet, vortex trapping can be absent if the qubit is well shielded from ambient magnetic field. A more detailed analysis is presented in Appendix~\ref{app:Vortex}. 

\section{Split transmon}
\label{sec:split}
\begin{figure}
    \centering
    \includegraphics{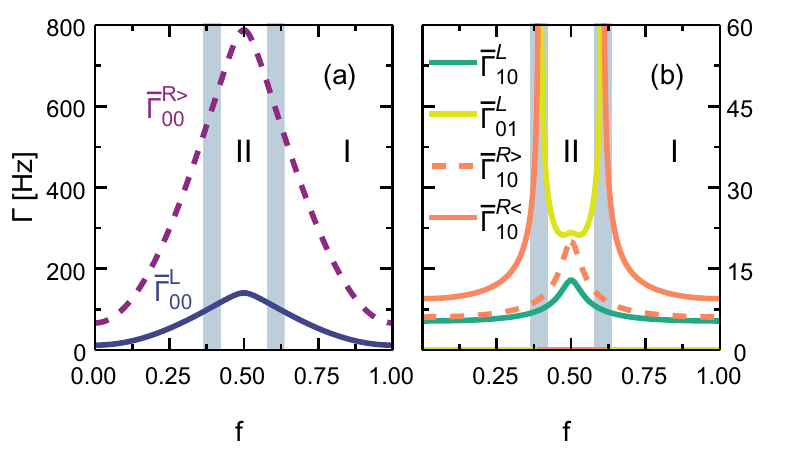}
    \caption{
    Quasiparticle rates vs normalized flux in a split transmon. (a) Parity switching rates. (b) Relaxation and excitation rates. (See Appendix~\ref{app:matSplit} for analytical expressions.) The shaded regions denote the region where the model cannot be applied. Parameters: $\dL/h=52$~GHz, $\dR/h=48$~GHz, $\omega_{10}(0)/(2\pi)=7$ GHz, $g_T^\Sigma=2\times 2.5 g_K$, $\nu_0=0.73\times10^{47}~{\rm J}^{-1}{\rm m}^{-3}$, $\mathcal V_L=\mathcal V_R=940~\mu$m$^3$, $T_\qp=20$~mK. }
    \label{fig:Figure9}
\end{figure}
In the previous section, we have discussed the steady-state values for the quasiparticle densities and the corresponding transition rates as a function of the gap difference $\wLR$. In particular, we have shown that quantities of experimental interest, such as the qubit excited state population or its inverse lifetime, may be significantly different depending on whether the qubit frequency $\w_{10}$ is higher or lower than the gap difference $\wLR$. The gap difference is determined by the nanofabrication process, through the thickness and the resistivity of the Al thin films composing the Josephson junctions, see Sec.~\ref{sec:intro}. The gaps can be controlled to some extent by applying a magnetic field parallel to the films, if their thickness is at least of order 7~nm~\cite{Adams2008}; the gap difference is expected to grow as the field increases, since thinner films have higher gap and  higher critical field. Although the overall suppression of the gaps also causes the qubit frequency to decrease, as confirmed in a recent experiment~\cite{Krause2022}, it should be possible to engineer devices in which the sign of $\w_{10} - \wLR$ can be flipped in parallel magnetic field.

Here we consider a qubit design in which the transition frequency $\w_{10}$ can be tuned on-chip. In the split transmon, a superconducting ring interrupted by two Josephson junctions replaces the single junction, so that the Josephson energy (and thus the frequency) can be controlled by threading a magnetic flux through the loop. More precisely, the Josephson energy $E_J(f)=E_J^\Sigma|\cos(\pi f)|\sqrt{1+d^2 \tan^2(\pi f)}$ and the qubit frequency $\w_{10}(f)=\sqrt{8E_J(f)E_C}-E_C$ are periodic functions of the normalized magnetic flux $f=\Phi/\Phi_0$ with period one, where $\Phi_0=h/2e$ is the flux quantum. The quantity $d=|E_J^1-E_J^2|/E_J^\Sigma$ accounts for the asymmetry between the two junctions and express the maximum relative suppression of the Josephson energy (occurring for $f=0.5$ modulo one) with respect to the zero-field value $E_J^\Sigma=E_J^1+E_J^2$ -- see also Appendix~\ref{app:Split}.

For a split transmon, the transition rates are properly computed by summing the contributions due to quasiparticles tunneling in each junction, as discussed in detail in Ref.~\cite{CatelaniPRB84}. As a result, the rates depend on the magnetic flux $f$ through both the frequency dependence of the spectral density [Eq.~\eqref{eq:Sqp_general}] and the matrix elements in Eq.~\eqref{wif_gen}, see Appendix~\ref{app:Split}. The qubit-quasiparticle dynamics is still described in terms of the generalized Rothwarf-Taylor equations Eqs.~\eqref{eq:dotxL}-\eqref{eq:dotxR<}.

Figure~\ref{fig:Figure9} displays the flux evolution of the transition rates conserving the total quasiparticle number in Eqs.~\eqref{eq:dotxL}-\eqref{eq:dotxR<} in a typical transmon device with $\w_{10}(0)>\wLR$ and focusing on a single period. The parity-switching rates, $\tilde\Gamma_{ii}^\alpha$, (see Fig.~\ref{fig:Figure9}a) and the relaxation rates due to quasiparticles located at energies larger than $\dL$, $\tilde\Gamma_{10}^L$ and $\tilde\Gamma_{10}^{R>}$ (Fig.~\ref{fig:Figure9}b), increase monotonically with $f$ in the half-period $[0,0.5]$. The relaxation rate associated to quasiparticle residing at energies $\sim\dR$ in the right lead $\tilde\Gamma_{10}^{R<}$ displays a monotonic growth for fluxes up to $f=f_c\approx 0.4$, where the BCS model predicts a divergence due to the resonant condition $\w_{10}(f_c)=\wLR$ (we recall however that our model cannot be applied here, see shaded regions in the plot), and it is zero for fluxes closer to the half-flux quantum. Similarly, the excitation rate $\tilde\Gamma_{01}^{L}$ is finite only for values of the flux where $\w_{10}(f)<\wLR$, i.e., $|f-0.5|< 0.5-f_c$. As mentioned above, a split transmon makes it in principle possible to switch between the two regimes, which we named case I and case II in the previous sections, if $\omega_{10}(0)> \wLR$ (case I) and $\omega_{10}(0.5) < \wLR$ (case II). 
\subsection{Steady state}

\begin{figure}
    \centering
    \includegraphics{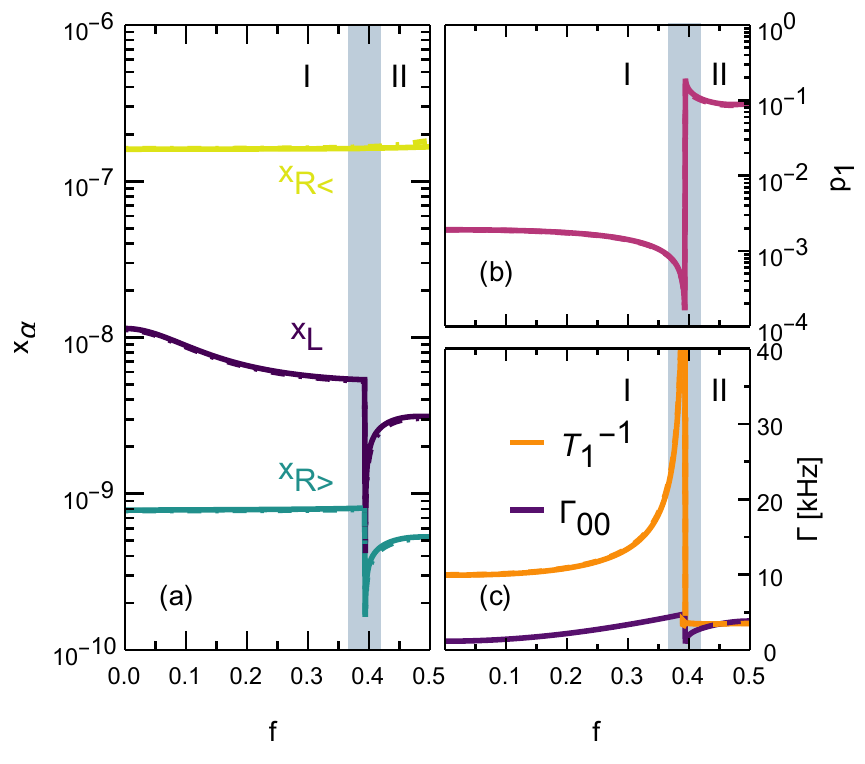}
    \caption{Qubit-quasiparticle steady state vs normalized flux for a split transmon. (a) Quasiparticle densities in the steady state. (b) Excited state population of the qubit. (c) Qubit inverse relaxation time and parity switching rate in the steady-state.
    Parameters: $\dL/h=52$~GHz, $\dR/h=48$~GHz, $\omega_{10}(0)/(2\pi)=7$~GHz, $g_T^\Sigma=2\times2.5 g_K$, $T_\qp=20$~mK, $\Gamma^{ph}=10$~Hz, $\mathcal V_L=\mathcal V_R=940~\mu$m$^3$, $\nu_0=0.73\times10^{47}~{\rm J}^{-1}{\rm m}^{-3}$, $\Gamma_{10}^{ee}=3$~kHz. The dashed lines, barely distinguishable from the solid ones, are obtained as in Fig.~\ref{fig:Figure6}, using the expressions for case I and the flux-dependent rates of Appendix~\ref{app:matSplit}.}
    \label{fig:Figure10}
\end{figure}

Here we briefly discuss the steady state of the qubit-quasiparticle system as a function of the magnetic flux $f$, for a split-transmon with $\w_{10}(0)>\wLR>\w_{10}(0.5)$. 
Figure~\ref{fig:Figure10} displays (a) the steady-state values of the quasiparticle densities, (b) the population of the excited state of the qubit and (c) the qubit relaxation and parity switching rates as a function of the reduced magnetic flux. Due to the sizable gap asymmetry, quasiparticles are mostly located in the lower-gap electrode at energies $\sim\dR$, in agreement with the discussion of Sec.~\ref{sec:steady}b. In particular, the density $\xRm\approx\sqrt{2g/r^{R<}}$ does not significantly depend on the magnetic flux, while the quasiparticle density in the left lead decreases monotonically by increasing the magnetic flux from zero up to the value $f_c$ where the resonant condition $\w_{10}(f_c) = \wLR$ is met.
In our modeling, the crossing of the resonant condition with flux significantly affects the population of the excited state of the qubit, which is enhanced for values of the flux where $\omega_{10}(f)<\wLR$~\footnote{In Ref.~\cite{Fedorov2020}, the excited state population was measured as function of flux, displaying a peak at an intermediate flux value. Its possible quasiparticle origin was excluded, based on the smooth dependence of the quasiparticle tunneling matrix elements on flux; however, at that time the role of gap difference was not considered. At those intermediate flux values, the qubit relaxation was Purcell-limited, so no corresponding peak in $T_1^{-1}$ could be observed.}. Similarly, the relaxation rate of the qubit is reduced for fluxes close to the half-quantum compared to fluxes such that $\omega_{10}(f) - \wLR \gg T_\qp$, and it is expected to grow as the resonance condition is approached in increasing flux due to the increase in the rate $\tilde\Gamma_{10}^{R<}$, cf. Fig.~\ref{fig:Figure9}b~\footnote{A peak in the total parity switching rate $\Gamma_{10}^{eo}+\Gamma_{11}^{eo}$ has been experimentally reported in Ref.~\onlinecite{DiamondArxiv2022} during the completion of this manuscript.} (we remind that our model cannot be quantitatively applied too close to the resonant flux value, where the condition $T_\qp\ll |\w_{10}-\wLR|$ is violated). The parity switching rate is less affected by the magnetic flux, and typically increases in the range $[0,f_c]$; the dip at $f>f_c$ reflects the decrease in the density $\xL$, which itself is a consequence of the finite value of $\bar\Gamma_{01}^L$ at those fluxes (cf. Sec.~\ref{sec:strong_asymm}).

Finally, we note that with our choice of parameters, namely photon energy large compared to $2\dL$ (see Sec.~\ref{sec:model}), the photon-assisted transition rate $\Gamma^{ph}$ and the generation rate $g$ depend only weakly on flux $f$, while a stronger dependence is expected for photons with energy closer to the threshold $\dL+\dR$, see Appendix~\ref{sec:gen_split}. Moreover, as in the case of the single-junction transmon, we expect that in the presence of vortices or engineered traps the quasiparticle densities, and hence the quasiparticle contributions to the transition rates, will be suppressed, cf. Appendix~\ref{app:Vortex}.

\section{Summary and outlook}
\label{sec:conclusions}

In this work we discuss the intrinsic decoherence in superconducting qubits due to quasiparticles. We generalize previous investigations to take into account the finite gap difference, or gap asymmetry, between the junction leads, see Sec.~\ref{sec:micromodel}. We model the coupled qubit-quasiparticle system, in which quasiparticles occupy narrow energy windows of width $T_\qp$ (the effective quasiparticle temperature) located just above the gaps $\dL$ and $\dR$ of the left and right leads. This approach can capture the presence of nonequilibrium quasiparticles in superconducting qubits operating at the base temperature of dilution refrigerators. For low temperatures $T_\qp$ compared to qubit frequency $\omega_{10}$ and gap difference $\wLR$, the model displays a weak dependence on $T_\qp$ and the quasiparticles can be described in terms of nonequilibrium densities, see Sec.~\ref{sec:model}. A graphic representation of the different physical processes we consider (tunneling, relaxation, generation, and recombination) is given in Fig.~\ref{fig:Figure3}. In Sec.~\ref{sec:steady} we discuss the steady-state values for the quasiparticle densities and their impact on the qubit, under the assumption that quasiparticle generation is dominated by pair-breaking at the junction due to the absorption of photons with energy well above the sum of the gaps $\dL+\dR$. Even though the gap asymmetry doesn't affect the total number of quasiparticles, we find that it has a strong impact on the qubit properties. We distinguish two cases, depending on the relation between qubit frequency and gap difference.

Case I: $\w_{10}-\wLR\gg T_{qp}$. Quasiparticles are present in both leads, so they can tunnel in both directions and absorb energy from the qubit, causing relaxation. In this regime the qubit lifetime can be significantly affected by nonequilibrium quasiparticles, if no or limited quasiparticle trapping is present. Indeed, by analyzing (Sec.~\ref{sec:small_asymm}) the experimental findings of Ref.~\onlinecite{Wang2014}, we find a good agreement with the reported data for devices where the quasiparticle dynamics is limited by recombination effects. 

Case II: $\wLR-\w_{10}\gg T_{qp}$.
In the case of qubit with large gap asymmetry, quasiparticle relaxation tends to be dominant, thus populating the states with energy close to the smaller gap $\dR$. Thus, quasiparticles are effectively trapped in the lower gap electrode and cannot tunnel to the higher-gap lead. Since a relaxation channel present in case I is not available in case II, typically the qubit lifetime improves for a given quasiparticle generation rate, but the population of the qubit excited state, and hence the qubit effective temperature, also increases, see Fig.~\ref{fig:Figure6}. Our investigation also shows that in general the effective qubit and quasiparticle temperatures are different, see Sec.~\ref{sec:strong_asymm}.

Quasiparticle effects can be mitigated through gap engineering. In this respect, the gap difference provides an intrinsic source of background trapping, as discussed in Sec.~\ref{sec:trapping}. Even with the improved filtering of state-of-the art setups, which lowers the generation rate by a few orders of magnitude, a full suppression of the nonequilibrium quasiparticle effects on the qubit is possibly achieved only through additional trapping (see also the discussion in Appendix~\ref{app:Vortex}).

While we specifically consider trasmon qubits, we expect our findings to be applicable to other qubit designs, such as the fluxonium~\cite{fluxonium}. Future analysis of our model could give insights into fundamental quantities in the superconducting state, such as the energy-dependent quasiparticle relaxation rates.  Hence, investigating the dynamics after stabilization (Sec.~\ref{sec:stabilize}), or taking advantage of the magnetic flux control in a split transmon (Sec.~\ref{sec:split}), may give information on such quantities. Our analysis also points toward the fact that the lifetime of the qubit can be severely limited when the qubit frequency is resonant with the gap difference. Therefore, from an experimental perspective, this situation should be avoided. On the theoretical side, the modeling in terms of quasiparticle densities cannot be adopted, and the predicted divergence within lowest-order perturbation theory applied to the BCS state needs to be properly renormalized, so further study of the resonant regime is needed.

\acknowledgments

We acknowledge S. Diamond, V. Fatemi, M. Hays, K. Serniak, and L. Glazman for discussions of their work~\cite{DiamondArxiv2022} prior to publication.
G.C. acknowledges support by the German Federal Ministry of Education and Research (BMBF), funding program ``Quantum technologies -- from basic research to market'', projects QSolid (Grant No. 13N16149) and GEQCOS (Grant No. 13N15685).

\appendix

\section{Matrix elements for single-junction transmon}
\label{app:matTra}

We denote the transmon states as $|ij\rangle$, with $i\in\{0,1\}$ for the logical value and $j\in\{e,\,o\}$ the parity. For both $\sin\hat\varphi/2$ and $\cos\hat\varphi/2$, matrix elements between states with the same parity vanish. For states of opposite parity ($\bar{e}=o$ and $\bar{o}=e$), 
in a single junction transmon the matrix elements associated with the transition between different qubit states are approximately~\cite{CatelaniPRB89}:
\begin{align}
\label{matSin10}
\langle 1j | \sin\frac{\hat\varphi}{2} | 0\bar{j} \rangle &\simeq \left(\frac{E_C}{8E_J}\right)^{1/4},\\
\label{matSin00}
\left| \langle ij | \sin\frac{\hat\varphi}{2} | i \bar{j} \rangle \right| &\simeq \left| \sin \left(2\pi n_g\right) \right| \left(\frac23\right)^{\frac{2}{3}}
\Gamma\left(\frac13\right) \left(\frac{E_C}{8E_J}\right)^{\frac{1}{6}} \frac{\varepsilon_i}{\omega_p} ,
\\
\label{matCos00}
\langle ij | \cos\frac{\hat\varphi}{2} | i\bar{j} \rangle &\simeq 1 - \left(i+\frac12\right)\sqrt{\frac{E_C}{8E_J}} - \frac32 \left(i+\frac14 \right)\frac{E_C}{8E_J},\\
\label{matCos10}
\left| \langle 1j | \cos\frac{\hat\varphi}{2} | 0 \bar{j} \rangle \right| &\propto \left| \sin \left(2\pi n_g\right) \right| \frac{\sqrt{|\varepsilon_0\varepsilon_1}|}{\omega_p} \left(\frac{E_C}{E_J}\right)^{1/3} \, .
\end{align}
Here $\Gamma$ denotes the gamma function and $\varepsilon_i$ the charge dispersion of level $i$ (that is, the energy difference between different parities):
\begin{equation}
    \varepsilon_i =(-1)^i E_C \sqrt{\frac{2}{\pi}}\frac{2^{2i+2}}{i!}\left(\frac{8E_J}{E_C}\right)^{(2i+3)/4}e^{-\sqrt{8E_J/E_C}}
\end{equation}
We note that at leading order in $E_J/E_C \gg 1$ the right-hand side of Eq.~\eqref{matCos00} is independent of $i$, and that due to their exponential suppression, the right-hand sides of Eqs.~\eqref{matSin00} and~\eqref{matCos10} can be set to zero. 

\section{Microscopic derivation of the transition rates}
\label{app:Micro}

The derivation of the transition rates for a qubit interacting with a quasiparticle bath has been reported in detail before, see \textit{e.g.}~\cite{CatelaniPRB84}. The extension to include the quasiparticle dynamics amounts to the derivation of the kinetic equation for the quasiparticle distribution function followed by integration over energy to arrive at equations for the quasiparticle densities. The quasiparticle kinetic equation has also been considered before, see \textit{e.g.}~\cite{Basko} and references therein, so here we only outline the main differences with previous treatments.

\subsection{Definitions of the densities}
\label{app:xRdef}

For the normalized density of quasiparticle in the left electrode our definition is the same as in the literature, namely
\begin{align}
\xL=\frac{2}{\dL}\int_{\dL}^{+\infty}d\epsilon \,
\frac{\epsilon f_L(\epsilon)}{\sqrt{\epsilon^2-\Delta_{L}^2}}.
\end{align}
where $f_L$ denotes the quasiparticle distribution function in the left lead.

For the right electrode we assume that, due to the competition between tunneling processes across the Josephson junction and  quasiparticle relaxation in the right electrode, the distribution function $f_R$ is the sum of two contributions with support in a narrow energy window  immediately above the two gaps $\dR$ and $\dL$; the width of this energy window is what we call the effective quasiparticle temperature $T_\qp$, and we assume $T_\qp\ll \wLR$. Therefore we write $f_R(\epsilon)=f_R^<(\epsilon)\theta(\dL-\epsilon)+f_R^>(\epsilon)\theta(\epsilon - \dL)$; correspondingly, we can identity two quasiparticle densities $\xRm$ and $\xRp$, where 
\begin{align}
\xRm&=\frac{2}{\dR}\int_{\dR}^{\dL}
d\epsilon \,
\frac{\epsilon f_R^<(\epsilon)}{\sqrt{\epsilon^2-\Delta_{R}^2}}\, ,\\
\xRp&=\frac{2}{\dR}\int_{\dL}^{+\infty}
d\epsilon \, \frac{\epsilon f_R^>(\epsilon)}{\sqrt{\epsilon^2-\Delta_{R}^2}}
 \\ & \simeq\frac{2}{\dR}\frac{\dL}{\sqrt{\Delta_{L}^2-\Delta_{R}^2}}\int_{\dL}^{+\infty}
d\epsilon \,
f_R^>(\epsilon) \, . \nonumber
\end{align}
The approximation in the last line follows from our assumption of low effective temperature compared to gap difference. 

\subsection{Transition rates for a single-junction transmon}
\label{app:qubitrates}

The golden-rule formula for the qubit transition rate $\Gamma_{if}$ is given in Eq.~\eqref{wif_gen}, with the matrix elements of Appendix~\ref{app:matTra}. Using the assumptions $T_\qp\ll \wLR,\,\omega_{10}, \, |\wLR-\omega_{10}|$, we can factorize the rate in the form $\Gamma_{if} = \tilde\Gamma_{if}^\alpha x_\alpha$. At leading order in $E_C/E_J \ll 1$, we find for the parity switching rates
\begin{align}
&\tilde\Gamma_{ii}^{L}
\simeq\frac{g_T\dL}{e^2}\frac{\dL-\dR}{\sqrt{\dL^2-\dR^2}}
\label{eq:GammaiiL}
\\
&\tilde\Gamma_{ii}^{R>}\simeq\frac{g_T\dR}{e^2}\sqrt{2}\frac{\dL-\dR}{\sqrt{T_{\rm qp}\dL}}
\label{eq:GammaiiR}
\end{align}
These expressions are valid for $T_\qp \ll \wLR$. We note that in writing $\tilde\Gamma_{ii}^{R>}$ we assume a constant value of $f_{R}^>(\e)$ in the narrow support $[\dL,\dL+T_\qp]$. Yet, we remark that our result is not strictly specific to this assumption and is expected to hold more generally within a numerical prefactor of order 1 (see also the discussion in Appendix A of Ref.~\onlinecite{RiwarPRB100}). Moreover, $\tilde\Gamma^{R<}_{ii} = 0$, as long as $\varepsilon_i < \omega_{LR}$. Introducing the short-hand notation $s_{10}=(E_C/8E_J)^{1/4}$, the relaxation rates are
\begin{align}
&\tilde\Gamma_{10}^{L}\simeq s_{10}^2\frac{g_T \dL}{e^2}\frac{\dL+\w_{10}+\dR}{\sqrt{(\dL+\w_{10})^2-\dR^2 }}
\label{eq:Gamma10L}
\\
&\tilde\Gamma_{10}^{R>}\simeq s_{10}^2\frac{g_T\dR}{e^2} \frac{\Delta_{L}+\w_{10}+\dR}{\sqrt{\omega_{10}(2\Delta_L+\omega_{10})}} 
\label{eq:Gamma10R>} \\
&\tilde\Gamma_{10}^{R<}\simeq
\left\{ \begin{array}{lc}
s_{10}^2 \frac{g_T\dR}{e^2} \frac{\dR+\w_{10}+\dL}{\sqrt{(\dR+\w_{10})^2-\dL^2 }} \phantom{m} & \mathrm{I}\\
0 & \mathrm{II} 
\end{array}\right. 
\label{eq:Gamma10R<}
\end{align}
where I and II refer to case I, $\omega_{10} - \wLR \gg T_\qp$, and case II, $\wLR - \omega_{10} \gg T_\qp$, respectively (cf. Sec.~\ref{sec:micromodel}). Finally the excitation rates are
\begin{align}
&\tilde\Gamma_{01}^{L}\simeq
\left\{ \begin{array}{lc}
0 & \mathrm{I} \\
s_{10}^2 \frac{g_T \dL}{e^2}\frac{\dL-\w_{10}+\dR}
{\sqrt{(\dL-\w_{10})^2-\dR^2 }} \phantom{m} &  \mathrm{II} 
\end{array} \right.
\label{eq:Gamma01L} \\
&\tilde\Gamma_{01}^{R>}\simeq 0
\label{eq:Gamma01R}
\end{align}
where the last line is a consequence of the assumption $T_\mathrm{qp} \ll \omega_{10}$. Finally, $\tilde\Gamma_{01}^{R<}$ = 0.

\subsection{Relaxation and recombination due to phonons}
\label{app:RelaxationRecombination}

The kinetic equation for the quasiparticle distribution function accounting for electron-phonon interaction can be written in the form $df_\alpha(\e)/dt = \mathrm{St}\{f_\alpha,n_\alpha\}$, where the collision integral $\mathrm{St}$ is in general a functional of the both the quasiparticle ($f_\alpha$) and phonon ($n_\alpha$) distribution functions in electrode $\alpha$. Explicit formulas for the collision integral can be found for instance in Refs.~\onlinecite{KaplanPRB14,ChangScalapinoPRB15}. Here we assume sufficiently low (effective) temperature, so that the phonons can be treated as a cold bath, $n_\alpha=0$, and the quasiparticle occupation probability is small, $f_\alpha \ll 1$. Then the collision integral consists of three terms, $\mathrm{St} = \mathrm{St}_{in} - \mathrm{St}_{out} - \mathrm{St}_{rec}$: an incoming term $\mathrm{St}_{in}$ due to relaxation of quasiparticle from higher energy, an outgoing term $\mathrm{St}_{out}$ due to  relaxation to lower energy, and a second outgoing term $\mathrm{St}_{rec}$ due to the recombination of two quasiparticles into a Cooper pair. All these processes take place via phonon emission.

With our assumptions, the three collision integrals are
\begin{align}
    &\mathrm{St}_{in} = 2\pi \int_0^\infty d\w F_\alpha(\w) \frac{\e(\e+\w)-\Delta_\alpha^2}{\e\sqrt{(\e+\w)^2-\Delta_\alpha^2}} f_\alpha (\e+\w)\\
    &\mathrm{St}_{out} = 2\pi \int_0^{\e-\Delta_\alpha}\!\!\!\! d\w F_\alpha(\w) \frac{\e(\e-\w)-\Delta_\alpha^2}{\e\sqrt{(\e-\w)^2-\Delta_\alpha^2}} f_\alpha (\e)\\
    &\mathrm{St}_{rec} = 2\pi \int_{\e+\Delta_\alpha}^{\infty} \!\!\!\!\!\!  d\w F_\alpha(\w) \frac{\e(\w-\e)+\Delta_\alpha^2}{\e\sqrt{(\w-\e)^2-\Delta_\alpha^2}}f_\alpha(\w-\e)f_\alpha(\e)
\end{align}
where the spectral function accounting for the matrix element of the electron-phonon interaction is
\be
F_\alpha(\w)=b_\alpha\w^2
\ee
with $b_\alpha$ a material-dependent parameter related to the electron-phonon coupling. The integral in the outgoing term can be carried out exactly to give
\be
\label{eq:StOut}
\mathrm{St}_{out} = \frac{1}{\tau_{0,\alpha}}\left(\frac{\Delta_\alpha}{T_{c,\alpha}}\right)^3 h(\epsilon/\Delta_\alpha) f_\alpha(\epsilon)
\ee
with $h(x)=(x^{2}-1)^{3/2}/3 + 5\sqrt{x^{2}-1}/2 - 
[2x+(2x)^{-1}]\ln(x + \sqrt{x^2 - 1})$. Using the notation of Ref.~\onlinecite{KaplanPRB14}, we have introduced the material-dependent rate $\tau_{0,\alpha}^{-1}=2\pi b_\alpha T_{c,\alpha}^3$, while the gap to critical temperature ratio to the cube is approximately 5.5 within BCS theory. The prefactor of $f_\alpha(\epsilon)$ in Eq.~\eqref{eq:StOut} expresses the energy-dependent relaxation rate in the superconducting state $\tau_{rel,\alpha}^{-1}(\epsilon)=\tau_{0,\alpha}^{-1}\left(\frac{\Delta_\alpha}{T_{c,\alpha}}\right)^3 h(\epsilon/\Delta_\alpha)$ (cf. Fig.~\ref{fig:Figure8}).

Considering first the left electrode, we multiply the kinetic equation by the density of states and integrate over all energies above $\dL$. The in and out contributions cancel out, while with our assumption of a narrow distribution function just above the gap, the recombination term gives
\be
\dot{x}_L = - r^L \xL^2
\ee
where $r^L= (4/\tau_{0,L})\left(\Delta_L/T_{c,L}\right)^3$.

For the right electrode, we consider separately the energies between $\dR$ and $\dL$ and above $\dL$, see Appendix~\ref{app:xRdef}. After energy integration we find
\begin{align}
    \dot{x}_{R>} & = -r^{R>} \xRp^2 - r^{<>} \xRm\xRp  -\tau_R^{-1} \xRp\\
    \dot{x}_{R<} & = -r^{R<} \xRm^2 - r^{<>} \xRm\xRp + \tau_R^{-1} \xRp
\end{align}
where $r^{R<}= (4/\tau_{0,R})\left(\Delta_R/T_{c,R}\right)^3$, $r^{<>}=r^{R<}(1+\delta)^3/8\delta^2$, and $r^{R>} = r^{R<}(1+\delta^2)/2\delta^2$. Note that by splitting the energy integrals, the cancellation between in and out terms is only partial, and the remaining contributions lead to the relaxation terms with $\tau_R^{-1} =\tau_{rel,R}^{-1}(\Delta_L)= r^{R<} h(\dL/\dR)/4$. In the calculations of the main text, we set $r^L=r^{R<}$. 
\subsection{Generation by photon-assisted tunneling}
\label{app:gen}

The rate of qubit transition in the presence of pair-breaking photons of frequency $\omega_\nu$ is given in Ref.~\cite{PRL123Photon}. It takes a structure similar to that in Eq.~\eqref{wif_gen}, namely the sum of two terms, each the product of a matrix element times a spectral density,
\begin{align}
    \Gamma_{if}^{ph} = \Gamma_{\nu}\frac{g_T\dL}{8g_K}\left[\left|\langle i | \cos\frac{\hat\varphi}{2} | f\rangle \right|^2 
S_-^{ph}\left(\frac{\w_\nu+\w_{if}}{\dL}; \frac{\dR}{\dL}\right) \right. \nonumber \\ + \left.
\left|\langle i | \sin\frac{\hat\varphi}{2} | f\rangle \right|^2 S_+^{ph}\left(\frac{\w_\nu+\w_{if}}{\dL};\frac{\dR}{\dL}\right)\right]
\label{eq:Gph}
\end{align}
For notational simplicity, compared to Appendix~\ref{app:matTra} we have dropped from the qubit state vector the parity label, since parity must change for the matrix element to be non-zero. The dimensionless prefactor $\Gamma_\nu$ accounts for the coupling strength between qubit and photons as well as the average number of photons~\cite{PRL123Photon}, and we remind that $\omega_{if}$ is the energy difference between initial and final qubit states. The spectral densities
\begin{equation}
S_\pm^{ph}(x;z)=\int_{1}^{\infty}dy\int_{z}^{\infty}dy'\frac{yy'\pm z}{\sqrt{y^2-1}\sqrt{y'^2-z^2}}\delta(x-y-y')
\label{eq:Sph_pm}
\end{equation}
generalize those of Ref.~\cite{PRL123Photon} to the case of unequal left/right superconducting gaps and have been defined taking the larger gap $\dL$ as reference for the energy scale. Clearly, $S_\pm^{ph}(x;z)=0$ for $x<1+z$.

The integral over $y'$ in Eq.~\eqref{eq:Sph_pm} represents the integral over quasiparticle energies in the right lead; more precisely, integration over the domain $z<y'<1$ corresponds to the contribution for energies between $\dR$ and $\dL$, while that for $y'>1$ to energies above $\dL$. Therefore, assuming $x>2$ the ratio between $g^{R<}$ and $g^{R>}$ is given by the ratio of the expressions obtained by restricting the integral over $y'$ to those two intervals. After integrating over $y'$ using the $\delta$-function and changing integration variable from $y$ to $u$ using $y=x-u$ for the low-energy integral and $y=1+xu$ in the high-energy one, we find that at leading order in $x \gg 1$ the ratio is given by $\sqrt{1-z^2}/x$. Therefore, for large photon energy we arrive at $g^{R<}/g^{R>}\simeq \sqrt{\dL^2 - \dR^2}/\omega_\nu \ll 1$.

\section{Vortex trapping}
\label{app:Vortex}

Thin superconducting films generally behaves as type II superconductors, so that in
the presence of out-of-plane magnetic fields normal-core vortices are generated in the films, introducing additional trapping for quasiparticle excitations. A detailed discussion of the vortex-trapping modeling can be found in the Supplementary Material of Ref.~\cite{Wang2014}. Effectively, vortex trapping induces outgoing terms in Eqs.\eqref{eq:dotxL}-\eqref{eq:dotxR<} proportional to the quasiparticle densities. For a small number of vortices~\cite{Wang2014}, vortex trapping adds a term $-N_L\tau_V^{-1} \xL$ to the right-hand side of Eq.~\eqref{eq:dotxL}, where $\tau_V^{-1} = P/A$ is the trapping rate due to a single vortex, and $N_L$ is the number of vortices in the left electrode. Here, $P$ is the vortex trapping power introduced in Ref.~\onlinecite{Wang2014} and $A$ is the area of the electrode. Assuming that the trapping into vortices is mediated by relaxation via phonon emission, the corresponding terms to add to the right-hand sides of Eqs.~(\ref{eq:dotxR>}) and (\ref{eq:dotxR<}) are $-N_R[1-(1-\delta)^3]\tau_V^{-1} \xRp$ and $-N_R\delta^3\tau_V^{-1} \xRm$, respectively (the cubic dependencies arise from integration over the phonon frequency of the electron-phonon matrix element, see Appendix~\ref{app:RelaxationRecombination}). In the presence of at least one vortex in both electrodes, all the non-linear terms in Eqs.~(\ref{eq:dotxL})-(\ref{eq:dotxR<}) can be dropped, since for typical experimental parameters $r^\alpha x_\alpha\ll \tau_V^{-1}$; the steady-state values of $x_\alpha$ are then obtained by solving the linear system. We find 
\begin{widetext}
\begin{align}
    x_L^{\rm vortex} & =
    g \,\frac{2\bar\Gamma_{00}^{R>}+ \tau_R^{-1}+N_R[1-(1-\delta)^3]\tau_V^{-1}}{\bar\Gamma_{00}^L\tau_R^{-1}+N_R[1-(1-\delta)^3]\bar\Gamma_{00}^L\tau_V^{-1}+
    (\bar\Gamma_{01}^L + N_L\tau_V^{-1}/\delta)(\bar\Gamma_{00}^{R>} +\tau_R^{-1}+N_R[1-(1-\delta)^3]\tau_V^{-1})} \, ,\label{eq:xLVortex} \\
    x_{R>}^{\rm vortex} & =  g \,\frac{2\bar\Gamma_{00}^L+\bar\Gamma_{01}^L + N_L\tau_V^{-1}/\delta}{\bar\Gamma_{00}^L\tau_R^{-1}+N_R[1-(1-\delta)^3]\bar\Gamma_{00}^L\tau_V^{-1}+(\bar\Gamma_{01}^L + N_L\tau_V^{-1}/\delta)(\bar\Gamma_{00}^{R>} +\tau_R^{-1}+N_R[1-(1-\delta)^3]\tau_V^{-1})} \, ,
    \label{eq:xR>Vortex} \\
    x_{R<}^{\rm vortex} & =\frac{\tau_V}{N_R\delta^3}(\bar\Gamma_{01}^L\xL^{\rm vortex}+\tau_R^{-1}\xRp^{\rm vortex})\, ,
\label{eq:xR<Vortex}
\end{align}
\end{widetext}
where the expressions for $\xL$ and $\xRp$ are readily obtained from Eqs.~\eqref{eq:xlss-II}-\eqref{eq:xrpss-II} with the replacements $\tau_R^{-1} \to \tau_R^{-1}+N_R[1-(1-\delta)^3]\tau_V^{-1}$ and $\bar\Gamma_{01}^L \to \bar\Gamma_{01}^L + N_L\tau_V^{-1}/\delta$. 

In Ref.~\onlinecite{Wang2014}, the vortex trapping power $P=0.067\,$cm$^2/$s was measured; for device B2 in that reference, $A=1.68\times 10^{-4}~$cm$^2$ and the single-vortex trapping rate is $\tau_V^{-1} \simeq 400\,$Hz. We note that the simplified approach adopted here overestimates the trapping rate, and a more accurate estimate can be obtained by taking into account the actual geometry of the device. Experimentally, the increase in the trapping rate by adding a single vortex was of the order 200~Hz, and we use this value for our estimates. Using the expressions Eqs.~\eqref{eq:xLVortex}-\eqref{eq:xR<Vortex}, the steady-state densities $\xL$ and $\xRp$ of Fig.~\ref{fig:Figure5} are reduced down to about $2\times 10^{-8}$ (or even less for the latter) when a single vortex is present in each electrode. Similar values are found for $\xRm$. As a consequence of the strong reduction of the densities in comparison with the estimates in Fig.~\ref{fig:Figure5}, $T_1^{-1}$ is reduced by approximately 30~kHz when a single vortex is included in the left electrode. Upon including an additional vortex in the right electrode, quasiparticle tunnelling would give a negligible contribution to the total relaxation and parity switching rates and to determining the qubit excited population [cf. Eq.~(\ref{p1ssI})]. 
Our findings are in reasonable agreement with the measurements on device B2 of Ref.~\onlinecite{Wang2014}, where upon the addition of a single vortex the decay rate fell by about 35~kHz, followed by another decrease  after the addition of a vortex in the opposite pad, and no more changes in $T_1$ (within error bars) with more vortices. In contrast, device B1 showed a more gradual reduction in the relaxation rate with increasing number of vortices~\cite{Wang2014}. The inclusion of a background source of quasiparticle generation could further improve the quantitative agreement between theory and experiment, and possibly reduce the discrepancy between our estimate of the generation rate and that in Ref.~\onlinecite{Wang2014} mentioned in Sec.~\ref{sec:small_asymm}. A more detailed comparison with experiments is left to future studies.

Finally, we comment briefly on the effect of vortices in qubits with large electrodes (type A devices in Ref.~\onlinecite{Wang2014} with electrode area $A = 250\times500\,\mu\mathrm{m}^2$). While the single-vortex trapping rate $\tau_V^{-1}$ decreases with increasing area $A$, the number of vortices in the electrode for a given magnetic field increases. In fact, it is likely that several vortices are always present in large electrodes, leading to a total trapping rate in each electrode 
likely larger than our estimate for a single vortex $\tau_V^{-1}\approx 200~$Hz. Then, based on our analysis, we expect that for values of $\Gamma^{ph}$ of order 10~Hz or lower, the quasiparticle tunneling contribution to the qubit transition rates is strongly suppressed by vortex trapping. Based on the results of Refs.~\cite{RiwarPRB94,RiwarPRB100}, we expect this conclusion to hold also in the presence of engineered normal or superconducting traps.

\section{Split transmon}
\label{app:Split}

A split transmon comprises a superconducting ring interrupted by two Josephson junctions. Nonetheless, the system has a single degree of freedom $\hat\phi$, due to the fluxoid quantization condition $\varphi_1 +\varphi_2=2\pi f$; here $f=\Phi/\Phi_0$, $\Phi$ is the external flux piercing the ring, $\Phi_0$ is the magnetic flux quantum, and $\varphi_1$ and $\varphi_2$ are the phase differences across the two junctions.
The qubit Hamiltonian can be written as
\be
\hat{H}_{\varphi}= 4E_C (\hat{N}-n_g)^2 -E_J(f) \cos (\hat\phi-\vartheta)
\label{eq:splitH}
\ee
which represents an effective single-junction qubit with flux-tunable Josephson energy
\be
E_J(f)=(E_{J}^1+E_{J}^2)|\cos\pi f|\sqrt{1+d^2\tan^2 (\pi f)} \equiv E_{J}^\Sigma\mathcal G(f,d)
\label{eq:splitTransmonEJ}
\ee
where in the last equality we defined $E_{J}^\Sigma=E_{J}^1+E_{J}^2$ and $\mathcal G(f,d)=|\cos\pi f|\sqrt{1+d^2\tan^2 \pi f}$. In these expressions, $d=(E_{J}^1-E_{J}^2)/(E_{J}^1+E_{J}^2)$ accounts for the asymmetry between the two junctions of the transmon, and $\tan \vartheta=d\tan(\pi f)$.

\subsection{Transition rates and matrix elements}
\label{app:matSplit}

The transition rates for a split-transmon are obtained through a generalization of the single-junction results~\cite{CatelaniPRB84},
\begin{align}
\label{wif_split_transmon}
\Gamma_{i\to f} = \sum_{n=1,2}&\left|\langle i|\sin \frac{\hat\varphi_n}{2}|f\rangle\right|^2
E_{J}^n\tilde S^{+,n}_\qp\left(\w_{if}\right)
\\
+&\left|\langle i|\cos \frac{\hat\varphi_n}{2}|f\rangle\right|^2
 E_{J}^n\tilde S^{-,n}_\qp\left(\w_{if}\right), \nonumber
\end{align}
where each spectral density has been scaled to the respective Josephson energy, that is, $\tilde S^{\pm,n}_\qp=S^{\pm,n}_\qp/E_{J}^n$. Similar to the effective Josephson energy, the matrix elements are functions of the scaled flux, and they can be calculated as in the single-junction transmon using the relations $\hat \varphi_n= \pi f\pm(\vartheta+\hat\phi)$ ($+$ for $n=2$ and $-$ for $n=1$). At leading order in $E_C/E_J(f)$ we find (with $j$ denoting parity)
\begin{align}
\label{matSin10Split}
\left|\langle 1j | \sin\frac{\hat\varphi_n}{2} | 0\bar{j} \rangle\right|^2 &\simeq \left[\frac{E_C}{8E_J(f)}\right]^{1/2}\frac{1+\cos(\pi f\pm\vartheta)}{2}\, ,\\
\label{matSin00Split}
 \left|\langle ij| \sin\frac{\hat\varphi_n}{2} | i\bar{j} \rangle\right|^2  &\simeq \frac{1-\cos(\pi f\pm\vartheta)}{2}\, , \\
\label{matCos00Split}
\left|\langle ij | \cos\frac{\hat\varphi_n}{2} | i\bar{j} \rangle\right|^2 &\simeq \frac{1+\cos(\pi f\pm\vartheta)}{2}\, ,\\
\label{matCos10Split}
\left|\langle 1j | \cos\frac{\hat\varphi_n}{2} | 0\bar{j} \rangle\right|^2  &\simeq\left[\frac{E_C}{8E_J(f)}\right]^{1/2}\frac{1-\cos(\pi f\pm\vartheta)}{2} \, ,
\end{align}
up to exponentially small corrections of the form given in Appendix~\ref{app:matTra}.

The rates entering Eqs.~\eqref{eq:dotxL}-\eqref{eq:dotxR<} for the split transmon are obtained after summing the contributions from the two junctions, Eq.~\eqref{wif_split_transmon}; they can be found from those of the single-junction transmon, Eqs.~\eqref{eq:GammaiiL}-\eqref{eq:Gamma01R}, upon the substitutions
\begin{align}
    & g_T \to g_T^\Sigma\,, \quad  \omega_{10} \to \omega_{10}(f) \\
    & s_{10} \to s_{10}(f) = [E_C/8E_J(f)]^{1/4}
\end{align}
where $g_T^\Sigma = g_T^1+g_T^2$ is the sum of the conductances of the two junctions, and upon multiplication of the last term in the right-most numerator by $\mathcal{G}(f,d)$. Taking Eq.~\eqref{eq:Gamma10L} as an example, this procedure gives
\begin{equation}
   \tilde\Gamma_{10}^{L}\simeq s_{10}^2(f) \frac{g_T^\Sigma \dL}{e^2}\frac{\dL+\w_{10}(f)+\mathcal{G}(f,d)\dR}{\sqrt{[\dL+\w_{10}(f)]^2-\dR^2 }}. 
\end{equation}

\subsection{Generation rates for a split transmon}
\label{sec:gen_split}

Analogously to the quasiparticle tunneling transition rates in Eq.~\eqref{wif_split_transmon},
the photon-assisted rates can be obtained through a generalization of Eq.~\eqref{eq:Gph},
\begin{align} \label{eq:GammaPhmn}
& \Gamma_{if}=\Gamma_{\nu}\sum_{n=1,2}\frac{g_T^n\dL}{8g_K}\left[ S_-^{ph}\left(\frac{\w_\nu+\w_{if}}{\dL};\frac{\dR}{\dL}\right)  \right.  \\ & \left. \left|\langle i | \cos\frac{\hat\varphi_n}{2} | f\rangle \right|^2\!\!+
 S_+^{ph}\left(\frac{\w_\nu+\w_{if}}{\dL};\frac{\dR}{\dL}\right)\left|\langle i | \sin\frac{\hat\varphi_n}{2} | f\rangle \right|^2\right]\!.
\nonumber
\end{align}

Using the expressions for the matrix elements of Sec.~\ref{app:matSplit}, we can write the parity switching rates as
\begin{equation}\label{eq:G00ph_split}
\Gamma_{00}^{ph}\approx \Gamma_{11}^{ph}\approx \Gamma_\nu\frac{g_T^\Sigma \Delta_L}{16 g_K}[S_+^{ph} + S_-^{ph} - (S_+^{ph} - S_-^{ph})\mathcal G(f,d)]
\end{equation}
where the arguments of the spectral functions are $\omega_\nu/\dL$ and $\dR/\dL$, and the relaxation/excitation rates as
\begin{align}\label{eq:G10ph_split}
\Gamma_{10}^{ph}\approx \Gamma_{01}^{ph}\approx&\,\Gamma_\nu\frac{E_C}{\omega_{10}(f)}\frac{g_T^\Sigma \Delta_L}{16 g_K} \\ &\, [S_+^{ph} + S_-^{ph} + (S_+^{ph} - S_-^{ph})\mathcal G(f,d)] \nonumber
\end{align}
where the arguments of the spectral functions are $[\omega_\nu\pm\omega_{10}(f)]/\dL$ and $\dR/\dL$ ($+$ for qubit relaxation, $-$ for excitation).

In the limit of high photon energy $\w_\nu\gg 2\Delta_L$, we can neglect the qubit frequency $\omega_{10}(f)$ in comparison with $\omega_\nu$, approximate $S_-^{ph}(\w_\nu/\dL,\dR/\dL)\approx S_+^{ph}(\w_\nu/\dL,\dR/\dL)$, and rewrite the total generation rate per single lead as 
\be\label{eq:gen_split}
\Gamma_{10}^{ph}+\Gamma_{00}^{ph}=\Gamma^{ph}\left[\frac{1}{\sqrt{\mathcal G(f,d)}}+\sqrt{\frac{8E_J^\Sigma}{E_C}}\right]
\ee
where
\begin{equation}
\Gamma^{ph}\equiv\Gamma_{10}^{ph}(\tilde\Phi=0)=\Gamma_\nu \frac{E_C}{ \w_{10}(0)} \frac{g_T^\Sigma \dL}{8 g_K} S_+^{ph}\left(\frac{\w_\nu}{\dL},\frac{\dR}{\dL}\right).
\end{equation}
Note that since $d\le\mathcal{G}(f,d)\le 1$, a strong dependence on flux in Eq.~\eqref{eq:gen_split} is possible only for small asymmetry $d\ll E_C/8E_J^\Sigma$. However, we remind that the approximation $S_-^{ph} \approx S_+^{ph}$ is valid only in the limit of large photon frequency; for frequencies closer to $\dL+\dR$, we cannot neglect the terms proportional to $\mathcal{G}$ in Eqs.~\eqref{eq:G00ph_split} and \eqref{eq:G10ph_split}, and hence a stronger dependence on flux is expected.


\begin{thebibliography}{79}%
	\makeatletter
	\providecommand \@ifxundefined [1]{%
		\@ifx{#1\undefined}
	}%
	\providecommand \@ifnum [1]{%
		\ifnum #1\expandafter \@firstoftwo
		\else \expandafter \@secondoftwo
		\fi
	}%
	\providecommand \@ifx [1]{%
		\ifx #1\expandafter \@firstoftwo
		\else \expandafter \@secondoftwo
		\fi
	}%
	\providecommand \natexlab [1]{#1}%
	\providecommand \enquote  [1]{``#1''}%
	\providecommand \bibnamefont  [1]{#1}%
	\providecommand \bibfnamefont [1]{#1}%
	\providecommand \citenamefont [1]{#1}%
	\providecommand \href@noop [0]{\@secondoftwo}%
	\providecommand \href [0]{\begingroup \@sanitize@url \@href}%
	\providecommand \@href[1]{\@@startlink{#1}\@@href}%
	\providecommand \@@href[1]{\endgroup#1\@@endlink}%
	\providecommand \@sanitize@url [0]{\catcode `\\12\catcode `\$12\catcode
		`\&12\catcode `\#12\catcode `\^12\catcode `\_12\catcode `\%12\relax}%
	\providecommand \@@startlink[1]{}%
	\providecommand \@@endlink[0]{}%
	\providecommand \url  [0]{\begingroup\@sanitize@url \@url }%
	\providecommand \@url [1]{\endgroup\@href {#1}{\urlprefix }}%
	\providecommand \urlprefix  [0]{URL }%
	\providecommand \Eprint [0]{\href }%
	\providecommand \doibase [0]{https://doi.org/}%
	\providecommand \selectlanguage [0]{\@gobble}%
	\providecommand \bibinfo  [0]{\@secondoftwo}%
	\providecommand \bibfield  [0]{\@secondoftwo}%
	\providecommand \translation [1]{[#1]}%
	\providecommand \BibitemOpen [0]{}%
	\providecommand \bibitemStop [0]{}%
	\providecommand \bibitemNoStop [0]{.\EOS\space}%
	\providecommand \EOS [0]{\spacefactor3000\relax}%
	\providecommand \BibitemShut  [1]{\csname bibitem#1\endcsname}%
	\let\auto@bib@innerbib\@empty
	\bibitem [{\citenamefont {Blais}\ \emph {et~al.}(2021)\citenamefont {Blais},
		\citenamefont {Grimsmo}, \citenamefont {Girvin},\ and\ \citenamefont
		{Wallraff}}]{BlaisRMP93}%
	\BibitemOpen
	\bibfield  {author} {\bibinfo {author} {\bibfnamefont {A.}~\bibnamefont
			{Blais}}, \bibinfo {author} {\bibfnamefont {A.~L.}\ \bibnamefont {Grimsmo}},
		\bibinfo {author} {\bibfnamefont {S.~M.}\ \bibnamefont {Girvin}},\ and\
		\bibinfo {author} {\bibfnamefont {A.}~\bibnamefont {Wallraff}},\ }\bibfield
	{title} {\bibinfo {title} {Circuit quantum electrodynamics},\ }\href
	{https://doi.org/10.1103/RevModPhys.93.025005} {\bibfield  {journal}
		{\bibinfo  {journal} {Rev. Mod. Phys.}\ }\textbf {\bibinfo {volume} {93}},\
		\bibinfo {pages} {025005} (\bibinfo {year} {2021})}\BibitemShut {NoStop}%
	\bibitem [{\citenamefont {Devoret}\ and\ \citenamefont
		{Schoelkopf}(2013)}]{DevoretScience339}%
	\BibitemOpen
	\bibfield  {author} {\bibinfo {author} {\bibfnamefont {M.~H.}\ \bibnamefont
			{Devoret}}\ and\ \bibinfo {author} {\bibfnamefont {R.~J.}\ \bibnamefont
			{Schoelkopf}},\ }\bibfield  {title} {\bibinfo {title} {Superconducting
			circuits for quantum information: An outlook},\ }\href
	{https://doi.org/10.1126/science.1231930} {\bibfield  {journal} {\bibinfo
			{journal} {Science}\ }\textbf {\bibinfo {volume} {339}},\ \bibinfo {pages}
		{1169} (\bibinfo {year} {2013})}\BibitemShut {NoStop}%
	\bibitem [{\citenamefont {Gao}\ \emph {et~al.}(2021)\citenamefont {Gao},
		\citenamefont {Rol}, \citenamefont {Touzard},\ and\ \citenamefont
		{Wang}}]{GaoPRXQuantum2}%
	\BibitemOpen
	\bibfield  {author} {\bibinfo {author} {\bibfnamefont {Y.~Y.}\ \bibnamefont
			{Gao}}, \bibinfo {author} {\bibfnamefont {M.~A.}\ \bibnamefont {Rol}},
		\bibinfo {author} {\bibfnamefont {S.}~\bibnamefont {Touzard}},\ and\ \bibinfo
		{author} {\bibfnamefont {C.}~\bibnamefont {Wang}},\ }\bibfield  {title}
	{\bibinfo {title} {Practical guide for building superconducting quantum
			devices},\ }\href {https://doi.org/10.1103/PRXQuantum.2.040202} {\bibfield
		{journal} {\bibinfo  {journal} {PRX Quantum}\ }\textbf {\bibinfo {volume}
			{2}},\ \bibinfo {pages} {040202} (\bibinfo {year} {2021})}\BibitemShut
	{NoStop}%
	\bibitem [{\citenamefont {Esmaeil~Zadeh}\ \emph {et~al.}(2021)\citenamefont
		{Esmaeil~Zadeh}, \citenamefont {Chang}, \citenamefont {Los}, \citenamefont
		{Gyger}, \citenamefont {Elshaari}, \citenamefont {Steinhauer}, \citenamefont
		{Dorenbos},\ and\ \citenamefont {Zwiller}}]{EsmaeilAPL118}%
	\BibitemOpen
	\bibfield  {author} {\bibinfo {author} {\bibfnamefont {I.}~\bibnamefont
			{Esmaeil~Zadeh}}, \bibinfo {author} {\bibfnamefont {J.}~\bibnamefont
			{Chang}}, \bibinfo {author} {\bibfnamefont {J.~W.~N.}\ \bibnamefont {Los}},
		\bibinfo {author} {\bibfnamefont {S.}~\bibnamefont {Gyger}}, \bibinfo
		{author} {\bibfnamefont {A.~W.}\ \bibnamefont {Elshaari}}, \bibinfo {author}
		{\bibfnamefont {S.}~\bibnamefont {Steinhauer}}, \bibinfo {author}
		{\bibfnamefont {S.~N.}\ \bibnamefont {Dorenbos}},\ and\ \bibinfo {author}
		{\bibfnamefont {V.}~\bibnamefont {Zwiller}},\ }\bibfield  {title} {\bibinfo
		{title} {Superconducting nanowire single-photon detectors: A perspective on
			evolution, state-of-the-art, future developments, and applications},\ }\href
	{https://doi.org/10.1063/5.0045990} {\bibfield  {journal} {\bibinfo
			{journal} {Appl. Phys. Lett.}\ }\textbf {\bibinfo {volume} {118}},\ \bibinfo
		{pages} {190502} (\bibinfo {year} {2021})}\BibitemShut {NoStop}%
	\bibitem [{\citenamefont {Arute}\ \emph {et~al.}(2019)\citenamefont {Arute},
		\citenamefont {Arya}, \citenamefont {Babbush}, \citenamefont {Bacon},
		\citenamefont {Bardin}, \citenamefont {Barends}, \citenamefont {Biswas},
		\citenamefont {Boixo}, \citenamefont {Brandao}, \citenamefont {Buell} \emph
		{et~al.}}]{Arute2019}%
	\BibitemOpen
	\bibfield  {author} {\bibinfo {author} {\bibfnamefont {F.}~\bibnamefont
			{Arute}}, \bibinfo {author} {\bibfnamefont {K.}~\bibnamefont {Arya}},
		\bibinfo {author} {\bibfnamefont {R.}~\bibnamefont {Babbush}}, \bibinfo
		{author} {\bibfnamefont {D.}~\bibnamefont {Bacon}}, \bibinfo {author}
		{\bibfnamefont {J.~C.}\ \bibnamefont {Bardin}}, \bibinfo {author}
		{\bibfnamefont {R.}~\bibnamefont {Barends}}, \bibinfo {author} {\bibfnamefont
			{R.}~\bibnamefont {Biswas}}, \bibinfo {author} {\bibfnamefont
			{S.}~\bibnamefont {Boixo}}, \bibinfo {author} {\bibfnamefont {F.~G. S.~L.}\
			\bibnamefont {Brandao}}, \bibinfo {author} {\bibfnamefont {D.~A.}\
			\bibnamefont {Buell}}, \emph {et~al.},\ }\bibfield  {title} {\bibinfo {title}
		{Quantum supremacy using a programmable superconducting processor},\ }\href
	{https://doi.org/10.1038/s41586-019-1666-5} {\bibfield  {journal} {\bibinfo
			{journal} {Nature}\ }\textbf {\bibinfo {volume} {574}},\ \bibinfo {pages}
		{505} (\bibinfo {year} {2019})}\BibitemShut {NoStop}%
	\bibitem [{\citenamefont {Wu}\ \emph {et~al.}(2021)\citenamefont {Wu},
		\citenamefont {Bao}, \citenamefont {Cao}, \citenamefont {Chen}, \citenamefont
		{Chen}, \citenamefont {Chen}, \citenamefont {Chung}, \citenamefont {Deng},
		\citenamefont {Du}, \citenamefont {Fan}, \citenamefont {Gong}, \citenamefont
		{Guo}, \citenamefont {Guo}, \citenamefont {Guo}, \citenamefont {Han},
		\citenamefont {Hong}, \citenamefont {Huang}, \citenamefont {Huo},
		\citenamefont {Li}, \citenamefont {Li}, \citenamefont {Li}, \citenamefont
		{Li}, \citenamefont {Liang}, \citenamefont {Lin}, \citenamefont {Lin},
		\citenamefont {Qian}, \citenamefont {Qiao}, \citenamefont {Rong},
		\citenamefont {Su}, \citenamefont {Sun}, \citenamefont {Wang}, \citenamefont
		{Wang}, \citenamefont {Wu}, \citenamefont {Xu}, \citenamefont {Yan},
		\citenamefont {Yang}, \citenamefont {Yang}, \citenamefont {Ye}, \citenamefont
		{Yin}, \citenamefont {Ying}, \citenamefont {Yu}, \citenamefont {Zha},
		\citenamefont {Zhang}, \citenamefont {Zhang}, \citenamefont {Zhang},
		\citenamefont {Zhang}, \citenamefont {Zhao}, \citenamefont {Zhao},
		\citenamefont {Zhou}, \citenamefont {Zhu}, \citenamefont {Lu}, \citenamefont
		{Peng}, \citenamefont {Zhu},\ and\ \citenamefont {Pan}}]{ZuchongzhiPRL127}%
	\BibitemOpen
	\bibfield  {author} {\bibinfo {author} {\bibfnamefont {Y.}~\bibnamefont
			{Wu}}, \bibinfo {author} {\bibfnamefont {W.-S.}\ \bibnamefont {Bao}},
		\bibinfo {author} {\bibfnamefont {S.}~\bibnamefont {Cao}}, \bibinfo {author}
		{\bibfnamefont {F.}~\bibnamefont {Chen}}, \bibinfo {author} {\bibfnamefont
			{M.-C.}\ \bibnamefont {Chen}}, \bibinfo {author} {\bibfnamefont
			{X.}~\bibnamefont {Chen}}, \bibinfo {author} {\bibfnamefont {T.-H.}\
			\bibnamefont {Chung}}, \bibinfo {author} {\bibfnamefont {H.}~\bibnamefont
			{Deng}}, \bibinfo {author} {\bibfnamefont {Y.}~\bibnamefont {Du}}, \bibinfo
		{author} {\bibfnamefont {D.}~\bibnamefont {Fan}}, \bibinfo {author}
		{\bibfnamefont {M.}~\bibnamefont {Gong}}, \bibinfo {author} {\bibfnamefont
			{C.}~\bibnamefont {Guo}}, \bibinfo {author} {\bibfnamefont {C.}~\bibnamefont
			{Guo}}, \bibinfo {author} {\bibfnamefont {S.}~\bibnamefont {Guo}}, \bibinfo
		{author} {\bibfnamefont {L.}~\bibnamefont {Han}}, \bibinfo {author}
		{\bibfnamefont {L.}~\bibnamefont {Hong}}, \bibinfo {author} {\bibfnamefont
			{H.-L.}\ \bibnamefont {Huang}}, \bibinfo {author} {\bibfnamefont {Y.-H.}\
			\bibnamefont {Huo}}, \bibinfo {author} {\bibfnamefont {L.}~\bibnamefont
			{Li}}, \bibinfo {author} {\bibfnamefont {N.}~\bibnamefont {Li}}, \bibinfo
		{author} {\bibfnamefont {S.}~\bibnamefont {Li}}, \bibinfo {author}
		{\bibfnamefont {Y.}~\bibnamefont {Li}}, \bibinfo {author} {\bibfnamefont
			{F.}~\bibnamefont {Liang}}, \bibinfo {author} {\bibfnamefont
			{C.}~\bibnamefont {Lin}}, \bibinfo {author} {\bibfnamefont {J.}~\bibnamefont
			{Lin}}, \bibinfo {author} {\bibfnamefont {H.}~\bibnamefont {Qian}}, \bibinfo
		{author} {\bibfnamefont {D.}~\bibnamefont {Qiao}}, \bibinfo {author}
		{\bibfnamefont {H.}~\bibnamefont {Rong}}, \bibinfo {author} {\bibfnamefont
			{H.}~\bibnamefont {Su}}, \bibinfo {author} {\bibfnamefont {L.}~\bibnamefont
			{Sun}}, \bibinfo {author} {\bibfnamefont {L.}~\bibnamefont {Wang}}, \bibinfo
		{author} {\bibfnamefont {S.}~\bibnamefont {Wang}}, \bibinfo {author}
		{\bibfnamefont {D.}~\bibnamefont {Wu}}, \bibinfo {author} {\bibfnamefont
			{Y.}~\bibnamefont {Xu}}, \bibinfo {author} {\bibfnamefont {K.}~\bibnamefont
			{Yan}}, \bibinfo {author} {\bibfnamefont {W.}~\bibnamefont {Yang}}, \bibinfo
		{author} {\bibfnamefont {Y.}~\bibnamefont {Yang}}, \bibinfo {author}
		{\bibfnamefont {Y.}~\bibnamefont {Ye}}, \bibinfo {author} {\bibfnamefont
			{J.}~\bibnamefont {Yin}}, \bibinfo {author} {\bibfnamefont {C.}~\bibnamefont
			{Ying}}, \bibinfo {author} {\bibfnamefont {J.}~\bibnamefont {Yu}}, \bibinfo
		{author} {\bibfnamefont {C.}~\bibnamefont {Zha}}, \bibinfo {author}
		{\bibfnamefont {C.}~\bibnamefont {Zhang}}, \bibinfo {author} {\bibfnamefont
			{H.}~\bibnamefont {Zhang}}, \bibinfo {author} {\bibfnamefont
			{K.}~\bibnamefont {Zhang}}, \bibinfo {author} {\bibfnamefont
			{Y.}~\bibnamefont {Zhang}}, \bibinfo {author} {\bibfnamefont
			{H.}~\bibnamefont {Zhao}}, \bibinfo {author} {\bibfnamefont {Y.}~\bibnamefont
			{Zhao}}, \bibinfo {author} {\bibfnamefont {L.}~\bibnamefont {Zhou}}, \bibinfo
		{author} {\bibfnamefont {Q.}~\bibnamefont {Zhu}}, \bibinfo {author}
		{\bibfnamefont {C.-Y.}\ \bibnamefont {Lu}}, \bibinfo {author} {\bibfnamefont
			{C.-Z.}\ \bibnamefont {Peng}}, \bibinfo {author} {\bibfnamefont
			{X.}~\bibnamefont {Zhu}},\ and\ \bibinfo {author} {\bibfnamefont {J.-W.}\
			\bibnamefont {Pan}},\ }\bibfield  {title} {\bibinfo {title} {Strong quantum
			computational advantage using a superconducting quantum processor},\ }\href
	{https://doi.org/10.1103/PhysRevLett.127.180501} {\bibfield  {journal}
		{\bibinfo  {journal} {Phys. Rev. Lett.}\ }\textbf {\bibinfo {volume} {127}},\
		\bibinfo {pages} {180501} (\bibinfo {year} {2021})}\BibitemShut {NoStop}%
	\bibitem [{\citenamefont {Krantz}\ \emph {et~al.}(2019)\citenamefont {Krantz},
		\citenamefont {Kjaergaard}, \citenamefont {Yan}, \citenamefont {Orlando},
		\citenamefont {Gustavsson},\ and\ \citenamefont {Oliver}}]{OliverReview}%
	\BibitemOpen
	\bibfield  {author} {\bibinfo {author} {\bibfnamefont {P.}~\bibnamefont
			{Krantz}}, \bibinfo {author} {\bibfnamefont {M.}~\bibnamefont {Kjaergaard}},
		\bibinfo {author} {\bibfnamefont {F.}~\bibnamefont {Yan}}, \bibinfo {author}
		{\bibfnamefont {T.~P.}\ \bibnamefont {Orlando}}, \bibinfo {author}
		{\bibfnamefont {S.}~\bibnamefont {Gustavsson}},\ and\ \bibinfo {author}
		{\bibfnamefont {W.~D.}\ \bibnamefont {Oliver}},\ }\bibfield  {title}
	{\bibinfo {title} {A quantum engineer's guide to superconducting qubits},\
	}\href {https://doi.org/10.1063/1.5089550} {\bibfield  {journal} {\bibinfo
			{journal} {Appl. Phys. Rev.}\ }\textbf {\bibinfo {volume} {6}},\ \bibinfo
		{pages} {021318} (\bibinfo {year} {2019})}\BibitemShut {NoStop}%
	\bibitem [{\citenamefont {Wang}\ \emph {et~al.}(2022)\citenamefont {Wang},
		\citenamefont {Li}, \citenamefont {Xu}, \citenamefont {Li}, \citenamefont
		{Wang}, \citenamefont {Yang}, \citenamefont {Mi}, \citenamefont {Liang},
		\citenamefont {Su}, \citenamefont {Yang}, \citenamefont {Wang}, \citenamefont
		{Wang}, \citenamefont {Li}, \citenamefont {Chen}, \citenamefont {Li},
		\citenamefont {Linghu}, \citenamefont {Han}, \citenamefont {Zhang},
		\citenamefont {Feng}, \citenamefont {Song}, \citenamefont {Ma}, \citenamefont
		{Zhang}, \citenamefont {Wang}, \citenamefont {Zhao}, \citenamefont {Liu},
		\citenamefont {Xue}, \citenamefont {Jin},\ and\ \citenamefont
		{Yu}}]{Wang2022}%
	\BibitemOpen
	\bibfield  {author} {\bibinfo {author} {\bibfnamefont {C.}~\bibnamefont
			{Wang}}, \bibinfo {author} {\bibfnamefont {X.}~\bibnamefont {Li}}, \bibinfo
		{author} {\bibfnamefont {H.}~\bibnamefont {Xu}}, \bibinfo {author}
		{\bibfnamefont {Z.}~\bibnamefont {Li}}, \bibinfo {author} {\bibfnamefont
			{J.}~\bibnamefont {Wang}}, \bibinfo {author} {\bibfnamefont {Z.}~\bibnamefont
			{Yang}}, \bibinfo {author} {\bibfnamefont {Z.}~\bibnamefont {Mi}}, \bibinfo
		{author} {\bibfnamefont {X.}~\bibnamefont {Liang}}, \bibinfo {author}
		{\bibfnamefont {T.}~\bibnamefont {Su}}, \bibinfo {author} {\bibfnamefont
			{C.}~\bibnamefont {Yang}}, \bibinfo {author} {\bibfnamefont {G.}~\bibnamefont
			{Wang}}, \bibinfo {author} {\bibfnamefont {W.}~\bibnamefont {Wang}}, \bibinfo
		{author} {\bibfnamefont {Y.}~\bibnamefont {Li}}, \bibinfo {author}
		{\bibfnamefont {M.}~\bibnamefont {Chen}}, \bibinfo {author} {\bibfnamefont
			{C.}~\bibnamefont {Li}}, \bibinfo {author} {\bibfnamefont {K.}~\bibnamefont
			{Linghu}}, \bibinfo {author} {\bibfnamefont {J.}~\bibnamefont {Han}},
		\bibinfo {author} {\bibfnamefont {Y.}~\bibnamefont {Zhang}}, \bibinfo
		{author} {\bibfnamefont {Y.}~\bibnamefont {Feng}}, \bibinfo {author}
		{\bibfnamefont {Y.}~\bibnamefont {Song}}, \bibinfo {author} {\bibfnamefont
			{T.}~\bibnamefont {Ma}}, \bibinfo {author} {\bibfnamefont {J.}~\bibnamefont
			{Zhang}}, \bibinfo {author} {\bibfnamefont {R.}~\bibnamefont {Wang}},
		\bibinfo {author} {\bibfnamefont {P.}~\bibnamefont {Zhao}}, \bibinfo {author}
		{\bibfnamefont {W.}~\bibnamefont {Liu}}, \bibinfo {author} {\bibfnamefont
			{G.}~\bibnamefont {Xue}}, \bibinfo {author} {\bibfnamefont {Y.}~\bibnamefont
			{Jin}},\ and\ \bibinfo {author} {\bibfnamefont {H.}~\bibnamefont {Yu}},\
	}\bibfield  {title} {\bibinfo {title} {Towards practical quantum computers:
			transmon qubit with a lifetime approaching 0.5 milliseconds},\ }\href
	{https://doi.org/10.1038/s41534-021-00510-2} {\bibfield  {journal} {\bibinfo
			{journal} {npj Quantum Inf.}\ }\textbf {\bibinfo {volume} {8}},\ \bibinfo
		{pages} {3} (\bibinfo {year} {2022})}\BibitemShut {NoStop}%
	\bibitem [{\citenamefont {Nakamura}\ \emph {et~al.}(1999)\citenamefont
		{Nakamura}, \citenamefont {Pashkin},\ and\ \citenamefont {Tsai}}]{Nakamura}%
	\BibitemOpen
	\bibfield  {author} {\bibinfo {author} {\bibfnamefont {Y.}~\bibnamefont
			{Nakamura}}, \bibinfo {author} {\bibfnamefont {Y.~A.}\ \bibnamefont
			{Pashkin}},\ and\ \bibinfo {author} {\bibfnamefont {J.~S.}\ \bibnamefont
			{Tsai}},\ }\bibfield  {title} {\bibinfo {title} {Coherent control of
			macroscopic quantum states in a single-{Cooper}-pair box},\ }\href
	{https://doi.org/10.1038/19718} {\bibfield  {journal} {\bibinfo  {journal}
			{Nature}\ }\textbf {\bibinfo {volume} {398}},\ \bibinfo {pages} {786}
		(\bibinfo {year} {1999})}\BibitemShut {NoStop}%
	\bibitem [{\citenamefont {Siddiqi}(2021)}]{SiddiqiReview}%
	\BibitemOpen
	\bibfield  {author} {\bibinfo {author} {\bibfnamefont {I.}~\bibnamefont
			{Siddiqi}},\ }\bibfield  {title} {\bibinfo {title} {Engineering
			high-coherence superconducting qubits},\ }\href
	{https://doi.org/10.1038/s41578-021-00370-4} {\bibfield  {journal} {\bibinfo
			{journal} {Nat. Rev. Mater.}\ }\textbf {\bibinfo {volume} {6}},\ \bibinfo
		{pages} {875} (\bibinfo {year} {2021})}\BibitemShut {NoStop}%
	\bibitem [{\citenamefont {Vissers}\ \emph {et~al.}(2010)\citenamefont
		{Vissers}, \citenamefont {Gao}, \citenamefont {Wisbey}, \citenamefont {Hite},
		\citenamefont {Tsuei}, \citenamefont {Corcoles}, \citenamefont {Steffen},\
		and\ \citenamefont {Pappas}}]{TiN2010}%
	\BibitemOpen
	\bibfield  {author} {\bibinfo {author} {\bibfnamefont {M.~R.}\ \bibnamefont
			{Vissers}}, \bibinfo {author} {\bibfnamefont {J.}~\bibnamefont {Gao}},
		\bibinfo {author} {\bibfnamefont {D.~S.}\ \bibnamefont {Wisbey}}, \bibinfo
		{author} {\bibfnamefont {D.~A.}\ \bibnamefont {Hite}}, \bibinfo {author}
		{\bibfnamefont {C.~C.}\ \bibnamefont {Tsuei}}, \bibinfo {author}
		{\bibfnamefont {A.~D.}\ \bibnamefont {Corcoles}}, \bibinfo {author}
		{\bibfnamefont {M.}~\bibnamefont {Steffen}},\ and\ \bibinfo {author}
		{\bibfnamefont {D.~P.}\ \bibnamefont {Pappas}},\ }\bibfield  {title}
	{\bibinfo {title} {Low loss superconducting titanium nitride coplanar
			waveguide resonators},\ }\href {https://doi.org/10.1063/1.3517252} {\bibfield
		{journal} {\bibinfo  {journal} {Appl. Phys. Lett.}\ }\textbf {\bibinfo
			{volume} {97}},\ \bibinfo {pages} {232509} (\bibinfo {year}
		{2010})}\BibitemShut {NoStop}%
	\bibitem [{\citenamefont {Barends}\ \emph {et~al.}(2010)\citenamefont
		{Barends}, \citenamefont {Vercruyssen}, \citenamefont {Endo}, \citenamefont
		{de~Visser}, \citenamefont {Zijlstra}, \citenamefont {Klapwijk},
		\citenamefont {Diener}, \citenamefont {Yates},\ and\ \citenamefont
		{Baselmans}}]{NbTiN2010}%
	\BibitemOpen
	\bibfield  {author} {\bibinfo {author} {\bibfnamefont {R.}~\bibnamefont
			{Barends}}, \bibinfo {author} {\bibfnamefont {N.}~\bibnamefont
			{Vercruyssen}}, \bibinfo {author} {\bibfnamefont {A.}~\bibnamefont {Endo}},
		\bibinfo {author} {\bibfnamefont {P.~J.}\ \bibnamefont {de~Visser}}, \bibinfo
		{author} {\bibfnamefont {T.}~\bibnamefont {Zijlstra}}, \bibinfo {author}
		{\bibfnamefont {T.~M.}\ \bibnamefont {Klapwijk}}, \bibinfo {author}
		{\bibfnamefont {P.}~\bibnamefont {Diener}}, \bibinfo {author} {\bibfnamefont
			{S.~J.~C.}\ \bibnamefont {Yates}},\ and\ \bibinfo {author} {\bibfnamefont
			{J.~J.~A.}\ \bibnamefont {Baselmans}},\ }\bibfield  {title} {\bibinfo {title}
		{Minimal resonator loss for circuit quantum electrodynamics},\ }\href
	{https://doi.org/10.1063/1.3458705} {\bibfield  {journal} {\bibinfo
			{journal} {Appl. Phys. Lett.}\ }\textbf {\bibinfo {volume} {97}},\ \bibinfo
		{pages} {023508} (\bibinfo {year} {2010})}\BibitemShut {NoStop}%
	\bibitem [{\citenamefont {Samkharadze}\ \emph {et~al.}(2016)\citenamefont
		{Samkharadze}, \citenamefont {Bruno}, \citenamefont {Scarlino}, \citenamefont
		{Zheng}, \citenamefont {DiVincenzo}, \citenamefont {DiCarlo},\ and\
		\citenamefont {Vandersypen}}]{NbTiN2016}%
	\BibitemOpen
	\bibfield  {author} {\bibinfo {author} {\bibfnamefont {N.}~\bibnamefont
			{Samkharadze}}, \bibinfo {author} {\bibfnamefont {A.}~\bibnamefont {Bruno}},
		\bibinfo {author} {\bibfnamefont {P.}~\bibnamefont {Scarlino}}, \bibinfo
		{author} {\bibfnamefont {G.}~\bibnamefont {Zheng}}, \bibinfo {author}
		{\bibfnamefont {D.~P.}\ \bibnamefont {DiVincenzo}}, \bibinfo {author}
		{\bibfnamefont {L.}~\bibnamefont {DiCarlo}},\ and\ \bibinfo {author}
		{\bibfnamefont {L.~M.~K.}\ \bibnamefont {Vandersypen}},\ }\bibfield  {title}
	{\bibinfo {title} {High-kinetic-inductance superconducting nanowire
			resonators for circuit qed in a magnetic field},\ }\href
	{https://doi.org/10.1103/PhysRevApplied.5.044004} {\bibfield  {journal}
		{\bibinfo  {journal} {Phys. Rev. Applied}\ }\textbf {\bibinfo {volume} {5}},\
		\bibinfo {pages} {044004} (\bibinfo {year} {2016})}\BibitemShut {NoStop}%
	\bibitem [{\citenamefont {Rotzinger}\ \emph {et~al.}(2016)\citenamefont
		{Rotzinger}, \citenamefont {Skacel}, \citenamefont {Pfirrmann}, \citenamefont
		{Voss}, \citenamefont {Münzberg}, \citenamefont {Probst}, \citenamefont
		{Bushev}, \citenamefont {Weides}, \citenamefont {Ustinov},\ and\
		\citenamefont {Mooij}}]{Rotzinger_2016}%
	\BibitemOpen
	\bibfield  {author} {\bibinfo {author} {\bibfnamefont {H.}~\bibnamefont
			{Rotzinger}}, \bibinfo {author} {\bibfnamefont {S.~T.}\ \bibnamefont
			{Skacel}}, \bibinfo {author} {\bibfnamefont {M.}~\bibnamefont {Pfirrmann}},
		\bibinfo {author} {\bibfnamefont {J.~N.}\ \bibnamefont {Voss}}, \bibinfo
		{author} {\bibfnamefont {J.}~\bibnamefont {Münzberg}}, \bibinfo {author}
		{\bibfnamefont {S.}~\bibnamefont {Probst}}, \bibinfo {author} {\bibfnamefont
			{P.}~\bibnamefont {Bushev}}, \bibinfo {author} {\bibfnamefont {M.~P.}\
			\bibnamefont {Weides}}, \bibinfo {author} {\bibfnamefont {A.~V.}\
			\bibnamefont {Ustinov}},\ and\ \bibinfo {author} {\bibfnamefont {J.~E.}\
			\bibnamefont {Mooij}},\ }\bibfield  {title} {\bibinfo {title}
		{Aluminium-oxide wires for superconducting high kinetic inductance
			circuits},\ }\href {https://doi.org/10.1088/0953-2048/30/2/025002} {\bibfield
		{journal} {\bibinfo  {journal} {Supercond. Sci. Technol.}\ }\textbf
		{\bibinfo {volume} {30}},\ \bibinfo {pages} {025002} (\bibinfo {year}
		{2016})}\BibitemShut {NoStop}%
	\bibitem [{\citenamefont {Gr\"unhaupt}\ \emph {et~al.}(2018)\citenamefont
		{Gr\"unhaupt}, \citenamefont {Maleeva}, \citenamefont {Skacel}, \citenamefont
		{Calvo}, \citenamefont {Levy-Bertrand}, \citenamefont {Ustinov},
		\citenamefont {Rotzinger}, \citenamefont {Monfardini}, \citenamefont
		{Catelani},\ and\ \citenamefont {Pop}}]{PRL121Res}%
	\BibitemOpen
	\bibfield  {author} {\bibinfo {author} {\bibfnamefont {L.}~\bibnamefont
			{Gr\"unhaupt}}, \bibinfo {author} {\bibfnamefont {N.}~\bibnamefont
			{Maleeva}}, \bibinfo {author} {\bibfnamefont {S.~T.}\ \bibnamefont {Skacel}},
		\bibinfo {author} {\bibfnamefont {M.}~\bibnamefont {Calvo}}, \bibinfo
		{author} {\bibfnamefont {F.}~\bibnamefont {Levy-Bertrand}}, \bibinfo {author}
		{\bibfnamefont {A.~V.}\ \bibnamefont {Ustinov}}, \bibinfo {author}
		{\bibfnamefont {H.}~\bibnamefont {Rotzinger}}, \bibinfo {author}
		{\bibfnamefont {A.}~\bibnamefont {Monfardini}}, \bibinfo {author}
		{\bibfnamefont {G.}~\bibnamefont {Catelani}},\ and\ \bibinfo {author}
		{\bibfnamefont {I.~M.}\ \bibnamefont {Pop}},\ }\bibfield  {title} {\bibinfo
		{title} {Loss mechanisms and quasiparticle dynamics in superconducting
			microwave resonators made of thin-film granular aluminum},\ }\href
	{https://doi.org/10.1103/PhysRevLett.121.117001} {\bibfield  {journal}
		{\bibinfo  {journal} {Phys. Rev. Lett.}\ }\textbf {\bibinfo {volume} {121}},\
		\bibinfo {pages} {117001} (\bibinfo {year} {2018})}\BibitemShut {NoStop}%
	\bibitem [{\citenamefont {Maleeva}\ \emph {et~al.}(2018)\citenamefont
		{Maleeva}, \citenamefont {Grünhaupt}, \citenamefont {Klein}, \citenamefont
		{Levy-Bertrand}, \citenamefont {Dupre}, \citenamefont {Calvo}, \citenamefont
		{Valenti}, \citenamefont {Winkel}, \citenamefont {Friedrich}, \citenamefont
		{Wernsdorfer}, \citenamefont {Ustinov}, \citenamefont {Rotzinger},
		\citenamefont {Monfardini}, \citenamefont {Fistul},\ and\ \citenamefont
		{Pop}}]{Maleeva}%
	\BibitemOpen
	\bibfield  {author} {\bibinfo {author} {\bibfnamefont {N.}~\bibnamefont
			{Maleeva}}, \bibinfo {author} {\bibfnamefont {L.}~\bibnamefont {Grünhaupt}},
		\bibinfo {author} {\bibfnamefont {T.}~\bibnamefont {Klein}}, \bibinfo
		{author} {\bibfnamefont {F.}~\bibnamefont {Levy-Bertrand}}, \bibinfo {author}
		{\bibfnamefont {O.}~\bibnamefont {Dupre}}, \bibinfo {author} {\bibfnamefont
			{M.}~\bibnamefont {Calvo}}, \bibinfo {author} {\bibfnamefont
			{F.}~\bibnamefont {Valenti}}, \bibinfo {author} {\bibfnamefont
			{P.}~\bibnamefont {Winkel}}, \bibinfo {author} {\bibfnamefont
			{F.}~\bibnamefont {Friedrich}}, \bibinfo {author} {\bibfnamefont
			{W.}~\bibnamefont {Wernsdorfer}}, \bibinfo {author} {\bibfnamefont {A.~V.}\
			\bibnamefont {Ustinov}}, \bibinfo {author} {\bibfnamefont {H.}~\bibnamefont
			{Rotzinger}}, \bibinfo {author} {\bibfnamefont {A.}~\bibnamefont
			{Monfardini}}, \bibinfo {author} {\bibfnamefont {M.~V.}\ \bibnamefont
			{Fistul}},\ and\ \bibinfo {author} {\bibfnamefont {I.~M.}\ \bibnamefont
			{Pop}},\ }\bibfield  {title} {\bibinfo {title} {Circuit quantum
			electrodynamics of granular aluminum resonators},\ }\href
	{https://doi.org/10.1038/s41467-018-06386-9} {\bibfield  {journal} {\bibinfo
			{journal} {Nat. Commun.}\ }\textbf {\bibinfo {volume} {9}},\ \bibinfo {pages}
		{3889} (\bibinfo {year} {2018})}\BibitemShut {NoStop}%
	\bibitem [{\citenamefont {Chang}\ \emph {et~al.}(2013)\citenamefont {Chang},
		\citenamefont {Vissers}, \citenamefont {Córcoles}, \citenamefont {Sandberg},
		\citenamefont {Gao}, \citenamefont {Abraham}, \citenamefont {Chow},
		\citenamefont {Gambetta}, \citenamefont {Beth~Rothwell}, \citenamefont
		{Keefe}, \citenamefont {Steffen},\ and\ \citenamefont {Pappas}}]{TiN2013}%
	\BibitemOpen
	\bibfield  {author} {\bibinfo {author} {\bibfnamefont {J.~B.}\ \bibnamefont
			{Chang}}, \bibinfo {author} {\bibfnamefont {M.~R.}\ \bibnamefont {Vissers}},
		\bibinfo {author} {\bibfnamefont {A.~D.}\ \bibnamefont {Córcoles}}, \bibinfo
		{author} {\bibfnamefont {M.}~\bibnamefont {Sandberg}}, \bibinfo {author}
		{\bibfnamefont {J.}~\bibnamefont {Gao}}, \bibinfo {author} {\bibfnamefont
			{D.~W.}\ \bibnamefont {Abraham}}, \bibinfo {author} {\bibfnamefont {J.~M.}\
			\bibnamefont {Chow}}, \bibinfo {author} {\bibfnamefont {J.~M.}\ \bibnamefont
			{Gambetta}}, \bibinfo {author} {\bibfnamefont {M.}~\bibnamefont
			{Beth~Rothwell}}, \bibinfo {author} {\bibfnamefont {G.~A.}\ \bibnamefont
			{Keefe}}, \bibinfo {author} {\bibfnamefont {M.}~\bibnamefont {Steffen}},\
		and\ \bibinfo {author} {\bibfnamefont {D.~P.}\ \bibnamefont {Pappas}},\
	}\bibfield  {title} {\bibinfo {title} {Improved superconducting qubit
			coherence using titanium nitride},\ }\href
	{https://doi.org/10.1063/1.4813269} {\bibfield  {journal} {\bibinfo
			{journal} {Appl. Phys. Lett.}\ }\textbf {\bibinfo {volume} {103}},\ \bibinfo
		{pages} {012602} (\bibinfo {year} {2013})}\BibitemShut {NoStop}%
	\bibitem [{\citenamefont {Christensen}\ \emph {et~al.}(2019)\citenamefont
		{Christensen}, \citenamefont {Wilen}, \citenamefont {Opremcak}, \citenamefont
		{Nelson}, \citenamefont {Schlenker}, \citenamefont {Zimonick}, \citenamefont
		{Faoro}, \citenamefont {Ioffe}, \citenamefont {Rosen}, \citenamefont
		{DuBois}, \citenamefont {Plourde},\ and\ \citenamefont
		{McDermott}}]{McDermott2019}%
	\BibitemOpen
	\bibfield  {author} {\bibinfo {author} {\bibfnamefont {B.~G.}\ \bibnamefont
			{Christensen}}, \bibinfo {author} {\bibfnamefont {C.~D.}\ \bibnamefont
			{Wilen}}, \bibinfo {author} {\bibfnamefont {A.}~\bibnamefont {Opremcak}},
		\bibinfo {author} {\bibfnamefont {J.}~\bibnamefont {Nelson}}, \bibinfo
		{author} {\bibfnamefont {F.}~\bibnamefont {Schlenker}}, \bibinfo {author}
		{\bibfnamefont {C.~H.}\ \bibnamefont {Zimonick}}, \bibinfo {author}
		{\bibfnamefont {L.}~\bibnamefont {Faoro}}, \bibinfo {author} {\bibfnamefont
			{L.~B.}\ \bibnamefont {Ioffe}}, \bibinfo {author} {\bibfnamefont {Y.~J.}\
			\bibnamefont {Rosen}}, \bibinfo {author} {\bibfnamefont {J.~L.}\ \bibnamefont
			{DuBois}}, \bibinfo {author} {\bibfnamefont {B.~L.~T.}\ \bibnamefont
			{Plourde}},\ and\ \bibinfo {author} {\bibfnamefont {R.}~\bibnamefont
			{McDermott}},\ }\bibfield  {title} {\bibinfo {title} {Anomalous charge noise
			in superconducting qubits},\ }\href
	{https://doi.org/10.1103/PhysRevB.100.140503} {\bibfield  {journal} {\bibinfo
			{journal} {Phys. Rev. B}\ }\textbf {\bibinfo {volume} {100}},\ \bibinfo
		{pages} {140503(R)} (\bibinfo {year} {2019})}\BibitemShut {NoStop}%
	\bibitem [{\citenamefont {Gordon}\ \emph {et~al.}(2022)\citenamefont {Gordon},
		\citenamefont {Murray}, \citenamefont {Kurter}, \citenamefont {Sandberg},
		\citenamefont {Hall}, \citenamefont {Balakrishnan}, \citenamefont {Shelby},
		\citenamefont {Wacaser}, \citenamefont {Stabile}, \citenamefont {Sleight},
		\citenamefont {Brink}, \citenamefont {Rothwell}, \citenamefont {Rodbell},
		\citenamefont {Dial},\ and\ \citenamefont {Steffen}}]{GordonAPL120}%
	\BibitemOpen
	\bibfield  {author} {\bibinfo {author} {\bibfnamefont {R.~T.}\ \bibnamefont
			{Gordon}}, \bibinfo {author} {\bibfnamefont {C.~E.}\ \bibnamefont {Murray}},
		\bibinfo {author} {\bibfnamefont {C.}~\bibnamefont {Kurter}}, \bibinfo
		{author} {\bibfnamefont {M.}~\bibnamefont {Sandberg}}, \bibinfo {author}
		{\bibfnamefont {S.~A.}\ \bibnamefont {Hall}}, \bibinfo {author}
		{\bibfnamefont {K.}~\bibnamefont {Balakrishnan}}, \bibinfo {author}
		{\bibfnamefont {R.}~\bibnamefont {Shelby}}, \bibinfo {author} {\bibfnamefont
			{B.}~\bibnamefont {Wacaser}}, \bibinfo {author} {\bibfnamefont {A.~A.}\
			\bibnamefont {Stabile}}, \bibinfo {author} {\bibfnamefont {J.~W.}\
			\bibnamefont {Sleight}}, \bibinfo {author} {\bibfnamefont {M.}~\bibnamefont
			{Brink}}, \bibinfo {author} {\bibfnamefont {M.~B.}\ \bibnamefont {Rothwell}},
		\bibinfo {author} {\bibfnamefont {K.~P.}\ \bibnamefont {Rodbell}}, \bibinfo
		{author} {\bibfnamefont {O.}~\bibnamefont {Dial}},\ and\ \bibinfo {author}
		{\bibfnamefont {M.}~\bibnamefont {Steffen}},\ }\bibfield  {title} {\bibinfo
		{title} {Environmental radiation impact on lifetimes and quasiparticle
			tunneling rates of fixed-frequency transmon qubits},\ }\href
	{https://doi.org/10.1063/5.0078785} {\bibfield  {journal} {\bibinfo
			{journal} {Appl. Phys. Lett.}\ }\textbf {\bibinfo {volume} {120}},\ \bibinfo
		{pages} {074002} (\bibinfo {year} {2022})}\BibitemShut {NoStop}%
	\bibitem [{\citenamefont {Place}\ \emph {et~al.}(2021)\citenamefont {Place},
		\citenamefont {Rodgers}, \citenamefont {Mundada}, \citenamefont {Smitham},
		\citenamefont {Fitzpatrick}, \citenamefont {Leng}, \citenamefont {Premkumar},
		\citenamefont {Bryon}, \citenamefont {Vrajitoarea}, \citenamefont {Sussman},
		\citenamefont {Cheng}, \citenamefont {Madhavan}, \citenamefont {Babla},
		\citenamefont {Le}, \citenamefont {Gang}, \citenamefont {Jäck},
		\citenamefont {Gyenis}, \citenamefont {Yao}, \citenamefont {Cava},
		\citenamefont {de~Leon},\ and\ \citenamefont {Houck}}]{Place2021}%
	\BibitemOpen
	\bibfield  {author} {\bibinfo {author} {\bibfnamefont {A.~P.~M.}\
			\bibnamefont {Place}}, \bibinfo {author} {\bibfnamefont {L.~V.~H.}\
			\bibnamefont {Rodgers}}, \bibinfo {author} {\bibfnamefont {P.}~\bibnamefont
			{Mundada}}, \bibinfo {author} {\bibfnamefont {B.~M.}\ \bibnamefont
			{Smitham}}, \bibinfo {author} {\bibfnamefont {M.}~\bibnamefont
			{Fitzpatrick}}, \bibinfo {author} {\bibfnamefont {Z.}~\bibnamefont {Leng}},
		\bibinfo {author} {\bibfnamefont {A.}~\bibnamefont {Premkumar}}, \bibinfo
		{author} {\bibfnamefont {J.}~\bibnamefont {Bryon}}, \bibinfo {author}
		{\bibfnamefont {A.}~\bibnamefont {Vrajitoarea}}, \bibinfo {author}
		{\bibfnamefont {S.}~\bibnamefont {Sussman}}, \bibinfo {author} {\bibfnamefont
			{G.}~\bibnamefont {Cheng}}, \bibinfo {author} {\bibfnamefont
			{T.}~\bibnamefont {Madhavan}}, \bibinfo {author} {\bibfnamefont {H.~K.}\
			\bibnamefont {Babla}}, \bibinfo {author} {\bibfnamefont {X.~H.}\ \bibnamefont
			{Le}}, \bibinfo {author} {\bibfnamefont {Y.}~\bibnamefont {Gang}}, \bibinfo
		{author} {\bibfnamefont {B.}~\bibnamefont {Jäck}}, \bibinfo {author}
		{\bibfnamefont {A.}~\bibnamefont {Gyenis}}, \bibinfo {author} {\bibfnamefont
			{N.}~\bibnamefont {Yao}}, \bibinfo {author} {\bibfnamefont {R.~J.}\
			\bibnamefont {Cava}}, \bibinfo {author} {\bibfnamefont {N.~P.}\ \bibnamefont
			{de~Leon}},\ and\ \bibinfo {author} {\bibfnamefont {A.~A.}\ \bibnamefont
			{Houck}},\ }\bibfield  {title} {\bibinfo {title} {New material platform for
			superconducting transmon qubits with coherence times exceeding 0.3
			milliseconds},\ }\href {https://doi.org/10.1038/s41467-021-22030-5}
	{\bibfield  {journal} {\bibinfo  {journal} {Nat. Commun.}\ }\textbf {\bibinfo
			{volume} {12}},\ \bibinfo {pages} {1779} (\bibinfo {year}
		{2021})}\BibitemShut {NoStop}%
	\bibitem [{\citenamefont {Tennant}\ \emph {et~al.}(2022)\citenamefont
		{Tennant}, \citenamefont {Martinez}, \citenamefont {Beck}, \citenamefont
		{O'Kelley}, \citenamefont {Wilen}, \citenamefont {McDermott}, \citenamefont
		{DuBois},\ and\ \citenamefont {Rosen}}]{tennant2021low}%
	\BibitemOpen
	\bibfield  {author} {\bibinfo {author} {\bibfnamefont {D.~M.}\ \bibnamefont
			{Tennant}}, \bibinfo {author} {\bibfnamefont {L.~A.}\ \bibnamefont
			{Martinez}}, \bibinfo {author} {\bibfnamefont {K.~M.}\ \bibnamefont {Beck}},
		\bibinfo {author} {\bibfnamefont {S.~R.}\ \bibnamefont {O'Kelley}}, \bibinfo
		{author} {\bibfnamefont {C.~D.}\ \bibnamefont {Wilen}}, \bibinfo {author}
		{\bibfnamefont {R.}~\bibnamefont {McDermott}}, \bibinfo {author}
		{\bibfnamefont {J.~L.}\ \bibnamefont {DuBois}},\ and\ \bibinfo {author}
		{\bibfnamefont {Y.~J.}\ \bibnamefont {Rosen}},\ }\bibfield  {title} {\bibinfo
		{title} {Low-frequency correlated charge-noise measurements across multiple
			energy transitions in a tantalum transmon},\ }\href
	{https://doi.org/10.1103/PRXQuantum.3.030307} {\bibfield  {journal} {\bibinfo
			{journal} {PRX Quantum}\ }\textbf {\bibinfo {volume} {3}},\ \bibinfo {pages}
		{030307} (\bibinfo {year} {2022})}\BibitemShut {NoStop}%
	\bibitem [{\citenamefont {Annunziata}\ \emph {et~al.}(2010)\citenamefont
		{Annunziata}, \citenamefont {Santavicca}, \citenamefont {Frunzio},
		\citenamefont {Catelani}, \citenamefont {Rooks}, \citenamefont {Frydman},\
		and\ \citenamefont {Prober}}]{Annunziata_2010}%
	\BibitemOpen
	\bibfield  {author} {\bibinfo {author} {\bibfnamefont {A.~J.}\ \bibnamefont
			{Annunziata}}, \bibinfo {author} {\bibfnamefont {D.~F.}\ \bibnamefont
			{Santavicca}}, \bibinfo {author} {\bibfnamefont {L.}~\bibnamefont {Frunzio}},
		\bibinfo {author} {\bibfnamefont {G.}~\bibnamefont {Catelani}}, \bibinfo
		{author} {\bibfnamefont {M.~J.}\ \bibnamefont {Rooks}}, \bibinfo {author}
		{\bibfnamefont {A.}~\bibnamefont {Frydman}},\ and\ \bibinfo {author}
		{\bibfnamefont {D.~E.}\ \bibnamefont {Prober}},\ }\bibfield  {title}
	{\bibinfo {title} {Tunable superconducting nanoinductors},\ }\href
	{https://doi.org/10.1088/0957-4484/21/44/445202} {\bibfield  {journal}
		{\bibinfo  {journal} {Nanotechnology}\ }\textbf {\bibinfo {volume} {21}},\
		\bibinfo {pages} {445202} (\bibinfo {year} {2010})}\BibitemShut {NoStop}%
	\bibitem [{\citenamefont {Niepce}\ \emph {et~al.}(2019)\citenamefont {Niepce},
		\citenamefont {Burnett},\ and\ \citenamefont {Bylander}}]{Bylander2019}%
	\BibitemOpen
	\bibfield  {author} {\bibinfo {author} {\bibfnamefont {D.}~\bibnamefont
			{Niepce}}, \bibinfo {author} {\bibfnamefont {J.}~\bibnamefont {Burnett}},\
		and\ \bibinfo {author} {\bibfnamefont {J.}~\bibnamefont {Bylander}},\
	}\bibfield  {title} {\bibinfo {title} {High kinetic inductance
			$\mathrm{Nb}\mathrm{N}$ nanowire superinductors},\ }\href
	{https://doi.org/10.1103/PhysRevApplied.11.044014} {\bibfield  {journal}
		{\bibinfo  {journal} {Phys. Rev. Applied}\ }\textbf {\bibinfo {volume}
			{11}},\ \bibinfo {pages} {044014} (\bibinfo {year} {2019})}\BibitemShut
	{NoStop}%
	\bibitem [{\citenamefont {Hazard}\ \emph {et~al.}(2019)\citenamefont {Hazard},
		\citenamefont {Gyenis}, \citenamefont {Di~Paolo}, \citenamefont {Asfaw},
		\citenamefont {Lyon}, \citenamefont {Blais},\ and\ \citenamefont
		{Houck}}]{Hazard2019}%
	\BibitemOpen
	\bibfield  {author} {\bibinfo {author} {\bibfnamefont {T.~M.}\ \bibnamefont
			{Hazard}}, \bibinfo {author} {\bibfnamefont {A.}~\bibnamefont {Gyenis}},
		\bibinfo {author} {\bibfnamefont {A.}~\bibnamefont {Di~Paolo}}, \bibinfo
		{author} {\bibfnamefont {A.~T.}\ \bibnamefont {Asfaw}}, \bibinfo {author}
		{\bibfnamefont {S.~A.}\ \bibnamefont {Lyon}}, \bibinfo {author}
		{\bibfnamefont {A.}~\bibnamefont {Blais}},\ and\ \bibinfo {author}
		{\bibfnamefont {A.~A.}\ \bibnamefont {Houck}},\ }\bibfield  {title} {\bibinfo
		{title} {Nanowire superinductance fluxonium qubit},\ }\href
	{https://doi.org/10.1103/PhysRevLett.122.010504} {\bibfield  {journal}
		{\bibinfo  {journal} {Phys. Rev. Lett.}\ }\textbf {\bibinfo {volume} {122}},\
		\bibinfo {pages} {010504} (\bibinfo {year} {2019})}\BibitemShut {NoStop}%
	\bibitem [{\citenamefont {Pita-Vidal}\ \emph {et~al.}(2020)\citenamefont
		{Pita-Vidal}, \citenamefont {Bargerbos}, \citenamefont {Yang}, \citenamefont
		{van Woerkom}, \citenamefont {Pfaff}, \citenamefont {Haider}, \citenamefont
		{Krogstrup}, \citenamefont {Kouwenhoven}, \citenamefont {de~Lange},\ and\
		\citenamefont {Kou}}]{Kou2020}%
	\BibitemOpen
	\bibfield  {author} {\bibinfo {author} {\bibfnamefont {M.}~\bibnamefont
			{Pita-Vidal}}, \bibinfo {author} {\bibfnamefont {A.}~\bibnamefont
			{Bargerbos}}, \bibinfo {author} {\bibfnamefont {C.-K.}\ \bibnamefont {Yang}},
		\bibinfo {author} {\bibfnamefont {D.~J.}\ \bibnamefont {van Woerkom}},
		\bibinfo {author} {\bibfnamefont {W.}~\bibnamefont {Pfaff}}, \bibinfo
		{author} {\bibfnamefont {N.}~\bibnamefont {Haider}}, \bibinfo {author}
		{\bibfnamefont {P.}~\bibnamefont {Krogstrup}}, \bibinfo {author}
		{\bibfnamefont {L.~P.}\ \bibnamefont {Kouwenhoven}}, \bibinfo {author}
		{\bibfnamefont {G.}~\bibnamefont {de~Lange}},\ and\ \bibinfo {author}
		{\bibfnamefont {A.}~\bibnamefont {Kou}},\ }\bibfield  {title} {\bibinfo
		{title} {Gate-tunable field-compatible fluxonium},\ }\href
	{https://doi.org/10.1103/PhysRevApplied.14.064038} {\bibfield  {journal}
		{\bibinfo  {journal} {Phys. Rev. Applied}\ }\textbf {\bibinfo {volume}
			{14}},\ \bibinfo {pages} {064038} (\bibinfo {year} {2020})}\BibitemShut
	{NoStop}%
	\bibitem [{\citenamefont {Gr\"unhaupt}\ \emph {et~al.}(2019)\citenamefont
		{Gr\"unhaupt}, \citenamefont {Spiecker}, \citenamefont {Gusenkova},
		\citenamefont {Maleeva}, \citenamefont {Skacel}, \citenamefont {Takmakov},
		\citenamefont {Valenti}, \citenamefont {Winkel}, \citenamefont {Rotzinger},
		\citenamefont {Wernsdorfer}, \citenamefont {Ustinov},\ and\ \citenamefont
		{Pop}}]{grAl2019}%
	\BibitemOpen
	\bibfield  {author} {\bibinfo {author} {\bibfnamefont {L.}~\bibnamefont
			{Gr\"unhaupt}}, \bibinfo {author} {\bibfnamefont {M.}~\bibnamefont
			{Spiecker}}, \bibinfo {author} {\bibfnamefont {D.}~\bibnamefont {Gusenkova}},
		\bibinfo {author} {\bibfnamefont {N.}~\bibnamefont {Maleeva}}, \bibinfo
		{author} {\bibfnamefont {S.~T.}\ \bibnamefont {Skacel}}, \bibinfo {author}
		{\bibfnamefont {I.}~\bibnamefont {Takmakov}}, \bibinfo {author}
		{\bibfnamefont {F.}~\bibnamefont {Valenti}}, \bibinfo {author} {\bibfnamefont
			{P.}~\bibnamefont {Winkel}}, \bibinfo {author} {\bibfnamefont
			{H.}~\bibnamefont {Rotzinger}}, \bibinfo {author} {\bibfnamefont
			{W.}~\bibnamefont {Wernsdorfer}}, \bibinfo {author} {\bibfnamefont {A.~V.}\
			\bibnamefont {Ustinov}},\ and\ \bibinfo {author} {\bibfnamefont {I.~M.}\
			\bibnamefont {Pop}},\ }\bibfield  {title} {\bibinfo {title} {Granular
			aluminium as a superconducting material for high-impedance quantum
			circuits},\ }\href {https://doi.org/10.1038/s41563-019-0350-3} {\bibfield
		{journal} {\bibinfo  {journal} {Nat. Mater.}\ }\textbf {\bibinfo {volume}
			{18}},\ \bibinfo {pages} {816} (\bibinfo {year} {2019})}\BibitemShut
	{NoStop}%
	\bibitem [{\citenamefont {Kreikebaum}\ \emph {et~al.}(2020)\citenamefont
		{Kreikebaum}, \citenamefont {O'Brien}, \citenamefont {Morvan},\ and\
		\citenamefont {Siddiqi}}]{Kreikebaum_2020}%
	\BibitemOpen
	\bibfield  {author} {\bibinfo {author} {\bibfnamefont {J.~M.}\ \bibnamefont
			{Kreikebaum}}, \bibinfo {author} {\bibfnamefont {K.~P.}\ \bibnamefont
			{O'Brien}}, \bibinfo {author} {\bibfnamefont {A.}~\bibnamefont {Morvan}},\
		and\ \bibinfo {author} {\bibfnamefont {I.}~\bibnamefont {Siddiqi}},\
	}\bibfield  {title} {\bibinfo {title} {Improving wafer-scale {Josephson}
			junction resistance variation in superconducting quantum coherent circuits},\
	}\href {https://doi.org/10.1088/1361-6668/ab8617} {\bibfield  {journal}
		{\bibinfo  {journal} {Supercond. Sci. Technol.}\ }\textbf {\bibinfo {volume}
			{33}},\ \bibinfo {pages} {06LT02} (\bibinfo {year} {2020})}\BibitemShut
	{NoStop}%
	\bibitem [{\citenamefont {Tinkham}(2004)}]{tinkham}%
	\BibitemOpen
	\bibfield  {author} {\bibinfo {author} {\bibfnamefont {M.}~\bibnamefont
			{Tinkham}},\ }\href {https://books.google.ae/books?id=VpUk3NfwDIkC} {\emph
		{\bibinfo {title} {Introduction to Superconductivity}}},\ Dover Books on
	Physics Series\ (\bibinfo  {publisher} {Dover Publications, New York},\
	\bibinfo {year} {2004})\BibitemShut {NoStop}%
	\bibitem [{\citenamefont {Catelani}\ \emph {et~al.}(2008)\citenamefont
		{Catelani}, \citenamefont {Wu},\ and\ \citenamefont {Adams}}]{Adams2008}%
	\BibitemOpen
	\bibfield  {author} {\bibinfo {author} {\bibfnamefont {G.}~\bibnamefont
			{Catelani}}, \bibinfo {author} {\bibfnamefont {X.~S.}\ \bibnamefont {Wu}},\
		and\ \bibinfo {author} {\bibfnamefont {P.~W.}\ \bibnamefont {Adams}},\
	}\bibfield  {title} {\bibinfo {title} {Fermi-liquid effects in the gapless
			state of marginally thin superconducting films},\ }\href
	{https://doi.org/10.1103/PhysRevB.78.104515} {\bibfield  {journal} {\bibinfo
			{journal} {Phys. Rev. B}\ }\textbf {\bibinfo {volume} {78}},\ \bibinfo
		{pages} {104515} (\bibinfo {year} {2008})}\BibitemShut {NoStop}%
	\bibitem [{\citenamefont {Ivry}\ \emph {et~al.}(2014)\citenamefont {Ivry},
		\citenamefont {Kim}, \citenamefont {Dane}, \citenamefont {De~Fazio},
		\citenamefont {McCaughan}, \citenamefont {Sunter}, \citenamefont {Zhao},\
		and\ \citenamefont {Berggren}}]{IvryPRB90}%
	\BibitemOpen
	\bibfield  {author} {\bibinfo {author} {\bibfnamefont {Y.}~\bibnamefont
			{Ivry}}, \bibinfo {author} {\bibfnamefont {C.-S.}\ \bibnamefont {Kim}},
		\bibinfo {author} {\bibfnamefont {A.~E.}\ \bibnamefont {Dane}}, \bibinfo
		{author} {\bibfnamefont {D.}~\bibnamefont {De~Fazio}}, \bibinfo {author}
		{\bibfnamefont {A.~N.}\ \bibnamefont {McCaughan}}, \bibinfo {author}
		{\bibfnamefont {K.~A.}\ \bibnamefont {Sunter}}, \bibinfo {author}
		{\bibfnamefont {Q.}~\bibnamefont {Zhao}},\ and\ \bibinfo {author}
		{\bibfnamefont {K.~K.}\ \bibnamefont {Berggren}},\ }\bibfield  {title}
	{\bibinfo {title} {Universal scaling of the critical temperature for thin
			films near the superconducting-to-insulating transition},\ }\href
	{https://doi.org/10.1103/PhysRevB.90.214515} {\bibfield  {journal} {\bibinfo
			{journal} {Phys. Rev. B}\ }\textbf {\bibinfo {volume} {90}},\ \bibinfo
		{pages} {214515} (\bibinfo {year} {2014})}\BibitemShut {NoStop}%
	\bibitem [{\citenamefont {Cherney}\ and\ \citenamefont
		{Shewchun}(1969)}]{CherneyThinAl}%
	\BibitemOpen
	\bibfield  {author} {\bibinfo {author} {\bibfnamefont {O.~A.~E.}\
			\bibnamefont {Cherney}}\ and\ \bibinfo {author} {\bibfnamefont
			{J.}~\bibnamefont {Shewchun}},\ }\bibfield  {title} {\bibinfo {title}
		{Enhancement of superconductivity in thin aluminium films},\ }\href
	{https://doi.org/10.1139/p69-138} {\bibfield  {journal} {\bibinfo  {journal}
			{Can. J. Phys.}\ }\textbf {\bibinfo {volume} {47}},\ \bibinfo {pages} {1101}
		(\bibinfo {year} {1969})}\BibitemShut {NoStop}%
	\bibitem [{\citenamefont {Court}\ \emph {et~al.}(2007)\citenamefont {Court},
		\citenamefont {Ferguson},\ and\ \citenamefont {Clark}}]{Court_2007}%
	\BibitemOpen
	\bibfield  {author} {\bibinfo {author} {\bibfnamefont {N.~A.}\ \bibnamefont
			{Court}}, \bibinfo {author} {\bibfnamefont {A.~J.}\ \bibnamefont
			{Ferguson}},\ and\ \bibinfo {author} {\bibfnamefont {R.~G.}\ \bibnamefont
			{Clark}},\ }\bibfield  {title} {\bibinfo {title} {Energy gap measurement of
			nanostructured aluminium thin films for single {Cooper}-pair devices},\
	}\href {https://doi.org/10.1088/0953-2048/21/01/015013} {\bibfield  {journal}
		{\bibinfo  {journal} {Supercond. Sci. Technol.}\ }\textbf {\bibinfo {volume}
			{21}},\ \bibinfo {pages} {015013} (\bibinfo {year} {2007})}\BibitemShut
	{NoStop}%
	\bibitem [{\citenamefont {Chubov}\ \emph {et~al.}(1969)\citenamefont {Chubov},
		\citenamefont {Eremenko},\ and\ \citenamefont {Pilipenko}}]{ChubovThinAl}%
	\BibitemOpen
	\bibfield  {author} {\bibinfo {author} {\bibfnamefont {P.}~\bibnamefont
			{Chubov}}, \bibinfo {author} {\bibfnamefont {V.}~\bibnamefont {Eremenko}},\
		and\ \bibinfo {author} {\bibfnamefont {Y.~A.}\ \bibnamefont {Pilipenko}},\
	}\bibfield  {title} {\bibinfo {title} {Dependence of the critical temperature
			and energy gap on the thickness of superconducting aluminum films},\
	}\href@noop {} {\bibfield  {journal} {\bibinfo  {journal} {Sov. Phys. JETP}\
		}\textbf {\bibinfo {volume} {28}},\ \bibinfo {pages} {389} (\bibinfo {year}
		{1969})}\BibitemShut {NoStop}%
	\bibitem [{\citenamefont {Meservey}\ and\ \citenamefont
		{Tedrow}(1971)}]{MeserveyThinAl}%
	\BibitemOpen
	\bibfield  {author} {\bibinfo {author} {\bibfnamefont {R.}~\bibnamefont
			{Meservey}}\ and\ \bibinfo {author} {\bibfnamefont {P.~M.}\ \bibnamefont
			{Tedrow}},\ }\bibfield  {title} {\bibinfo {title} {Properties of very thin
			aluminum films},\ }\href {https://doi.org/10.1063/1.1659648} {\bibfield
		{journal} {\bibinfo  {journal} {J. Appl. Phys.}\ }\textbf {\bibinfo {volume}
			{42}},\ \bibinfo {pages} {51} (\bibinfo {year} {1971})}\BibitemShut {NoStop}%
	\bibitem [{\citenamefont {Wang}\ \emph {et~al.}(2014)\citenamefont {Wang},
		\citenamefont {Gao}, \citenamefont {Pop}, \citenamefont {Vool}, \citenamefont
		{Axline}, \citenamefont {Brecht}, \citenamefont {Heeres}, \citenamefont
		{Frunzio}, \citenamefont {Devoret}, \citenamefont {Catelani}, \citenamefont
		{Glazman},\ and\ \citenamefont {Schoelkopf}}]{Wang2014}%
	\BibitemOpen
	\bibfield  {author} {\bibinfo {author} {\bibfnamefont {C.}~\bibnamefont
			{Wang}}, \bibinfo {author} {\bibfnamefont {Y.~Y.}\ \bibnamefont {Gao}},
		\bibinfo {author} {\bibfnamefont {I.~M.}\ \bibnamefont {Pop}}, \bibinfo
		{author} {\bibfnamefont {U.}~\bibnamefont {Vool}}, \bibinfo {author}
		{\bibfnamefont {C.}~\bibnamefont {Axline}}, \bibinfo {author} {\bibfnamefont
			{T.}~\bibnamefont {Brecht}}, \bibinfo {author} {\bibfnamefont {R.~W.}\
			\bibnamefont {Heeres}}, \bibinfo {author} {\bibfnamefont {L.}~\bibnamefont
			{Frunzio}}, \bibinfo {author} {\bibfnamefont {M.~H.}\ \bibnamefont
			{Devoret}}, \bibinfo {author} {\bibfnamefont {G.}~\bibnamefont {Catelani}},
		\bibinfo {author} {\bibfnamefont {L.~I.}\ \bibnamefont {Glazman}},\ and\
		\bibinfo {author} {\bibfnamefont {R.~J.}\ \bibnamefont {Schoelkopf}},\ }\href
	{https://doi.org/10.1038/ncomms6836} {\bibfield  {journal} {\bibinfo
			{journal} {Nat. Commun.}\ }\textbf {\bibinfo {volume} {5}},\ \bibinfo {pages}
		{5836} (\bibinfo {year} {2014})}\BibitemShut {NoStop}%
	\bibitem [{\citenamefont {Catelani}\ \emph
		{et~al.}(2011{\natexlab{a}})\citenamefont {Catelani}, \citenamefont {Koch},
		\citenamefont {Frunzio}, \citenamefont {Schoelkopf}, \citenamefont
		{Devoret},\ and\ \citenamefont {Glazman}}]{CatelaniPRL106}%
	\BibitemOpen
	\bibfield  {author} {\bibinfo {author} {\bibfnamefont {G.}~\bibnamefont
			{Catelani}}, \bibinfo {author} {\bibfnamefont {J.}~\bibnamefont {Koch}},
		\bibinfo {author} {\bibfnamefont {L.}~\bibnamefont {Frunzio}}, \bibinfo
		{author} {\bibfnamefont {R.~J.}\ \bibnamefont {Schoelkopf}}, \bibinfo
		{author} {\bibfnamefont {M.~H.}\ \bibnamefont {Devoret}},\ and\ \bibinfo
		{author} {\bibfnamefont {L.~I.}\ \bibnamefont {Glazman}},\ }\bibfield
	{title} {\bibinfo {title} {Quasiparticle relaxation of superconducting qubits
			in the presence of flux},\ }\href
	{https://doi.org/10.1103/PhysRevLett.106.077002} {\bibfield  {journal}
		{\bibinfo  {journal} {Phys. Rev. Lett.}\ }\textbf {\bibinfo {volume} {106}},\
		\bibinfo {pages} {077002} (\bibinfo {year} {2011}{\natexlab{a}})}\BibitemShut
	{NoStop}%
	\bibitem [{\citenamefont {Catelani}\ \emph
		{et~al.}(2011{\natexlab{b}})\citenamefont {Catelani}, \citenamefont
		{Schoelkopf}, \citenamefont {Devoret},\ and\ \citenamefont
		{Glazman}}]{CatelaniPRB84}%
	\BibitemOpen
	\bibfield  {author} {\bibinfo {author} {\bibfnamefont {G.}~\bibnamefont
			{Catelani}}, \bibinfo {author} {\bibfnamefont {R.~J.}\ \bibnamefont
			{Schoelkopf}}, \bibinfo {author} {\bibfnamefont {M.~H.}\ \bibnamefont
			{Devoret}},\ and\ \bibinfo {author} {\bibfnamefont {L.~I.}\ \bibnamefont
			{Glazman}},\ }\bibfield  {title} {\bibinfo {title} {Relaxation and frequency
			shifts induced by quasiparticles in superconducting qubits},\ }\href
	{https://doi.org/10.1103/PhysRevB.84.064517} {\bibfield  {journal} {\bibinfo
			{journal} {Phys. Rev. B}\ }\textbf {\bibinfo {volume} {84}},\ \bibinfo
		{pages} {064517} (\bibinfo {year} {2011}{\natexlab{b}})}\BibitemShut
	{NoStop}%
	\bibitem [{\citenamefont {Koch}\ \emph {et~al.}(2007)\citenamefont {Koch},
		\citenamefont {Yu}, \citenamefont {Gambetta}, \citenamefont {Houck},
		\citenamefont {Schuster}, \citenamefont {Majer}, \citenamefont {Blais},
		\citenamefont {Devoret}, \citenamefont {Girvin},\ and\ \citenamefont
		{Schoelkopf}}]{transmonPhysRevA76}%
	\BibitemOpen
	\bibfield  {author} {\bibinfo {author} {\bibfnamefont {J.}~\bibnamefont
			{Koch}}, \bibinfo {author} {\bibfnamefont {T.~M.}\ \bibnamefont {Yu}},
		\bibinfo {author} {\bibfnamefont {J.}~\bibnamefont {Gambetta}}, \bibinfo
		{author} {\bibfnamefont {A.~A.}\ \bibnamefont {Houck}}, \bibinfo {author}
		{\bibfnamefont {D.~I.}\ \bibnamefont {Schuster}}, \bibinfo {author}
		{\bibfnamefont {J.}~\bibnamefont {Majer}}, \bibinfo {author} {\bibfnamefont
			{A.}~\bibnamefont {Blais}}, \bibinfo {author} {\bibfnamefont {M.~H.}\
			\bibnamefont {Devoret}}, \bibinfo {author} {\bibfnamefont {S.~M.}\
			\bibnamefont {Girvin}},\ and\ \bibinfo {author} {\bibfnamefont {R.~J.}\
			\bibnamefont {Schoelkopf}},\ }\bibfield  {title} {\bibinfo {title}
		{Charge-insensitive qubit design derived from the {Cooper} pair box},\ }\href
	{https://doi.org/10.1103/PhysRevA.76.042319} {\bibfield  {journal} {\bibinfo
			{journal} {Phys. Rev. A}\ }\textbf {\bibinfo {volume} {76}},\ \bibinfo
		{pages} {042319} (\bibinfo {year} {2007})}\BibitemShut {NoStop}%
	\bibitem [{\citenamefont {Barone}\ and\ \citenamefont
		{Patern{\`o}}(1982)}]{Barone}%
	\BibitemOpen
	\bibfield  {author} {\bibinfo {author} {\bibfnamefont {A.}~\bibnamefont
			{Barone}}\ and\ \bibinfo {author} {\bibfnamefont {G.}~\bibnamefont
			{Patern{\`o}}},\ }\href@noop {} {\emph {\bibinfo {title} {Physics and
				applications of the Josephson effect}}}\ (\bibinfo  {publisher} {Wiley, New
		York},\ \bibinfo {year} {1982})\BibitemShut {NoStop}%
	\bibitem [{\citenamefont {Catelani}(2014)}]{CatelaniPRB89}%
	\BibitemOpen
	\bibfield  {author} {\bibinfo {author} {\bibfnamefont {G.}~\bibnamefont
			{Catelani}},\ }\bibfield  {title} {\bibinfo {title} {Parity switching and
			decoherence by quasiparticles in single-junction transmons},\ }\href
	{https://doi.org/10.1103/PhysRevB.89.094522} {\bibfield  {journal} {\bibinfo
			{journal} {Phys. Rev. B}\ }\textbf {\bibinfo {volume} {89}},\ \bibinfo
		{pages} {094522} (\bibinfo {year} {2014})}\BibitemShut {NoStop}%
	\bibitem [{\citenamefont {Glazman}\ and\ \citenamefont
		{Catelani}(2021)}]{CatelaniSciPostReview}%
	\BibitemOpen
	\bibfield  {author} {\bibinfo {author} {\bibfnamefont {L.~I.}\ \bibnamefont
			{Glazman}}\ and\ \bibinfo {author} {\bibfnamefont {G.}~\bibnamefont
			{Catelani}},\ }\bibfield  {title} {\bibinfo {title} {{Bogoliubov
				Quasiparticles in Superconducting Qubits}},\ }\href
	{https://doi.org/10.21468/SciPostPhysLectNotes.31} {\bibfield  {journal}
		{\bibinfo  {journal} {SciPost Phys. Lect. Notes}\ ,\ \bibinfo {pages} {31}}
		(\bibinfo {year} {2021})}\BibitemShut {NoStop}%
	\bibitem [{Note1()}]{Note1}%
	\BibitemOpen
	\bibinfo {note} {This expression is derived from the critical current ($I_c$)
		of an asymmetric Josephson junction~\cite {Barone}, through the expression
		$E_J=\hbar I_C/(2e)$, and it is valid in the low quasiparticle density
		limit.}\BibitemShut {Stop}%
	\bibitem [{\citenamefont {Prudnikov}\ \emph {et~al.}(1986)\citenamefont
		{Prudnikov}, \citenamefont {Brychkov}, \citenamefont {Brychkov},\ and\
		\citenamefont {Marichev}}]{PrudnikovBook}%
	\BibitemOpen
	\bibfield  {author} {\bibinfo {author} {\bibfnamefont {A.}~\bibnamefont
			{Prudnikov}}, \bibinfo {author} {\bibfnamefont {I.}~\bibnamefont {Brychkov}},
		\bibinfo {author} {\bibfnamefont {I.}~\bibnamefont {Brychkov}},\ and\
		\bibinfo {author} {\bibfnamefont {O.}~\bibnamefont {Marichev}},\ }\href
	{https://books.google.com/books?id=2t2cNs00aTgC} {\emph {\bibinfo {title}
			{Integrals and Series: Special functions}}},\ Integrals and Series\ (\bibinfo
	{publisher} {Gordon and Breach Science Publishers},\ \bibinfo {year}
	{1986})\BibitemShut {NoStop}%
	\bibitem [{\citenamefont {Catelani}\ and\ \citenamefont
		{Basko}(2019{\natexlab{a}})}]{CatelaniSciPost6}%
	\BibitemOpen
	\bibfield  {author} {\bibinfo {author} {\bibfnamefont {G.}~\bibnamefont
			{Catelani}}\ and\ \bibinfo {author} {\bibfnamefont {D.~M.}\ \bibnamefont
			{Basko}},\ }\bibfield  {title} {\bibinfo {title} {{Non-equilibrium
				quasiparticles in superconducting circuits: photons vs. phonons}},\ }\href
	{https://doi.org/10.21468/SciPostPhys.6.1.013} {\bibfield  {journal}
		{\bibinfo  {journal} {SciPost Phys.}\ }\textbf {\bibinfo {volume} {6}},\
		\bibinfo {pages} {13} (\bibinfo {year} {2019}{\natexlab{a}})}\BibitemShut
	{NoStop}%
	\bibitem [{\citenamefont {Serniak}\ \emph {et~al.}(2018)\citenamefont
		{Serniak}, \citenamefont {Hays}, \citenamefont {de~Lange}, \citenamefont
		{Diamond}, \citenamefont {Shankar}, \citenamefont {Burkhart}, \citenamefont
		{Frunzio}, \citenamefont {Houzet},\ and\ \citenamefont
		{Devoret}}]{PRL121Exp}%
	\BibitemOpen
	\bibfield  {author} {\bibinfo {author} {\bibfnamefont {K.}~\bibnamefont
			{Serniak}}, \bibinfo {author} {\bibfnamefont {M.}~\bibnamefont {Hays}},
		\bibinfo {author} {\bibfnamefont {G.}~\bibnamefont {de~Lange}}, \bibinfo
		{author} {\bibfnamefont {S.}~\bibnamefont {Diamond}}, \bibinfo {author}
		{\bibfnamefont {S.}~\bibnamefont {Shankar}}, \bibinfo {author} {\bibfnamefont
			{L.~D.}\ \bibnamefont {Burkhart}}, \bibinfo {author} {\bibfnamefont
			{L.}~\bibnamefont {Frunzio}}, \bibinfo {author} {\bibfnamefont
			{M.}~\bibnamefont {Houzet}},\ and\ \bibinfo {author} {\bibfnamefont {M.~H.}\
			\bibnamefont {Devoret}},\ }\bibfield  {title} {\bibinfo {title} {Hot
			nonequilibrium quasiparticles in transmon qubits},\ }\href
	{https://doi.org/10.1103/PhysRevLett.121.157701} {\bibfield  {journal}
		{\bibinfo  {journal} {Phys. Rev. Lett.}\ }\textbf {\bibinfo {volume} {121}},\
		\bibinfo {pages} {157701} (\bibinfo {year} {2018})}\BibitemShut {NoStop}%
	\bibitem [{\citenamefont {Houzet}\ \emph {et~al.}(2019)\citenamefont {Houzet},
		\citenamefont {Serniak}, \citenamefont {Catelani}, \citenamefont {Devoret},\
		and\ \citenamefont {Glazman}}]{PRL123Photon}%
	\BibitemOpen
	\bibfield  {author} {\bibinfo {author} {\bibfnamefont {M.}~\bibnamefont
			{Houzet}}, \bibinfo {author} {\bibfnamefont {K.}~\bibnamefont {Serniak}},
		\bibinfo {author} {\bibfnamefont {G.}~\bibnamefont {Catelani}}, \bibinfo
		{author} {\bibfnamefont {M.~H.}\ \bibnamefont {Devoret}},\ and\ \bibinfo
		{author} {\bibfnamefont {L.~I.}\ \bibnamefont {Glazman}},\ }\bibfield
	{title} {\bibinfo {title} {Photon-assisted charge-parity jumps in a
			superconducting qubit},\ }\href
	{https://doi.org/10.1103/PhysRevLett.123.107704} {\bibfield  {journal}
		{\bibinfo  {journal} {Phys. Rev. Lett.}\ }\textbf {\bibinfo {volume} {123}},\
		\bibinfo {pages} {107704} (\bibinfo {year} {2019})}\BibitemShut {NoStop}%
	\bibitem [{\citenamefont {Rothwarf}\ and\ \citenamefont
		{Taylor}(1967)}]{RothwarfTaylor}%
	\BibitemOpen
	\bibfield  {author} {\bibinfo {author} {\bibfnamefont {A.}~\bibnamefont
			{Rothwarf}}\ and\ \bibinfo {author} {\bibfnamefont {B.~N.}\ \bibnamefont
			{Taylor}},\ }\bibfield  {title} {\bibinfo {title} {Measurement of
			recombination lifetimes in superconductors},\ }\href
	{https://doi.org/10.1103/PhysRevLett.19.27} {\bibfield  {journal} {\bibinfo
			{journal} {Phys. Rev. Lett.}\ }\textbf {\bibinfo {volume} {19}},\ \bibinfo
		{pages} {27} (\bibinfo {year} {1967})}\BibitemShut {NoStop}%
	\bibitem [{\citenamefont {Jin}\ \emph {et~al.}(2015)\citenamefont {Jin},
		\citenamefont {Kamal}, \citenamefont {Sears}, \citenamefont {Gudmundsen},
		\citenamefont {Hover}, \citenamefont {Miloshi}, \citenamefont {Slattery},
		\citenamefont {Yan}, \citenamefont {Yoder}, \citenamefont {Orlando},
		\citenamefont {Gustavsson},\ and\ \citenamefont {Oliver}}]{PRL114MIT}%
	\BibitemOpen
	\bibfield  {author} {\bibinfo {author} {\bibfnamefont {X.~Y.}\ \bibnamefont
			{Jin}}, \bibinfo {author} {\bibfnamefont {A.}~\bibnamefont {Kamal}}, \bibinfo
		{author} {\bibfnamefont {A.~P.}\ \bibnamefont {Sears}}, \bibinfo {author}
		{\bibfnamefont {T.}~\bibnamefont {Gudmundsen}}, \bibinfo {author}
		{\bibfnamefont {D.}~\bibnamefont {Hover}}, \bibinfo {author} {\bibfnamefont
			{J.}~\bibnamefont {Miloshi}}, \bibinfo {author} {\bibfnamefont
			{R.}~\bibnamefont {Slattery}}, \bibinfo {author} {\bibfnamefont
			{F.}~\bibnamefont {Yan}}, \bibinfo {author} {\bibfnamefont {J.}~\bibnamefont
			{Yoder}}, \bibinfo {author} {\bibfnamefont {T.~P.}\ \bibnamefont {Orlando}},
		\bibinfo {author} {\bibfnamefont {S.}~\bibnamefont {Gustavsson}},\ and\
		\bibinfo {author} {\bibfnamefont {W.~D.}\ \bibnamefont {Oliver}},\ }\bibfield
	{title} {\bibinfo {title} {Thermal and residual excited-state population in
			a 3d transmon qubit},\ }\href
	{https://doi.org/10.1103/PhysRevLett.114.240501} {\bibfield  {journal}
		{\bibinfo  {journal} {Phys. Rev. Lett.}\ }\textbf {\bibinfo {volume} {114}},\
		\bibinfo {pages} {240501} (\bibinfo {year} {2015})}\BibitemShut {NoStop}%
	\bibitem [{\citenamefont {Vool}\ \emph {et~al.}(2014)\citenamefont {Vool},
		\citenamefont {Pop}, \citenamefont {Sliwa}, \citenamefont {Abdo},
		\citenamefont {Wang}, \citenamefont {Brecht}, \citenamefont {Gao},
		\citenamefont {Shankar}, \citenamefont {Hatridge}, \citenamefont {Catelani},
		\citenamefont {Mirrahimi}, \citenamefont {Frunzio}, \citenamefont
		{Schoelkopf}, \citenamefont {Glazman},\ and\ \citenamefont
		{Devoret}}]{Vool2014}%
	\BibitemOpen
	\bibfield  {author} {\bibinfo {author} {\bibfnamefont {U.}~\bibnamefont
			{Vool}}, \bibinfo {author} {\bibfnamefont {I.~M.}\ \bibnamefont {Pop}},
		\bibinfo {author} {\bibfnamefont {K.}~\bibnamefont {Sliwa}}, \bibinfo
		{author} {\bibfnamefont {B.}~\bibnamefont {Abdo}}, \bibinfo {author}
		{\bibfnamefont {C.}~\bibnamefont {Wang}}, \bibinfo {author} {\bibfnamefont
			{T.}~\bibnamefont {Brecht}}, \bibinfo {author} {\bibfnamefont {Y.~Y.}\
			\bibnamefont {Gao}}, \bibinfo {author} {\bibfnamefont {S.}~\bibnamefont
			{Shankar}}, \bibinfo {author} {\bibfnamefont {M.}~\bibnamefont {Hatridge}},
		\bibinfo {author} {\bibfnamefont {G.}~\bibnamefont {Catelani}}, \bibinfo
		{author} {\bibfnamefont {M.}~\bibnamefont {Mirrahimi}}, \bibinfo {author}
		{\bibfnamefont {L.}~\bibnamefont {Frunzio}}, \bibinfo {author} {\bibfnamefont
			{R.~J.}\ \bibnamefont {Schoelkopf}}, \bibinfo {author} {\bibfnamefont
			{L.~I.}\ \bibnamefont {Glazman}},\ and\ \bibinfo {author} {\bibfnamefont
			{M.~H.}\ \bibnamefont {Devoret}},\ }\bibfield  {title} {\bibinfo {title}
		{Non-{Poissonian} quantum jumps of a fluxonium qubit due to quasiparticle
			excitations},\ }\href {https://doi.org/10.1103/PhysRevLett.113.247001}
	{\bibfield  {journal} {\bibinfo  {journal} {Phys. Rev. Lett.}\ }\textbf
		{\bibinfo {volume} {113}},\ \bibinfo {pages} {247001} (\bibinfo {year}
		{2014})}\BibitemShut {NoStop}%
	\bibitem [{\citenamefont {Kulikov}\ \emph {et~al.}(2020)\citenamefont
		{Kulikov}, \citenamefont {Navarathna},\ and\ \citenamefont
		{Fedorov}}]{Fedorov2020}%
	\BibitemOpen
	\bibfield  {author} {\bibinfo {author} {\bibfnamefont {A.}~\bibnamefont
			{Kulikov}}, \bibinfo {author} {\bibfnamefont {R.}~\bibnamefont
			{Navarathna}},\ and\ \bibinfo {author} {\bibfnamefont {A.}~\bibnamefont
			{Fedorov}},\ }\bibfield  {title} {\bibinfo {title} {Measuring effective
			temperatures of qubits using correlations},\ }\href
	{https://doi.org/10.1103/PhysRevLett.124.240501} {\bibfield  {journal}
		{\bibinfo  {journal} {Phys. Rev. Lett.}\ }\textbf {\bibinfo {volume} {124}},\
		\bibinfo {pages} {240501} (\bibinfo {year} {2020})}\BibitemShut {NoStop}%
	\bibitem [{\citenamefont {Catelani}\ and\ \citenamefont
		{Basko}(2019{\natexlab{b}})}]{Basko}%
	\BibitemOpen
	\bibfield  {author} {\bibinfo {author} {\bibfnamefont {G.}~\bibnamefont
			{Catelani}}\ and\ \bibinfo {author} {\bibfnamefont {D.~M.}\ \bibnamefont
			{Basko}},\ }\bibfield  {title} {\bibinfo {title} {Non-equilibrium
			quasiparticles in superconducting circuits: photons vs. phonons},\ }\href
	{https://doi.org/10.21468/SciPostPhys.6.1.013} {\bibfield  {journal}
		{\bibinfo  {journal} {SciPost Phys.}\ }\textbf {\bibinfo {volume} {6}},\
		\bibinfo {pages} {13} (\bibinfo {year} {2019}{\natexlab{b}})}\BibitemShut
	{NoStop}%
	\bibitem [{\citenamefont {Veps\"{a}l\"{a}inen}\ \emph
		{et~al.}(2020)\citenamefont {Veps\"{a}l\"{a}inen}, \citenamefont {Karamlou},
		\citenamefont {Orrell}, \citenamefont {Dogra}, \citenamefont {Loer},
		\citenamefont {Vasconcelos}, \citenamefont {Kim}, \citenamefont {Melville},
		\citenamefont {Niedzielski}, \citenamefont {Yoder}, \citenamefont
		{Gustavsson}, \citenamefont {Formaggio}, \citenamefont {VanDevender},\ and\
		\citenamefont {Oliver}}]{Vepslinen2020}%
	\BibitemOpen
	\bibfield  {author} {\bibinfo {author} {\bibfnamefont {A.~P.}\ \bibnamefont
			{Veps\"{a}l\"{a}inen}}, \bibinfo {author} {\bibfnamefont {A.~H.}\
			\bibnamefont {Karamlou}}, \bibinfo {author} {\bibfnamefont {J.~L.}\
			\bibnamefont {Orrell}}, \bibinfo {author} {\bibfnamefont {A.~S.}\
			\bibnamefont {Dogra}}, \bibinfo {author} {\bibfnamefont {B.}~\bibnamefont
			{Loer}}, \bibinfo {author} {\bibfnamefont {F.}~\bibnamefont {Vasconcelos}},
		\bibinfo {author} {\bibfnamefont {D.~K.}\ \bibnamefont {Kim}}, \bibinfo
		{author} {\bibfnamefont {A.~J.}\ \bibnamefont {Melville}}, \bibinfo {author}
		{\bibfnamefont {B.~M.}\ \bibnamefont {Niedzielski}}, \bibinfo {author}
		{\bibfnamefont {J.~L.}\ \bibnamefont {Yoder}}, \bibinfo {author}
		{\bibfnamefont {S.}~\bibnamefont {Gustavsson}}, \bibinfo {author}
		{\bibfnamefont {J.~A.}\ \bibnamefont {Formaggio}}, \bibinfo {author}
		{\bibfnamefont {B.~A.}\ \bibnamefont {VanDevender}},\ and\ \bibinfo {author}
		{\bibfnamefont {W.~D.}\ \bibnamefont {Oliver}},\ }\bibfield  {title}
	{\bibinfo {title} {Impact of ionizing radiation on superconducting qubit
			coherence},\ }\href {https://doi.org/10.1038/s41586-020-2619-8} {\bibfield
		{journal} {\bibinfo  {journal} {Nature}\ }\textbf {\bibinfo {volume} {584}},\
		\bibinfo {pages} {551} (\bibinfo {year} {2020})}\BibitemShut {NoStop}%
	\bibitem [{\citenamefont {Cardani}\ \emph {et~al.}(2021)\citenamefont
		{Cardani}, \citenamefont {Valenti}, \citenamefont {Casali}, \citenamefont
		{Catelani}, \citenamefont {Charpentier}, \citenamefont {Clemenza},
		\citenamefont {Colantoni}, \citenamefont {Cruciani}, \citenamefont
		{D'Imperio}, \citenamefont {Gironi}, \citenamefont {Gr\"{u}nhaupt},
		\citenamefont {Gusenkova}, \citenamefont {Henriques}, \citenamefont {Lagoin},
		\citenamefont {Martinez}, \citenamefont {Pettinari}, \citenamefont {Rusconi},
		\citenamefont {Sander}, \citenamefont {Tomei}, \citenamefont {Ustinov},
		\citenamefont {Weber}, \citenamefont {Wernsdorfer}, \citenamefont {Vignati},
		\citenamefont {Pirro},\ and\ \citenamefont {Pop}}]{Cardani2021}%
	\BibitemOpen
	\bibfield  {author} {\bibinfo {author} {\bibfnamefont {L.}~\bibnamefont
			{Cardani}}, \bibinfo {author} {\bibfnamefont {F.}~\bibnamefont {Valenti}},
		\bibinfo {author} {\bibfnamefont {N.}~\bibnamefont {Casali}}, \bibinfo
		{author} {\bibfnamefont {G.}~\bibnamefont {Catelani}}, \bibinfo {author}
		{\bibfnamefont {T.}~\bibnamefont {Charpentier}}, \bibinfo {author}
		{\bibfnamefont {M.}~\bibnamefont {Clemenza}}, \bibinfo {author}
		{\bibfnamefont {I.}~\bibnamefont {Colantoni}}, \bibinfo {author}
		{\bibfnamefont {A.}~\bibnamefont {Cruciani}}, \bibinfo {author}
		{\bibfnamefont {G.}~\bibnamefont {D'Imperio}}, \bibinfo {author}
		{\bibfnamefont {L.}~\bibnamefont {Gironi}}, \bibinfo {author} {\bibfnamefont
			{L.}~\bibnamefont {Gr\"{u}nhaupt}}, \bibinfo {author} {\bibfnamefont
			{D.}~\bibnamefont {Gusenkova}}, \bibinfo {author} {\bibfnamefont
			{F.}~\bibnamefont {Henriques}}, \bibinfo {author} {\bibfnamefont
			{M.}~\bibnamefont {Lagoin}}, \bibinfo {author} {\bibfnamefont
			{M.}~\bibnamefont {Martinez}}, \bibinfo {author} {\bibfnamefont
			{G.}~\bibnamefont {Pettinari}}, \bibinfo {author} {\bibfnamefont
			{C.}~\bibnamefont {Rusconi}}, \bibinfo {author} {\bibfnamefont
			{O.}~\bibnamefont {Sander}}, \bibinfo {author} {\bibfnamefont
			{C.}~\bibnamefont {Tomei}}, \bibinfo {author} {\bibfnamefont {A.~V.}\
			\bibnamefont {Ustinov}}, \bibinfo {author} {\bibfnamefont {M.}~\bibnamefont
			{Weber}}, \bibinfo {author} {\bibfnamefont {W.}~\bibnamefont {Wernsdorfer}},
		\bibinfo {author} {\bibfnamefont {M.}~\bibnamefont {Vignati}}, \bibinfo
		{author} {\bibfnamefont {S.}~\bibnamefont {Pirro}},\ and\ \bibinfo {author}
		{\bibfnamefont {I.~M.}\ \bibnamefont {Pop}},\ }\bibfield  {title} {\bibinfo
		{title} {Reducing the impact of radioactivity on quantum circuits in a
			deep-underground facility},\ }\href
	{https://doi.org/10.1038/s41467-021-23032-z} {\bibfield  {journal} {\bibinfo
			{journal} {Nat. Commun.}\ }\textbf {\bibinfo {volume} {12}},\ \bibinfo
		{pages} {2733} (\bibinfo {year} {2021})}\BibitemShut {NoStop}%
	\bibitem [{\citenamefont {McEwen}\ \emph {et~al.}(2021)\citenamefont {McEwen},
		\citenamefont {Faoro}, \citenamefont {Arya}, \citenamefont {Dunsworth},
		\citenamefont {Huang}, \citenamefont {Kim}, \citenamefont {Burkett},
		\citenamefont {Fowler}, \citenamefont {Arute}, \citenamefont {Bardin},
		\citenamefont {Bengtsson}, \citenamefont {Bilmes}, \citenamefont {Buckley},
		\citenamefont {Bushnell}, \citenamefont {Chen}, \citenamefont {Collins},
		\citenamefont {Demura}, \citenamefont {Derk}, \citenamefont {Erickson},
		\citenamefont {Giustina}, \citenamefont {Harrington}, \citenamefont {Hong},
		\citenamefont {Jeffrey}, \citenamefont {Kelly}, \citenamefont {Klimov},
		\citenamefont {Kostritsa}, \citenamefont {Laptev}, \citenamefont {Locharla},
		\citenamefont {Mi}, \citenamefont {Miao}, \citenamefont {Montazeri},
		\citenamefont {Mutus}, \citenamefont {Naaman}, \citenamefont {Neeley},
		\citenamefont {Neill}, \citenamefont {Opremcak}, \citenamefont {Quintana},
		\citenamefont {Redd}, \citenamefont {Roushan}, \citenamefont {Sank},
		\citenamefont {Satzinger}, \citenamefont {Shvarts}, \citenamefont {White},
		\citenamefont {Yao}, \citenamefont {Yeh}, \citenamefont {Yoo}, \citenamefont
		{Chen}, \citenamefont {Smelyanskiy}, \citenamefont {Martinis}, \citenamefont
		{Neven}, \citenamefont {Megrant}, \citenamefont {Ioffe},\ and\ \citenamefont
		{Barends}}]{McEwen2021}%
	\BibitemOpen
	\bibfield  {author} {\bibinfo {author} {\bibfnamefont {M.}~\bibnamefont
			{McEwen}}, \bibinfo {author} {\bibfnamefont {L.}~\bibnamefont {Faoro}},
		\bibinfo {author} {\bibfnamefont {K.}~\bibnamefont {Arya}}, \bibinfo {author}
		{\bibfnamefont {A.}~\bibnamefont {Dunsworth}}, \bibinfo {author}
		{\bibfnamefont {T.}~\bibnamefont {Huang}}, \bibinfo {author} {\bibfnamefont
			{S.}~\bibnamefont {Kim}}, \bibinfo {author} {\bibfnamefont {B.}~\bibnamefont
			{Burkett}}, \bibinfo {author} {\bibfnamefont {A.}~\bibnamefont {Fowler}},
		\bibinfo {author} {\bibfnamefont {F.}~\bibnamefont {Arute}}, \bibinfo
		{author} {\bibfnamefont {J.~C.}\ \bibnamefont {Bardin}}, \bibinfo {author}
		{\bibfnamefont {A.}~\bibnamefont {Bengtsson}}, \bibinfo {author}
		{\bibfnamefont {A.}~\bibnamefont {Bilmes}}, \bibinfo {author} {\bibfnamefont
			{B.~B.}\ \bibnamefont {Buckley}}, \bibinfo {author} {\bibfnamefont
			{N.}~\bibnamefont {Bushnell}}, \bibinfo {author} {\bibfnamefont
			{Z.}~\bibnamefont {Chen}}, \bibinfo {author} {\bibfnamefont {R.}~\bibnamefont
			{Collins}}, \bibinfo {author} {\bibfnamefont {S.}~\bibnamefont {Demura}},
		\bibinfo {author} {\bibfnamefont {A.~R.}\ \bibnamefont {Derk}}, \bibinfo
		{author} {\bibfnamefont {C.}~\bibnamefont {Erickson}}, \bibinfo {author}
		{\bibfnamefont {M.}~\bibnamefont {Giustina}}, \bibinfo {author}
		{\bibfnamefont {S.~D.}\ \bibnamefont {Harrington}}, \bibinfo {author}
		{\bibfnamefont {S.}~\bibnamefont {Hong}}, \bibinfo {author} {\bibfnamefont
			{E.}~\bibnamefont {Jeffrey}}, \bibinfo {author} {\bibfnamefont
			{J.}~\bibnamefont {Kelly}}, \bibinfo {author} {\bibfnamefont {P.~V.}\
			\bibnamefont {Klimov}}, \bibinfo {author} {\bibfnamefont {F.}~\bibnamefont
			{Kostritsa}}, \bibinfo {author} {\bibfnamefont {P.}~\bibnamefont {Laptev}},
		\bibinfo {author} {\bibfnamefont {A.}~\bibnamefont {Locharla}}, \bibinfo
		{author} {\bibfnamefont {X.}~\bibnamefont {Mi}}, \bibinfo {author}
		{\bibfnamefont {K.~C.}\ \bibnamefont {Miao}}, \bibinfo {author}
		{\bibfnamefont {S.}~\bibnamefont {Montazeri}}, \bibinfo {author}
		{\bibfnamefont {J.}~\bibnamefont {Mutus}}, \bibinfo {author} {\bibfnamefont
			{O.}~\bibnamefont {Naaman}}, \bibinfo {author} {\bibfnamefont
			{M.}~\bibnamefont {Neeley}}, \bibinfo {author} {\bibfnamefont
			{C.}~\bibnamefont {Neill}}, \bibinfo {author} {\bibfnamefont
			{A.}~\bibnamefont {Opremcak}}, \bibinfo {author} {\bibfnamefont
			{C.}~\bibnamefont {Quintana}}, \bibinfo {author} {\bibfnamefont
			{N.}~\bibnamefont {Redd}}, \bibinfo {author} {\bibfnamefont {P.}~\bibnamefont
			{Roushan}}, \bibinfo {author} {\bibfnamefont {D.}~\bibnamefont {Sank}},
		\bibinfo {author} {\bibfnamefont {K.~J.}\ \bibnamefont {Satzinger}}, \bibinfo
		{author} {\bibfnamefont {V.}~\bibnamefont {Shvarts}}, \bibinfo {author}
		{\bibfnamefont {T.}~\bibnamefont {White}}, \bibinfo {author} {\bibfnamefont
			{Z.~J.}\ \bibnamefont {Yao}}, \bibinfo {author} {\bibfnamefont
			{P.}~\bibnamefont {Yeh}}, \bibinfo {author} {\bibfnamefont {J.}~\bibnamefont
			{Yoo}}, \bibinfo {author} {\bibfnamefont {Y.}~\bibnamefont {Chen}}, \bibinfo
		{author} {\bibfnamefont {V.}~\bibnamefont {Smelyanskiy}}, \bibinfo {author}
		{\bibfnamefont {J.~M.}\ \bibnamefont {Martinis}}, \bibinfo {author}
		{\bibfnamefont {H.}~\bibnamefont {Neven}}, \bibinfo {author} {\bibfnamefont
			{A.}~\bibnamefont {Megrant}}, \bibinfo {author} {\bibfnamefont
			{L.}~\bibnamefont {Ioffe}},\ and\ \bibinfo {author} {\bibfnamefont
			{R.}~\bibnamefont {Barends}},\ }\bibfield  {title} {\bibinfo {title}
		{Resolving catastrophic error bursts from cosmic rays in large arrays of
			superconducting qubits},\ }\href {https://doi.org/10.1038/s41567-021-01432-8}
	{\bibfield  {journal} {\bibinfo  {journal} {Nat. Phys.}\ }\textbf {\bibinfo
			{volume} {18}},\ \bibinfo {pages} {107} (\bibinfo {year} {2021})}\BibitemShut
	{NoStop}%
	\bibitem [{\citenamefont {Wilen}\ \emph {et~al.}(2021)\citenamefont {Wilen},
		\citenamefont {Abdullah}, \citenamefont {Kurinsky}, \citenamefont {Stanford},
		\citenamefont {Cardani}, \citenamefont {D'Imperio}, \citenamefont {Tomei},
		\citenamefont {Faoro}, \citenamefont {Ioffe}, \citenamefont {Liu},
		\citenamefont {Opremcak}, \citenamefont {Christensen}, \citenamefont
		{DuBois},\ and\ \citenamefont {McDermott}}]{Wilen2021}%
	\BibitemOpen
	\bibfield  {author} {\bibinfo {author} {\bibfnamefont {C.~D.}\ \bibnamefont
			{Wilen}}, \bibinfo {author} {\bibfnamefont {S.}~\bibnamefont {Abdullah}},
		\bibinfo {author} {\bibfnamefont {N.~A.}\ \bibnamefont {Kurinsky}}, \bibinfo
		{author} {\bibfnamefont {C.}~\bibnamefont {Stanford}}, \bibinfo {author}
		{\bibfnamefont {L.}~\bibnamefont {Cardani}}, \bibinfo {author} {\bibfnamefont
			{G.}~\bibnamefont {D'Imperio}}, \bibinfo {author} {\bibfnamefont
			{C.}~\bibnamefont {Tomei}}, \bibinfo {author} {\bibfnamefont
			{L.}~\bibnamefont {Faoro}}, \bibinfo {author} {\bibfnamefont {L.~B.}\
			\bibnamefont {Ioffe}}, \bibinfo {author} {\bibfnamefont {C.~H.}\ \bibnamefont
			{Liu}}, \bibinfo {author} {\bibfnamefont {A.}~\bibnamefont {Opremcak}},
		\bibinfo {author} {\bibfnamefont {B.~G.}\ \bibnamefont {Christensen}},
		\bibinfo {author} {\bibfnamefont {J.~L.}\ \bibnamefont {DuBois}},\ and\
		\bibinfo {author} {\bibfnamefont {R.}~\bibnamefont {McDermott}},\ }\bibfield
	{title} {\bibinfo {title} {Correlated charge noise and relaxation errors in
			superconducting qubits},\ }\href {https://doi.org/10.1038/s41586-021-03557-5}
	{\bibfield  {journal} {\bibinfo  {journal} {Nature}\ }\textbf {\bibinfo
			{volume} {594}},\ \bibinfo {pages} {369} (\bibinfo {year}
		{2021})}\BibitemShut {NoStop}%
	\bibitem [{\citenamefont {Henriques}\ \emph {et~al.}(2019)\citenamefont
		{Henriques}, \citenamefont {Valenti}, \citenamefont {Charpentier},
		\citenamefont {Lagoin}, \citenamefont {Gouriou}, \citenamefont {Martínez},
		\citenamefont {Cardani}, \citenamefont {Vignati}, \citenamefont {Grünhaupt},
		\citenamefont {Gusenkova}, \citenamefont {Ferrero}, \citenamefont {Skacel},
		\citenamefont {Wernsdorfer}, \citenamefont {Ustinov}, \citenamefont
		{Catelani}, \citenamefont {Sander},\ and\ \citenamefont {Pop}}]{APLphonon}%
	\BibitemOpen
	\bibfield  {author} {\bibinfo {author} {\bibfnamefont {F.}~\bibnamefont
			{Henriques}}, \bibinfo {author} {\bibfnamefont {F.}~\bibnamefont {Valenti}},
		\bibinfo {author} {\bibfnamefont {T.}~\bibnamefont {Charpentier}}, \bibinfo
		{author} {\bibfnamefont {M.}~\bibnamefont {Lagoin}}, \bibinfo {author}
		{\bibfnamefont {C.}~\bibnamefont {Gouriou}}, \bibinfo {author} {\bibfnamefont
			{M.}~\bibnamefont {Martínez}}, \bibinfo {author} {\bibfnamefont
			{L.}~\bibnamefont {Cardani}}, \bibinfo {author} {\bibfnamefont
			{M.}~\bibnamefont {Vignati}}, \bibinfo {author} {\bibfnamefont
			{L.}~\bibnamefont {Grünhaupt}}, \bibinfo {author} {\bibfnamefont
			{D.}~\bibnamefont {Gusenkova}}, \bibinfo {author} {\bibfnamefont
			{J.}~\bibnamefont {Ferrero}}, \bibinfo {author} {\bibfnamefont {S.~T.}\
			\bibnamefont {Skacel}}, \bibinfo {author} {\bibfnamefont {W.}~\bibnamefont
			{Wernsdorfer}}, \bibinfo {author} {\bibfnamefont {A.~V.}\ \bibnamefont
			{Ustinov}}, \bibinfo {author} {\bibfnamefont {G.}~\bibnamefont {Catelani}},
		\bibinfo {author} {\bibfnamefont {O.}~\bibnamefont {Sander}},\ and\ \bibinfo
		{author} {\bibfnamefont {I.~M.}\ \bibnamefont {Pop}},\ }\bibfield  {title}
	{\bibinfo {title} {Phonon traps reduce the quasiparticle density in
			superconducting circuits},\ }\href {https://doi.org/10.1063/1.5124967}
	{\bibfield  {journal} {\bibinfo  {journal} {Appl. Phys. Lett.}\ }\textbf
		{\bibinfo {volume} {115}},\ \bibinfo {pages} {212601} (\bibinfo {year}
		{2019})}\BibitemShut {NoStop}%
	\bibitem [{\citenamefont {Iaia}\ \emph {et~al.}(2022)\citenamefont {Iaia},
		\citenamefont {Ku}, \citenamefont {Ballard}, \citenamefont {Larson},
		\citenamefont {Yelton}, \citenamefont {Liu}, \citenamefont {Patel},
		\citenamefont {McDermott},\ and\ \citenamefont {Plourde}}]{PlourdePhonons}%
	\BibitemOpen
	\bibfield  {author} {\bibinfo {author} {\bibfnamefont {V.}~\bibnamefont
			{Iaia}}, \bibinfo {author} {\bibfnamefont {J.}~\bibnamefont {Ku}}, \bibinfo
		{author} {\bibfnamefont {A.}~\bibnamefont {Ballard}}, \bibinfo {author}
		{\bibfnamefont {C.~P.}\ \bibnamefont {Larson}}, \bibinfo {author}
		{\bibfnamefont {E.}~\bibnamefont {Yelton}}, \bibinfo {author} {\bibfnamefont
			{C.~H.}\ \bibnamefont {Liu}}, \bibinfo {author} {\bibfnamefont
			{S.}~\bibnamefont {Patel}}, \bibinfo {author} {\bibfnamefont
			{R.}~\bibnamefont {McDermott}},\ and\ \bibinfo {author} {\bibfnamefont
			{B.~L.~T.}\ \bibnamefont {Plourde}},\ }\href@noop {} {\bibinfo {title}
		{Phonon downconversion to suppress correlated errors in superconducting
			qubits}} (\bibinfo {year} {2022}),\ \Eprint
	{https://arxiv.org/abs/2203.06586} {arXiv:2203.06586 [quant-ph]} \BibitemShut
	{NoStop}%
	\bibitem [{\citenamefont {Spring}\ \emph {et~al.}(2022)\citenamefont {Spring},
		\citenamefont {Cao}, \citenamefont {Tsunoda}, \citenamefont {Campanaro},
		\citenamefont {Fasciati}, \citenamefont {Wills}, \citenamefont {Bakr},
		\citenamefont {Chidambaram}, \citenamefont {Shteynas}, \citenamefont
		{Carpenter}, \citenamefont {Gow}, \citenamefont {Gates}, \citenamefont
		{Vlastakis},\ and\ \citenamefont {Leek}}]{spring2021high}%
	\BibitemOpen
	\bibfield  {author} {\bibinfo {author} {\bibfnamefont {P.~A.}\ \bibnamefont
			{Spring}}, \bibinfo {author} {\bibfnamefont {S.}~\bibnamefont {Cao}},
		\bibinfo {author} {\bibfnamefont {T.}~\bibnamefont {Tsunoda}}, \bibinfo
		{author} {\bibfnamefont {G.}~\bibnamefont {Campanaro}}, \bibinfo {author}
		{\bibfnamefont {S.}~\bibnamefont {Fasciati}}, \bibinfo {author}
		{\bibfnamefont {J.}~\bibnamefont {Wills}}, \bibinfo {author} {\bibfnamefont
			{M.}~\bibnamefont {Bakr}}, \bibinfo {author} {\bibfnamefont {V.}~\bibnamefont
			{Chidambaram}}, \bibinfo {author} {\bibfnamefont {B.}~\bibnamefont
			{Shteynas}}, \bibinfo {author} {\bibfnamefont {L.}~\bibnamefont {Carpenter}},
		\bibinfo {author} {\bibfnamefont {P.}~\bibnamefont {Gow}}, \bibinfo {author}
		{\bibfnamefont {J.}~\bibnamefont {Gates}}, \bibinfo {author} {\bibfnamefont
			{B.}~\bibnamefont {Vlastakis}},\ and\ \bibinfo {author} {\bibfnamefont
			{P.~J.}\ \bibnamefont {Leek}},\ }\bibfield  {title} {\bibinfo {title} {High
			coherence and low cross-talk in a tileable 3d integrated superconducting
			circuit architecture},\ }\href {https://doi.org/10.1126/sciadv.abl6698}
	{\bibfield  {journal} {\bibinfo  {journal} {Sci. Adv.}\ }\textbf {\bibinfo
			{volume} {8}},\ \bibinfo {pages} {eabl6698} (\bibinfo {year}
		{2022})}\BibitemShut {NoStop}%
	\bibitem [{\citenamefont {Kurter}\ \emph {et~al.}(2022)\citenamefont {Kurter},
		\citenamefont {Murray}, \citenamefont {Gordon}, \citenamefont {Wymore},
		\citenamefont {Sandberg}, \citenamefont {Shelby}, \citenamefont {Eddins},
		\citenamefont {Adiga}, \citenamefont {Finck}, \citenamefont {Rivera},
		\citenamefont {Stabile}, \citenamefont {Trimm}, \citenamefont {Wacaser},
		\citenamefont {Balakrishnan}, \citenamefont {Pyzyna}, \citenamefont
		{Sleight}, \citenamefont {Steffen},\ and\ \citenamefont
		{Rodbell}}]{kurter2022quasiparticle}%
	\BibitemOpen
	\bibfield  {author} {\bibinfo {author} {\bibfnamefont {C.}~\bibnamefont
			{Kurter}}, \bibinfo {author} {\bibfnamefont {C.~E.}\ \bibnamefont {Murray}},
		\bibinfo {author} {\bibfnamefont {R.~T.}\ \bibnamefont {Gordon}}, \bibinfo
		{author} {\bibfnamefont {B.~B.}\ \bibnamefont {Wymore}}, \bibinfo {author}
		{\bibfnamefont {M.}~\bibnamefont {Sandberg}}, \bibinfo {author}
		{\bibfnamefont {R.~M.}\ \bibnamefont {Shelby}}, \bibinfo {author}
		{\bibfnamefont {A.}~\bibnamefont {Eddins}}, \bibinfo {author} {\bibfnamefont
			{V.~P.}\ \bibnamefont {Adiga}}, \bibinfo {author} {\bibfnamefont {A.~D.~K.}\
			\bibnamefont {Finck}}, \bibinfo {author} {\bibfnamefont {E.}~\bibnamefont
			{Rivera}}, \bibinfo {author} {\bibfnamefont {A.~A.}\ \bibnamefont {Stabile}},
		\bibinfo {author} {\bibfnamefont {B.}~\bibnamefont {Trimm}}, \bibinfo
		{author} {\bibfnamefont {B.}~\bibnamefont {Wacaser}}, \bibinfo {author}
		{\bibfnamefont {K.}~\bibnamefont {Balakrishnan}}, \bibinfo {author}
		{\bibfnamefont {A.}~\bibnamefont {Pyzyna}}, \bibinfo {author} {\bibfnamefont
			{J.}~\bibnamefont {Sleight}}, \bibinfo {author} {\bibfnamefont
			{M.}~\bibnamefont {Steffen}},\ and\ \bibinfo {author} {\bibfnamefont
			{K.}~\bibnamefont {Rodbell}},\ }\bibfield  {title} {\bibinfo {title}
		{Quasiparticle tunneling as a probe of {Josephson} junction barrier and
			capacitor material in superconducting qubits},\ }\href
	{https://doi.org/10.1038/s41534-022-00542-2} {\bibfield  {journal} {\bibinfo
			{journal} {npj Quantum Inf}\ }\textbf {\bibinfo {volume} {8}},\ \bibinfo
		{pages} {31} (\bibinfo {year} {2022})}\BibitemShut {NoStop}%
	\bibitem [{Note2()}]{Note2}%
	\BibitemOpen
	\bibinfo {note} {Note that the approximation $g^{R<}\approx 0$ is
		particularly accurate in this case, due to the limited width of the
		quasiparticle energy window between $\Delta _R$ and $\Delta _L$}\BibitemShut
	{NoStop}%
	\bibitem [{Note3()}]{Note3}%
	\BibitemOpen
	\bibinfo {note} {In our modeling $\Gamma _{\protect \rm ex}=2\Gamma
		^{ph}+\Gamma _{10}^{ee}$ is the inverse lifetime of the qubit in the absence
		of nonequilibrium quasiparticles.}\BibitemShut {Stop}%
	\bibitem [{Note4()}]{Note4}%
	\BibitemOpen
	\bibinfo {note} {In the calculations, we consider a fixed electrode volume
		$\protect \mathcal V$ while decreasing the smaller gap $\Delta _R$. This
		variation can be obtained by increasing the thickness of the right junction
		lead. By limiting the change of thickness to the junction region, as done for
		example in Ref.~\cite {Fei2022}, the total electrode volume is not
		significantly affected.}\BibitemShut {Stop}%
	\bibitem [{Note5()}]{Note5}%
	\BibitemOpen
	\bibinfo {note} {The solutions are obtained through a find-root procedure,
		setting to zero time derivative terms}\BibitemShut {NoStop}%
	\bibitem [{Note6()}]{Note6}%
	\BibitemOpen
	\bibinfo {note} {This value is computed using Eq.~(3) in Ref.~\cite
		{Wang2014}, using the parameters $\omega _{10}$, $T_1=T_1(B\sim 0$~mG) given
		in the supplementary material for device B2, and $\Gamma _{\protect \rm ex}=
		T_1^{-1}(B\sim 100$~mG)}\BibitemShut {NoStop}%
	\bibitem [{Note7()}]{Note7}%
	\BibitemOpen
	\bibinfo {note} {For the consistency of the set of equations, the term
		$r^{R<>}$ must be dropped also in the last equation. This approximation works
		well for $x_{R>}\ll x_{R<}$.}\BibitemShut {Stop}%
	\bibitem [{\citenamefont {Serniak}\ \emph {et~al.}(2019)\citenamefont
		{Serniak}, \citenamefont {Diamond}, \citenamefont {Hays}, \citenamefont
		{Fatemi}, \citenamefont {Shankar}, \citenamefont {Frunzio}, \citenamefont
		{Schoelkopf},\ and\ \citenamefont {Devoret}}]{SerniakPRApp}%
	\BibitemOpen
	\bibfield  {author} {\bibinfo {author} {\bibfnamefont {K.}~\bibnamefont
			{Serniak}}, \bibinfo {author} {\bibfnamefont {S.}~\bibnamefont {Diamond}},
		\bibinfo {author} {\bibfnamefont {M.}~\bibnamefont {Hays}}, \bibinfo {author}
		{\bibfnamefont {V.}~\bibnamefont {Fatemi}}, \bibinfo {author} {\bibfnamefont
			{S.}~\bibnamefont {Shankar}}, \bibinfo {author} {\bibfnamefont
			{L.}~\bibnamefont {Frunzio}}, \bibinfo {author} {\bibfnamefont {R.~J.}\
			\bibnamefont {Schoelkopf}},\ and\ \bibinfo {author} {\bibfnamefont {M.~H.}\
			\bibnamefont {Devoret}},\ }\bibfield  {title} {\bibinfo {title} {Direct
			dispersive monitoring of charge parity in offset-charge-sensitive
			transmons},\ }\href {https://doi.org/10.1103/PhysRevApplied.12.014052}
	{\bibfield  {journal} {\bibinfo  {journal} {Phys. Rev. Applied}\ }\textbf
		{\bibinfo {volume} {12}},\ \bibinfo {pages} {014052} (\bibinfo {year}
		{2019})}\BibitemShut {NoStop}%
	\bibitem [{\citenamefont {Pan}\ \emph {et~al.}(2022)\citenamefont {Pan},
		\citenamefont {Yuan}, \citenamefont {Zhou}, \citenamefont {Zhang},
		\citenamefont {Li}, \citenamefont {Liu}, \citenamefont {Jiang}, \citenamefont
		{Catelani}, \citenamefont {Hu},\ and\ \citenamefont {Yan}}]{Fei2022}%
	\BibitemOpen
	\bibfield  {author} {\bibinfo {author} {\bibfnamefont {X.}~\bibnamefont
			{Pan}}, \bibinfo {author} {\bibfnamefont {H.}~\bibnamefont {Yuan}}, \bibinfo
		{author} {\bibfnamefont {Y.}~\bibnamefont {Zhou}}, \bibinfo {author}
		{\bibfnamefont {L.}~\bibnamefont {Zhang}}, \bibinfo {author} {\bibfnamefont
			{J.}~\bibnamefont {Li}}, \bibinfo {author} {\bibfnamefont {S.}~\bibnamefont
			{Liu}}, \bibinfo {author} {\bibfnamefont {Z.~H.}\ \bibnamefont {Jiang}},
		\bibinfo {author} {\bibfnamefont {G.}~\bibnamefont {Catelani}}, \bibinfo
		{author} {\bibfnamefont {L.}~\bibnamefont {Hu}},\ and\ \bibinfo {author}
		{\bibfnamefont {F.}~\bibnamefont {Yan}},\ }\href@noop {} {\bibinfo {title}
		{Engineering superconducting qubits to reduce quasiparticles and charge
			noise}} (\bibinfo {year} {2022}),\ \Eprint {https://arxiv.org/abs/2202.01435}
	{arXiv:2202.01435 [quant-ph]} \BibitemShut {NoStop}%
	\bibitem [{\citenamefont {Liu}\ \emph {et~al.}(2022)\citenamefont {Liu},
		\citenamefont {Harrison}, \citenamefont {Patel}, \citenamefont {Wilen},
		\citenamefont {Rafferty}, \citenamefont {Shearrow}, \citenamefont {Ballard},
		\citenamefont {Iaia}, \citenamefont {Ku}, \citenamefont {Plourde},\ and\
		\citenamefont {McDermott}}]{McDermott2022}%
	\BibitemOpen
	\bibfield  {author} {\bibinfo {author} {\bibfnamefont {C.-H.}\ \bibnamefont
			{Liu}}, \bibinfo {author} {\bibfnamefont {D.~C.}\ \bibnamefont {Harrison}},
		\bibinfo {author} {\bibfnamefont {S.}~\bibnamefont {Patel}}, \bibinfo
		{author} {\bibfnamefont {C.~D.}\ \bibnamefont {Wilen}}, \bibinfo {author}
		{\bibfnamefont {O.}~\bibnamefont {Rafferty}}, \bibinfo {author}
		{\bibfnamefont {A.}~\bibnamefont {Shearrow}}, \bibinfo {author}
		{\bibfnamefont {A.}~\bibnamefont {Ballard}}, \bibinfo {author} {\bibfnamefont
			{V.}~\bibnamefont {Iaia}}, \bibinfo {author} {\bibfnamefont {J.}~\bibnamefont
			{Ku}}, \bibinfo {author} {\bibfnamefont {B.~L.~T.}\ \bibnamefont {Plourde}},\
		and\ \bibinfo {author} {\bibfnamefont {R.}~\bibnamefont {McDermott}},\
	}\href@noop {} {\bibinfo {title} {Quasiparticle poisoning of superconducting
			qubits from resonant absorption of pair-breaking photons}} (\bibinfo {year}
	{2022}),\ \Eprint {https://arxiv.org/abs/2203.06577} {arXiv:2203.06577
		[quant-ph]} \BibitemShut {NoStop}%
	\bibitem [{\citenamefont {Spiecker}\ \emph {et~al.}(2022)\citenamefont
		{Spiecker}, \citenamefont {Paluch}, \citenamefont {Drucker}, \citenamefont
		{Matityahu}, \citenamefont {Gusenkova}, \citenamefont {Gosling},
		\citenamefont {Günzler}, \citenamefont {Rieger}, \citenamefont {Takmakov},
		\citenamefont {Valenti}, \citenamefont {Winkel}, \citenamefont {Gebauer},
		\citenamefont {Sander}, \citenamefont {Catelani}, \citenamefont {Shnirman},
		\citenamefont {Ustinov}, \citenamefont {Wernsdorfer}, \citenamefont {Cohen},\
		and\ \citenamefont {Pop}}]{PopFluxonium2022}%
	\BibitemOpen
	\bibfield  {author} {\bibinfo {author} {\bibfnamefont {M.}~\bibnamefont
			{Spiecker}}, \bibinfo {author} {\bibfnamefont {P.}~\bibnamefont {Paluch}},
		\bibinfo {author} {\bibfnamefont {N.}~\bibnamefont {Drucker}}, \bibinfo
		{author} {\bibfnamefont {S.}~\bibnamefont {Matityahu}}, \bibinfo {author}
		{\bibfnamefont {D.}~\bibnamefont {Gusenkova}}, \bibinfo {author}
		{\bibfnamefont {N.}~\bibnamefont {Gosling}}, \bibinfo {author} {\bibfnamefont
			{S.}~\bibnamefont {Günzler}}, \bibinfo {author} {\bibfnamefont
			{D.}~\bibnamefont {Rieger}}, \bibinfo {author} {\bibfnamefont
			{I.}~\bibnamefont {Takmakov}}, \bibinfo {author} {\bibfnamefont
			{F.}~\bibnamefont {Valenti}}, \bibinfo {author} {\bibfnamefont
			{P.}~\bibnamefont {Winkel}}, \bibinfo {author} {\bibfnamefont
			{R.}~\bibnamefont {Gebauer}}, \bibinfo {author} {\bibfnamefont
			{O.}~\bibnamefont {Sander}}, \bibinfo {author} {\bibfnamefont
			{G.}~\bibnamefont {Catelani}}, \bibinfo {author} {\bibfnamefont
			{A.}~\bibnamefont {Shnirman}}, \bibinfo {author} {\bibfnamefont {A.~V.}\
			\bibnamefont {Ustinov}}, \bibinfo {author} {\bibfnamefont {W.}~\bibnamefont
			{Wernsdorfer}}, \bibinfo {author} {\bibfnamefont {Y.}~\bibnamefont {Cohen}},\
		and\ \bibinfo {author} {\bibfnamefont {I.~M.}\ \bibnamefont {Pop}},\
	}\href@noop {} {\bibinfo {title} {A quantum szilard engine for two-level
			systems coupled to a qubit}} (\bibinfo {year} {2022}),\ \Eprint
	{https://arxiv.org/abs/2204.00499} {arXiv:2204.00499 [quant-ph]} \BibitemShut
	{NoStop}%
	\bibitem [{\citenamefont {Riwar}\ \emph {et~al.}(2016)\citenamefont {Riwar},
		\citenamefont {Hosseinkhani}, \citenamefont {Burkhart}, \citenamefont {Gao},
		\citenamefont {Schoelkopf}, \citenamefont {Glazman},\ and\ \citenamefont
		{Catelani}}]{RiwarPRB94}%
	\BibitemOpen
	\bibfield  {author} {\bibinfo {author} {\bibfnamefont {R.-P.}\ \bibnamefont
			{Riwar}}, \bibinfo {author} {\bibfnamefont {A.}~\bibnamefont {Hosseinkhani}},
		\bibinfo {author} {\bibfnamefont {L.~D.}\ \bibnamefont {Burkhart}}, \bibinfo
		{author} {\bibfnamefont {Y.~Y.}\ \bibnamefont {Gao}}, \bibinfo {author}
		{\bibfnamefont {R.~J.}\ \bibnamefont {Schoelkopf}}, \bibinfo {author}
		{\bibfnamefont {L.~I.}\ \bibnamefont {Glazman}},\ and\ \bibinfo {author}
		{\bibfnamefont {G.}~\bibnamefont {Catelani}},\ }\bibfield  {title} {\bibinfo
		{title} {Normal-metal quasiparticle traps for superconducting qubits},\
	}\href {https://doi.org/10.1103/PhysRevB.94.104516} {\bibfield  {journal}
		{\bibinfo  {journal} {Phys. Rev. B}\ }\textbf {\bibinfo {volume} {94}},\
		\bibinfo {pages} {104516} (\bibinfo {year} {2016})}\BibitemShut {NoStop}%
	\bibitem [{Note8()}]{Note8}%
	\BibitemOpen
	\bibinfo {note} {As temperature increases, quasiparticle can escape the trap
		due to thermal activation by phonons, see~\cite {Fei2022}.}\BibitemShut
	{Stop}%
	\bibitem [{\citenamefont {Krause}\ \emph {et~al.}(2022)\citenamefont {Krause},
		\citenamefont {Dickel}, \citenamefont {Vaal}, \citenamefont {Vielmetter},
		\citenamefont {Feng}, \citenamefont {Bounds}, \citenamefont {Catelani},
		\citenamefont {Fink},\ and\ \citenamefont {Ando}}]{Krause2022}%
	\BibitemOpen
	\bibfield  {author} {\bibinfo {author} {\bibfnamefont {J.}~\bibnamefont
			{Krause}}, \bibinfo {author} {\bibfnamefont {C.}~\bibnamefont {Dickel}},
		\bibinfo {author} {\bibfnamefont {E.}~\bibnamefont {Vaal}}, \bibinfo {author}
		{\bibfnamefont {M.}~\bibnamefont {Vielmetter}}, \bibinfo {author}
		{\bibfnamefont {J.}~\bibnamefont {Feng}}, \bibinfo {author} {\bibfnamefont
			{R.}~\bibnamefont {Bounds}}, \bibinfo {author} {\bibfnamefont
			{G.}~\bibnamefont {Catelani}}, \bibinfo {author} {\bibfnamefont {J.~M.}\
			\bibnamefont {Fink}},\ and\ \bibinfo {author} {\bibfnamefont
			{Y.}~\bibnamefont {Ando}},\ }\bibfield  {title} {\bibinfo {title} {Magnetic
			field resilience of three-dimensional transmons with thin-film
			$\text{Al/AlO}_{x}/\text{Al}$ {Josephson} junctions approaching 1~{T}},\
	}\href {https://doi.org/10.1103/PhysRevApplied.17.034032} {\bibfield
		{journal} {\bibinfo  {journal} {Phys. Rev. Applied}\ }\textbf {\bibinfo
			{volume} {17}},\ \bibinfo {pages} {034032} (\bibinfo {year}
		{2022})}\BibitemShut {NoStop}%
	\bibitem [{Note9()}]{Note9}%
	\BibitemOpen
	\bibinfo {note} {In Ref.~\cite {Fedorov2020}, the excited state population
		was measured as function of flux, displaying a peak at an intermediate flux
		value. Its possible quasiparticle origin was excluded, based on the smooth
		dependence of the quasiparticle tunneling matrix elements on flux; however,
		at that time the role of gap difference was not considered. At those
		intermediate flux values, the qubit relaxation was Purcell-limited, so no
		corresponding peak in $T_1^{-1}$ could be observed.}\BibitemShut {Stop}%
	\bibitem [{Note10()}]{Note10}%
	\BibitemOpen
	\bibinfo {note} {A peak in the total parity switching rate $\Gamma
		_{10}^{eo}+\Gamma _{11}^{eo}$ has been experimentally reported in
		Ref.~\protect \rev@citealp {DiamondArxiv2022} during the completion of this
		manuscript.}\BibitemShut {Stop}%
	\bibitem [{\citenamefont {Manucharyan}\ \emph {et~al.}(2009)\citenamefont
		{Manucharyan}, \citenamefont {Koch}, \citenamefont {Glazman},\ and\
		\citenamefont {Devoret}}]{fluxonium}%
	\BibitemOpen
	\bibfield  {author} {\bibinfo {author} {\bibfnamefont {V.~E.}\ \bibnamefont
			{Manucharyan}}, \bibinfo {author} {\bibfnamefont {J.}~\bibnamefont {Koch}},
		\bibinfo {author} {\bibfnamefont {L.~I.}\ \bibnamefont {Glazman}},\ and\
		\bibinfo {author} {\bibfnamefont {M.~H.}\ \bibnamefont {Devoret}},\
	}\bibfield  {title} {\bibinfo {title} {Fluxonium: Single {Cooper}-pair
			circuit free of charge offsets},\ }\href
	{https://doi.org/10.1126/science.1175552} {\bibfield  {journal} {\bibinfo
			{journal} {Science}\ }\textbf {\bibinfo {volume} {326}},\ \bibinfo {pages}
		{113} (\bibinfo {year} {2009})}\BibitemShut {NoStop}%
	\bibitem [{\citenamefont {Diamond}\ \emph {et~al.}(2022)\citenamefont
		{Diamond}, \citenamefont {Fatemi}, \citenamefont {Hays}, \citenamefont {Nho},
		\citenamefont {Kurilovich}, \citenamefont {Connolly}, \citenamefont {Joshi},
		\citenamefont {Serniak}, \citenamefont {Frunzio}, \citenamefont {Glazman},\
		and\ \citenamefont {Devoret}}]{DiamondArxiv2022}%
	\BibitemOpen
	\bibfield  {author} {\bibinfo {author} {\bibfnamefont {S.}~\bibnamefont
			{Diamond}}, \bibinfo {author} {\bibfnamefont {V.}~\bibnamefont {Fatemi}},
		\bibinfo {author} {\bibfnamefont {M.}~\bibnamefont {Hays}}, \bibinfo {author}
		{\bibfnamefont {H.}~\bibnamefont {Nho}}, \bibinfo {author} {\bibfnamefont
			{P.~D.}\ \bibnamefont {Kurilovich}}, \bibinfo {author} {\bibfnamefont
			{T.}~\bibnamefont {Connolly}}, \bibinfo {author} {\bibfnamefont {V.~R.}\
			\bibnamefont {Joshi}}, \bibinfo {author} {\bibfnamefont {K.}~\bibnamefont
			{Serniak}}, \bibinfo {author} {\bibfnamefont {L.}~\bibnamefont {Frunzio}},
		\bibinfo {author} {\bibfnamefont {L.~I.}\ \bibnamefont {Glazman}},\ and\
		\bibinfo {author} {\bibfnamefont {M.~H.}\ \bibnamefont {Devoret}},\
	}\href@noop {} {\bibinfo {title} {Distinguishing parity-switching mechanisms
			in a superconducting qubit}} (\bibinfo {year} {2022}),\ \Eprint
	{https://arxiv.org/abs/2204.07458} {arXiv:2204.07458 [quant-ph]} \BibitemShut
	{NoStop}%
	\bibitem [{\citenamefont {Riwar}\ and\ \citenamefont
		{Catelani}(2019)}]{RiwarPRB100}%
	\BibitemOpen
	\bibfield  {author} {\bibinfo {author} {\bibfnamefont {R.-P.}\ \bibnamefont
			{Riwar}}\ and\ \bibinfo {author} {\bibfnamefont {G.}~\bibnamefont
			{Catelani}},\ }\bibfield  {title} {\bibinfo {title} {Efficient quasiparticle
			traps with low dissipation through gap engineering},\ }\href
	{https://doi.org/10.1103/PhysRevB.100.144514} {\bibfield  {journal} {\bibinfo
			{journal} {Phys. Rev. B}\ }\textbf {\bibinfo {volume} {100}},\ \bibinfo
		{pages} {144514} (\bibinfo {year} {2019})}\BibitemShut {NoStop}%
	\bibitem [{\citenamefont {Kaplan}\ \emph {et~al.}(1976)\citenamefont {Kaplan},
		\citenamefont {Chi}, \citenamefont {Langenberg}, \citenamefont {Chang},
		\citenamefont {Jafarey},\ and\ \citenamefont {Scalapino}}]{KaplanPRB14}%
	\BibitemOpen
	\bibfield  {author} {\bibinfo {author} {\bibfnamefont {S.~B.}\ \bibnamefont
			{Kaplan}}, \bibinfo {author} {\bibfnamefont {C.~C.}\ \bibnamefont {Chi}},
		\bibinfo {author} {\bibfnamefont {D.~N.}\ \bibnamefont {Langenberg}},
		\bibinfo {author} {\bibfnamefont {J.~J.}\ \bibnamefont {Chang}}, \bibinfo
		{author} {\bibfnamefont {S.}~\bibnamefont {Jafarey}},\ and\ \bibinfo {author}
		{\bibfnamefont {D.~J.}\ \bibnamefont {Scalapino}},\ }\bibfield  {title}
	{\bibinfo {title} {Quasiparticle and phonon lifetimes in superconductors},\
	}\href {https://doi.org/10.1103/PhysRevB.14.4854} {\bibfield  {journal}
		{\bibinfo  {journal} {Phys. Rev. B}\ }\textbf {\bibinfo {volume} {14}},\
		\bibinfo {pages} {4854} (\bibinfo {year} {1976})}\BibitemShut {NoStop}%
	\bibitem [{\citenamefont {Chang}\ and\ \citenamefont
		{Scalapino}(1977)}]{ChangScalapinoPRB15}%
	\BibitemOpen
	\bibfield  {author} {\bibinfo {author} {\bibfnamefont {J.-J.}\ \bibnamefont
			{Chang}}\ and\ \bibinfo {author} {\bibfnamefont {D.~J.}\ \bibnamefont
			{Scalapino}},\ }\bibfield  {title} {\bibinfo {title} {Kinetic-equation
			approach to nonequilibrium superconductivity},\ }\href
	{https://doi.org/10.1103/PhysRevB.15.2651} {\bibfield  {journal} {\bibinfo
			{journal} {Phys. Rev. B}\ }\textbf {\bibinfo {volume} {15}},\ \bibinfo
		{pages} {2651} (\bibinfo {year} {1977})}\BibitemShut {NoStop}%
\end{thebibliography}
\end{document}